\documentclass[final,1p,times]{elsarticle}

\usepackage{graphicx,psfrag}
\usepackage{times}
\usepackage{color}
\usepackage{amsfonts,amssymb,stmaryrd,latexsym,amsmath,dsfont,slashed}

\numberwithin{equation}{section}
\newcommand{\beq}{\begin{equation}}
\newcommand{\eeq}{\end{equation}}
\newcommand{\Order}{\mathcal{O}}
\newcommand{\mpp}{m_{\rm p}}
\newcommand{\mn}{m_{\rm n}}
\newcommand{\mpi}{M_{\pi}}
\newcommand{\mpii}{M_{\pi^0}}
\newcommand{\md}{m_{\rm d}}
\newcommand{\muu}{m_{\rm u}}
\newcommand{\Fpi}{F_\pi}
\newcommand{\ga}{g_{\rm A}}
\newcommand{\eps}{\epsilon}
\newcommand{\qq}{\mathbf{q}}
\newcommand{\pp}{\mathbf{p}}
\newcommand{\PP}{\mathbf{P}}
\newcommand{\kk}{\mathbf{k}}
\newcommand{\bis}{\text{--}}
\newcommand{\xip}{\xi_{\rm p}}
\newcommand{\xid}{\xi_{\rm d}}
\newcommand{\diff}{\text{d}}
\newcommand{\lbf}{\mathbf{l}}
\newcommand{\unity}{\mathds{1}}
\newcommand{\M}{\mathcal{M}}
\newcommand{\Lagr}{\mathcal{L}}
\renewcommand{\Re}{\text{Re}\,}

\journal{Nuclear Physics A}

\begin{document}

\begin{frontmatter}

\title{Precision calculation of threshold $\pi^-d$ scattering,  
$\pi N$ scattering lengths, and  the GMO sum rule}

\author[bochum,juelich,moscow]{V.\ Baru} 
\address[bochum]{Institut f\"ur Theoretische Physik II, Ruhr-Universit\"at Bochum, D--44870 Bochum, Germany}
\address[juelich]{Institut f\"{u}r Kernphysik and J\"ulich Center
        for Hadron Physics, Forschungszentrum J\"{u}lich,
        D--52425 J\"{u}lich, Germany}
\address[moscow]{Institute for Theoretical and Experimental Physics, B.\ Cheremushinskaya 25, 
        117218 Moscow, Russia}

\author[juelich,ias]{C.\ Hanhart} 
\address[ias]{Institute for Advanced Simulation,
        Forschungszentrum J\"{u}lich, D--52425 J\"{u}lich, Germany}

\author[hiskp,ohio]{M.\ Hoferichter} 
\address[hiskp]{Helmholtz-Institut f\"ur Strahlen- und
         Kernphysik and Bethe Center for Theoretical Physics,\\ Universit\"at
        Bonn,  D--53115 Bonn, Germany}

\author[hiskp]{B.\ Kubis} 

\author[juelich,ias]{A.\ Nogga} 

\author[hiskp,ohio]{D.\ R.\ Phillips
}
\address[ohio]{Institute of Nuclear and Particle Physics and Department of Physics and Astronomy, \\
        Ohio University, Athens, OH 45701, USA}

\begin{abstract}
We use chiral perturbation theory (ChPT) to calculate the $\pi^- d$ scattering
length with an accuracy of a few percent, including isospin-violating corrections in both the two- and three-body sectors. In particular, we provide the technical details of a recent letter~\cite{JOB},
where we used data on pionic deuterium and pionic hydrogen atoms to extract  the isoscalar and isovector pion--nucleon scattering lengths
$a^+$ and $a^-$. We study isospin-breaking contributions to the three-body part of $a_{\pi^-d}$ due to mass differences, isospin violation in the $\pi N$ scattering lengths, and virtual photons. This last class of effects is ostensibly infrared enhanced due to the smallness of the deuteron binding energy. However, we show that the leading virtual-photon effects that might undergo such enhancement cancel, and hence the standard ChPT counting provides a reliable estimate of isospin violation in $a_{\pi^- d}$ due to virtual photons. Finally, we discuss the validity of the Goldberger--Miyazawa--Oehme sum rule in the presence of isospin violation, and use it to determine the charged-pion--nucleon coupling constant.
\end{abstract}


\end{frontmatter}

\section{Introduction}

Hadron--hadron scattering lengths are fundamental quantities
characterizing the strong interaction, and are slowly becoming accessible to
{\it ab initio} calculations in QCD~\cite{Beane:Lattice,Torok}.
Among them, of particular
interest are pion--hadron scattering lengths: the chiral symmetry of QCD and
the Goldstone-boson nature of the pions dictate that they are
small~\cite{Weinberg66}, while their non-vanishing size is linked to fundamental
quantities like the light quark masses and condensates. For example, the combination
of two-loop chiral perturbation theory (ChPT) and Roy equations resulted in very precise predictions 
for the pion--pion scattering lengths~\cite{CGL00}
\beq
a_0^0=(0.220\pm 0.005)\mpi^{-1},\quad a_0^2 = (-0.0444\pm 0.0010)\mpi^{-1},
\eeq
which were essential to confirm the role of the quark condensate as the leading order parameter in the
spontaneous breaking of chiral symmetry~\cite{CGL01}. 

In the case of pion--nucleon scattering, chiral symmetry predicts that the isoscalar scattering length $a^+$
is suppressed compared to its isovector counterpart $a^-$. In particular, the low-energy theorem for $a^-$~\cite{Weinberg66,BKM93}
\beq
a^-=\frac{\mpi}{8\pi(1+\mpi/\mpp)\Fpi^2}+\Order(\mpi^3)\approx 80\cdot 10^{-3}\mpi^{-1}
\eeq
receives corrections only at third order in the pion mass and its prediction is numerically very close to the full result.
Meanwhile, the expansion of the isoscalar scattering length~\cite{BKM93}
\beq
\label{aplus_exp}
a^+=0+\frac{\mpi^2}{4\pi(1+\mpi/\mpp)\Fpi^2}\bigg\{-\frac{\ga^2}{4\mpp}
+2(c_2+c_3-2c_1)\bigg\}+\Order(\mpi^3)\approx 0,
\eeq
with the pion decay constant $\Fpi$, the axial charge of the nucleon $\ga$, and low-energy constants (LECs) $c_i$, stands in marked contrast: the leading order vanishes---leaving $a^+$ as a measure of the explicit breaking of chiral symmetry---and at sub-leading orders poorly determined LECs and 
huge cancellations between individual terms limit the predictive power of the expansion. Experimentally, 
lack of $\pi^0$ beams and neutron targets makes direct
pion--nucleon scattering experiments impossible in some charge channels,
complicating a measurement of $a^+$.
In the isospin limit  the $\pi^0 p$ scattering length is purely isoscalar, and corrections to the isospin limit are well-controlled for this quantity. The best hope for access to $a^+$ in the $\pi N$ sector therefore lies in precision measurements of threshold neutral-pion photoproduction~\cite{Bernstein1,Bernstein2}.  But, until the advent of such measurements, extractions of $a^+$ from $\pi N$ scattering data suffer from large uncertainties. Different phase-shift analyses 
 yield values covering a wide range from  $-10\cdot 10^{-3}\mpi^{-1}$ to $+5\cdot 10^{-3}\mpi^{-1}$~\cite{FM00}. Indeed,
 the combination of data
and theory has, until now, lacked sufficient accuracy to even establish
definitively that $a^+ \neq 0$.

A precise determination of $\pi N$ scattering lengths improves our knowledge in many areas;
two particularly important examples of this are the following. First, $a^+$ is one of several inputs to dispersive
analyses of the pion--nucleon $\sigma$-term~\cite{sigmaterm}, which measures
the explicit chiral symmetry breaking in the nucleon mass due to up and down
quark masses, and is, in turn, connected to the strangeness content of the
nucleon.   Second, $a^-$ serves as a vital
input to a determination of the pion--nucleon coupling constant via the
Goldberger--Miyazawa--Oehme (GMO) sum rule~\cite{GMO}. 
While the uncertainty
in $a^-$ is much smaller than that in $a^+$, it still contributes significantly
to the overall error bar on the sum-rule evaluation~\cite{ELT,AMS}.
This latter example is thus one of several where data on pion--nucleon scattering affects
more complicated systems like the nucleon--nucleon ($NN$) interaction, and
hence has an impact on nuclear physics.

In view of the difficulties concerning both direct experimental access and the convergence of its chiral expansion~\eqref{aplus_exp}, data on hadronic atoms have become the primary source of information on $a^+$~\cite{hadatoms}. In these systems, the strong interaction modifies the spectrum compared to pure QED by shifting the energy levels and introducing a finite width to the states. Both effects are sensitive to threshold pion--nucleon scattering.
In this way, new information on pion--nucleon scattering 
lengths has become available due to recent high-accuracy
measurements of pionic hydrogen ($\pi H$) and pionic deuterium ($\pi D$). In the case of $\pi H$, the latest experimental
results~\cite{Gottawidth} are
\beq
\eps_{1s}=(-7.120\pm 0.012)\,{\rm eV},\quad \Gamma_{1s}=(0.823\pm 0.019)\,{\rm eV},
\label{eq:piH}
\eeq 
for the (attractive) shift of the $1s$ level of $\pi H$ due to strong
interactions and its width. The shift of the ground state is related to the $\pi^-p$ scattering length $a_{\pi^- p}$, while the width gives access to the charge-exchange scattering length $a_{\pi^-p}^{\rm cex}\equiv a_{\pi^-p\to \pi^0 n}$~\cite{hadatoms}. More precisely, $\epsilon_{1s}$ is
related to $a_{\pi^- p}$ through an improved Deser formula~\cite{LR00} 
\beq
\eps_{1s}=-2\alpha^3 \mu_H^2 a_{\pi^-p}(1+K_\eps+\delta_\eps^{\rm vac}),
\label{eq:eps1s}
\eeq
where 
$\alpha=e^2/4\pi$, $\mu_H$ is the reduced mass of $\pi H$, $K_\eps=2\alpha(1-\log \alpha)\mu_Ha_{\pi^-p}$, and 
$\delta_{\eps}^{\rm vac}=2\delta \Psi_H(0)/\Psi_H(0)=0.48\%$ is the effect of 
vacuum polarization on the wave function at the origin~\cite{vac_pol}. 
The width, in turn, determines $a_{\pi^-p}^{\rm cex}$ via~\cite{zemp}
\beq
\Gamma_{1s}=4\alpha^3\mu_H^2p_1\Big(1+\frac{1}{P}\Big)\big(a_{\pi^-p}^{\rm cex}\big)^2\big(1+K_\Gamma+\delta_\eps^{\rm vac}\big),
\label{eq:Gamma1s}
\eeq
with $K_\Gamma=4\alpha(1-\log\alpha) \mu_Ha_{\pi^-p}+2\mu_H(\mpp+\mpi-\mn-\mpii)(a_{\pi^0n})^2$, 
$\mpp$, $\mn$, $\mpi$, and $\mpii$ the masses of proton, neutron, charged and neutral pions, respectively,
$p_1$ the momentum of the outgoing $n \pi^0$ pair, and the Panofsky ratio~\cite{Panofsky}
\beq
P=\frac{\sigma(\pi^- p\rightarrow \pi^0 n )}{\sigma(\pi^- p\rightarrow n \gamma )}=1.546\pm 0.009.
\eeq 
Similarly, the (repulsive) strong shift $\eps_{1s}^D$ of the $1s$ level of $\pi D$ yields the real part of the $\pi^-d$ scattering length $\Re a_{\pi^-d}$ via~\cite{mrr1}
\beq
\eps_{1s}^{D}=-2\alpha^3 \mu_D^2 \Re a_{\pi^-d}(1+K_{\eps^D}+\delta_{\eps^D}^{\rm vac})\label{pid_Deser},
\eeq
where $\mu_D$ is the reduced mass of $\pi D$, $K_{\eps^D}=2\alpha(1-\log \alpha)\mu_D\Re a_{\pi^-d}$, and $\delta_{\eps^D}^{\rm vac}=2\delta \Psi_D(0)/\Psi_D(0)=0.51\%$~\cite{vac_pol}.

In the isospin limit, the level shift of $\pi H$ is sensitive to $a^++a^-$, whereas the width is solely determined by $a^-$. In this way, data from $\pi H$ alone permit, in principle, an extraction of the $\pi N$ scattering lengths. 
However, the chiral suppression of $a^+$ makes it very sensitive to isospin-violating corrections (see Sect.~\ref{sec:IV_scatt}), such that
additional experimental information---in particular from isoscalar nuclei as they provide better access to $a^+$---are essential to check the systematics and potentially improve the accuracy of the scattering-length determination. To this end, we split the $\pi^-d$ scattering length into its two-($\pi N$) and three-($\pi NN$) body contributions
\beq
\Re a_{\pi^- d}= a_{\pi^- d}^{(2)}+ a_{\pi^- d}^{(3)},
\label{a23}
\eeq
where the former is related to $a^+$ via
\beq
\quad a_{\pi^- d}^{(2)}=\frac{2\xip}{\xid}(\tilde a^++\Delta\tilde a^+).
\label{a2}
\eeq 
Here the difference between $a^+$ and $\tilde a^+$ as well as $\Delta\tilde a^+$ are determined by isospin-violating corrections (Sect.~\ref{sec:IV_scatt}) and
\beq
\xip=1+\frac{\mpi}{\mpp},\quad \xid=1+\frac{\mpi}{\md},
\eeq
with the deuteron mass $\md$. Once isospin breaking in the two-body sector is under control, we 
therefore have to develop a theoretical description of $a_{\pi^- d}^{(3)}$ that finally allows one to exploit information on $\pi D$ at the same level of accuracy as in $\pi H$, which requires that we can compute $a_{\pi^- d}^{(3)}$ to an accuracy of better than $10\,\%$. As we shall discuss in this work, this proves to be possible, and a combined analysis of the data~\eqref{eq:piH} on $\pi H$ and the recently remeasured level shift in $\pi D$~\cite{Gottapid}
\beq
\eps_{1s}^D=(2.356\pm 0.031)\,{\rm eV}
\eeq
then yields the determination of $a^+$ and $a^-$ of unprecedented accuracy in~\cite{JOB}.  (The width of $\pi D$ is governed by $\pi^-d\to nn$ ($\text{BR}=73.9\,\%$) and $\pi^-d\to nn\gamma$ ($\text{BR}=26.1\,\%$)~\cite{pid_channel}, such that no additional information on threshold $\pi N$ physics is provided.) The main purpose of this paper is to provide the details of the calculation of the three-body part of $a_{\pi^-d}$, which we decompose as
\beq
a_{\pi^- d}^{(3)}=a^{\rm str}+a^{{\rm disp}+\Delta}+a^{\rm EM},
\label{eq:apid3}
\eeq
where 
$a^{{\rm disp} + \Delta}$ involves two-nucleon or $\Delta$-isobar intermediate states, $a^{\rm EM}$ represents virtual-photon
corrections, and $a^{\rm str}$ denotes ``strong'' diagrams, i.e.~essentially all other contributions in the chiral expansion (the definition of each class of diagrams can be found in Sects.~\ref{sec:strong}\bis\ref{dispdel}).

The paper is organized as follows: we first briefly review isospin-violating corrections to the $\pi N$ scattering lengths in Sect.~\ref{sec:IV_scatt}. Then, we summarize the hierarchy of diagrams contributing to $a_{\pi^-d}^{(3)}$ in both the isospin-conserving and the isospin-violating sector in Sect.~\ref{counting}, before discussing strong, virtual-photon, and dispersive $+\,\Delta$ contributions in detail in Sects.~\ref{sec:strong}, \ref{sec:virt}, and \ref{dispdel}. A reader not interested in the details of the calculation may skip Sects.~\ref{sec:strong}\bis\ref{dispdel} and proceed to Sect.~\ref{sec:summary}, where we summarize our main conclusions concerning three-body contributions to the $\pi^-d$ scattering length. The consequences for the $\pi N$ scattering lengths and the $\pi NN$ coupling constant are presented in Sects.~\ref{sec:results} and \ref{sec:GMO}. We conclude in Sect.~\ref{sec:conc}. Various details of the calculation are provided in the appendices.

\section{Isospin violation in the $\pi N$ scattering lengths}
\label{sec:IV_scatt}

Before turning to the calculation of $a_{\pi^- d}^{(3)}$, we review isospin-violating  corrections to the $\pi N$ scattering lengths, which provide an essential input to the present analysis. The scattering lengths in the isospin limit for all eight channels can be written in terms of $a^+$ and $a^-$ as
\begin{align}
a_{\pi^- p}&\equiv a_{\pi^- p\rightarrow \pi^- p}=a_{\pi^+ n}\equiv a_{\pi^+ n\rightarrow \pi^+ n}=a^++a^-,\notag\\
a_{\pi^+ p}&\equiv a_{\pi^+ p\rightarrow \pi^+ p}=a_{\pi^- n}\equiv a_{\pi^- n\rightarrow \pi^- n}=a^+-a^-,\notag\\
a_{\pi^- p}^{\rm cex}&\equiv a_{\pi^- p\rightarrow \pi^0 n}=a_{\pi^+ n}^{\rm cex}\equiv a_{\pi^+ n\rightarrow \pi^0 p}=-\sqrt{2}\,a^-,\notag\\
a_{\pi^0 p}&\equiv a_{\pi^0 p\rightarrow \pi^0 p}=a_{\pi^0 n}\equiv a_{\pi^0 n\rightarrow \pi^0 n}=a^+.
\end{align}
To extract $a^+$ and $a^-$ from hadronic-atom data, we need to relate the scattering lengths in particular charge channels to those in the isospin limit, i.e.~we need the corrections
\beq
\Delta a_{\pi^- p}= a_{\pi^- p}- (a^++a^-),\quad \Delta a_{\pi^- n}= a_{\pi^- n}- (a^+-a^-),\quad \Delta a_{\pi^- p}^{\rm cex}=a_{\pi^- p}^{\rm cex}+\sqrt{2}\,a^-.
\eeq 
These corrections are generated by the quark mass difference $\md-\muu$ and electromagnetic interactions. They can be calculated systematically in ChPT, and have been worked out at next-to-leading order (NLO) in the chiral expansion in~\cite{HKM,HKMlong,GR02}. 

In those works, and throughout this study, the counting $\md-\muu\sim e^2$ is used, i.e.~electromagnetic and quark-mass effects are assumed to contribute at the same order. This counting is phenomenologically rather successful. The prime example is the nucleon mass difference, to which---according to the evaluation of the Cottingham sum rule in~\cite{GL82}---the quark mass difference and electromagnetic interactions contribute $(2.1\pm 0.3)\,{\rm MeV}$ and $(-0.8\pm 0.3)\,{\rm MeV}$, respectively. (This result is consistent with recent determinations from the lattice~\cite{Beane06,Blum10} and from charge symmetry breaking in $pn\to d  \pi^0$~\cite{filin}.) A similar picture emerges from the kaon mass difference, where---depending on the assumptions about violation of Dashen's theorem~\cite{FLAG,BP97,AM04}---quark-mass effects are a factor $2$\bis$3$ larger than electromagnetic ones. It is also instructive to look at tree-level contributions to isospin violation in $\pi N$ scattering~\cite{HKM}: $a_{\pi^- p}-a_{\pi^+ n}$ and $a_{\pi^+ p}-a_{\pi^- n}$ are purely electromagnetic, $a_{\pi^0 p}-a_{\pi^0 n}$ is solely due to $\md-\muu$, while both effects are of the same size in $a_{\pi^- p}^{\rm cex}-a_{\pi^+ n}^{\rm cex}$. Similar conclusions can be drawn from tree-level isospin breaking in the $\pi K$ scattering lengths~\cite{piK_IV}, where the corrections for some channels are purely electromagnetic, for some purely quark-mass induced, and for some due to both effects, sometimes the former being a factor of $2$ larger, sometimes the latter.

First of all, the major consequence of the leading-order (LO) isospin breaking in ChPT~\cite{mrr2}
\begin{align}
\label{piN_IV_LO}
\Delta a_{\pi^- p}^{\rm LO}&=\frac{1}{4\pi\xip}\bigg\{\frac{4\Delta_\pi}{\Fpi^2}c_1-\frac{e^2}{2}(4f_1+f_2)\bigg\},\quad \Delta a_{\pi^- p}^{\rm cex\, LO}=\frac{\sqrt{2}}{4\pi\xip}\bigg\{\frac{e^2f_2}{2}+\frac{\ga^2\Delta_\pi}{4\Fpi^2\mpp}\bigg\},\notag\\
\Delta a_{\pi^- n}^{\rm LO}&=\frac{1}{4\pi\xip}\bigg\{\frac{4\Delta_\pi}{\Fpi^2}c_1-\frac{e^2}{2}(4f_1-f_2)\bigg\},\quad \Delta_\pi=\mpi^2-\mpii^2,
\end{align}
is that
it is impossible to directly extract $a^+$ from hadronic atoms. Only the combination
\beq
\tilde a^+ \equiv a^+ + \frac{1}{4\pi\xip}
\bigg\{\frac{4\Delta_\pi}{\Fpi^2}c_1-2e^2 f_1\bigg\}\label{atilde_def}
\eeq
is accessible, and $a^+$ itself cannot be obtained absent input on the LECs $c_1$ and $f_1$ from other sources (the full list of LECs relevant for the present work is given in~\ref{app:lagr}). If the standard single-nucleon-sector counting $e \sim p$ is employed, then these isospin-violating effects are actually of the same size as the piece of $\tilde a^+$ that would be present in the isospin limit. $c_1$~enters these effects because its contribution to $a^+$ is proportional to $\mpii^2$, and $f_1$ features in the electromagnetic contributions to $\mpp$ and $\mn$. Estimates of these constants will be discussed in Sect.~\ref{sec:results}. 

Since only $\tilde a^+$ can directly be extracted, it is convenient to work with
\beq
\Delta \tilde a_{\pi^-p}=a_{\pi^-p}-(\tilde a^++a^-),\quad \Delta \tilde a_{\pi^-n}=a_{\pi^-n}-(\tilde a^+-a^-)
\eeq
instead of $\Delta a_{\pi^-p}$ and $\Delta a_{\pi^-n}$.   
The results relevant in the present context may then be written as
{\allowdisplaybreaks
\begin{align}
\label{IV_piN}
\Delta \tilde a_{\pi^-p}&=\Delta \tilde a^+ + \Delta a^-, \quad \Delta \tilde a_{\pi^-n}=\Delta \tilde a^+ -\Delta a^-,\notag\\
\Delta \tilde a^+ &= \frac{1}{4\pi\xip}\bigg\{e^2\mpi\big(2g^{\rm r}_6+g^{\rm r}_8\big)-\frac{\ga^2\mpi}{32\pi\Fpi^2}\bigg(\frac{33\Delta_\pi}{4\Fpi^2} +e^2\bigg)\bigg\},\notag\\
\Delta a^- &=-\frac{e^2 f_2}{8\pi\xip} -\frac{\mpi}{4\pi\xip}
\bigg\{\frac{\Delta_\pi}{32\pi^2\Fpi^4}\left(3+\log\frac{\mpi^2}{\mu^2}\right)
+\frac{8\Delta_\pi}{\Fpi^2}d^{\rm r}_5+\frac{e^2\ga^2}{16\pi^2\Fpi^2}\left(1+4\log 2+3\log\frac{\mpi^2}{\mu^2}\right) \notag\\
&-e^2g^{\rm r}_8+\frac{10}{9}\frac{e^2}{\Fpi^2}\big(k^{\rm r}_1+k^{\rm r}_2\big)\bigg\},\notag\\
\Delta a^{\rm cex}_{\pi^-p}&=\frac{\sqrt{2}}{4\pi\xip}
\bigg\{\frac{e^2f_2}{2}+\frac{\ga^2\Delta_\pi}{4\Fpi^2\mpp}-\frac{3\mpi\Delta_\pi}{16\Fpi^2\mpp^2}-\frac{\mpi\Delta_{\rm N}}{4\Fpi^2\mpp}\big(1+2\ga^2\big)+\frac{\mpi\Delta_\pi}{8\Fpi^2\mpp^2}(1+4\mpp c_4)\notag\\
&+\frac{\mpi\Delta_\pi}{192\pi^2\Fpi^4}\left(2-7\ga^2+\big(2-5\ga^2\big)\log\frac{\mpi^2}{\mu^2}\right)+\frac{e^2\mpi}{32\pi^2\Fpi^2}\left(5+3\log\frac{\mpi^2}{\mu^2}\right)\notag\\
&+\frac{8\mpi\Delta_\pi}{\Fpi^2}d^{\rm r}_5+\frac{e^2\mpi}{2\Fpi^2}\left(\Fpi^2 g^{\rm r}_7-2k^{\rm r}_3+k^{\rm r}_4+\frac{20}{9}\big(k^{\rm r}_1+k^{\rm r}_2\big)\right)\bigg\},
\end{align}
}\noindent
where
\beq
\Delta_{\rm N}=\mn-\mpp.
\eeq
The apparent dependence on the renormalization scale $\mu$ is canceled by the scale dependence of the LECs, whose definition is briefly reviewed in~\ref{app:lagr} (for more details we refer to~\cite{GR02}).  Estimating the LECs as in~\cite{HKM} yields
\begin{align}
 \Delta \tilde a^+&=(-3.3\pm 0.3)\cdot 10^{-3}\mpi^{-1},\quad \Delta a^-=(1.4\pm 1.3)\cdot 10^{-3}\mpi^{-1},\notag\\
\Delta \tilde a_{\pi^- p}&=(-2.0\pm 1.3)\cdot 10^{-3}\mpi^{-1},\quad \Delta a^{\rm cex}_{\pi^-p}=(0.4\pm 0.9)\cdot 10^{-3}\mpi^{-1}.
\end{align}
In principle, one could also define an $\tilde a^-$ in which some of the LECs appearing in $\Delta a^-$ and $\Delta a^{\rm cex}_{\pi^-p}$ in the same way can be absorbed (namely $f_2$, $d_5^{\rm r}$, $k_1^{\rm r}$, and $k_2^{\rm r}$), similarly to the definition of $\tilde a^+$ with respect to $c_1$ and $f_1$. In this way, the constraints of $\pi H$ and $\pi D$ on $\tilde a^+$ and $\tilde a^-$ would be considered and only in the end the estimates for the LECs inserted. The advantage of this alternative procedure is that the dependence on the LECs is more transparent and correlations between the three constraints better under control. However, we have checked that the results obtained in such an approach differ only marginally from what we present here. In order to keep the discussion as simple as possible we work in terms of $\tilde a^+$, $ \Delta \tilde a^+$, $a^-$, and $\Delta a^-$.  

\section{Hierarchy of three-body operators and  Weinberg power counting}
\label{counting}

\subsection{Isospin-conserving operators}
\label{ICO}

So far no counting scheme is known that permits consistent,
realistic, and simultaneous consideration of the two- and three-body operators that contribute to 
$\pi^-d$ scattering. For example, in the original power  counting by Weinberg ~\cite{weinberg,beane98} 
the leading two-body operator (proportional to $a^+$) 
 appears formally at one order lower than the leading three-body terms  shown in Table~\ref{threebody}.
\begin{table}
\begin{center}
\begin{tabular}{llr}
Chiral order &  \hspace*{2cm}Three-body operator & Reference\\
\hline\\[-2mm]
{$\text{LO}=\Order(1)$} &
 \parbox[c]{7cm}{\includegraphics[width=\linewidth]{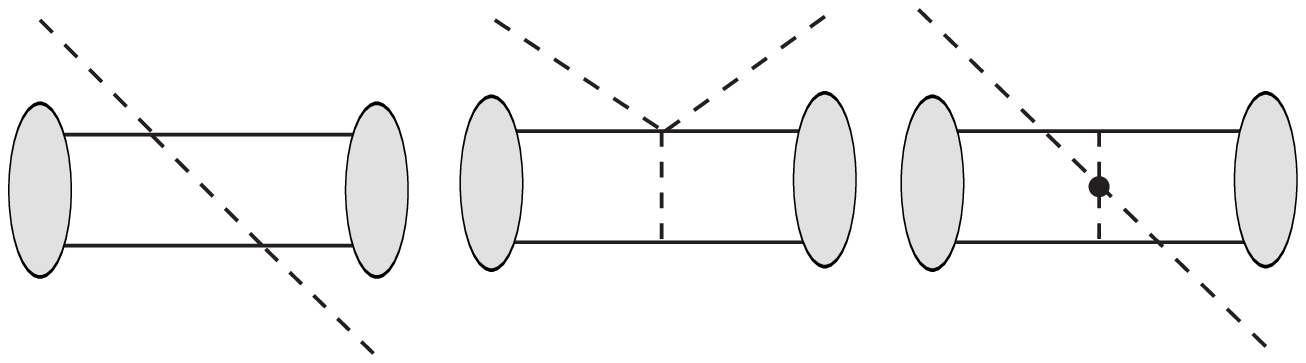}} 
& \hspace*{1.cm} {  \cite{weinberg,beane98}} \\[10mm]
\hline\\[-2mm]
{$\text{NLO}=\Order(p)$} & 
 \parbox[c]{7cm}{\includegraphics[width=\linewidth]{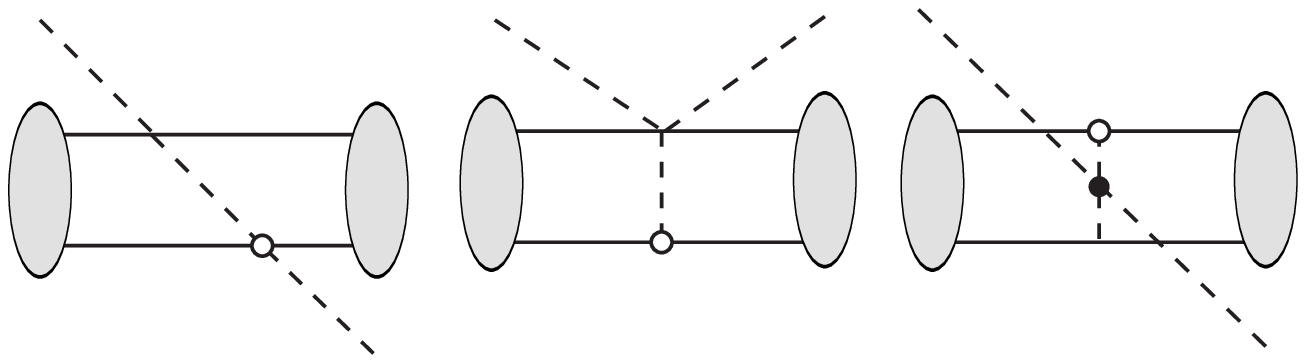}}  & 
 \cite{beane} 
\\[10mm]
&   \hspace*{2.3cm} \parbox[c]{2.3cm}{\includegraphics[width=\linewidth]{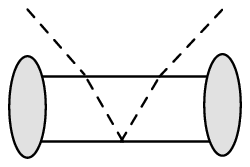}} 
& \hspace*{-1.5cm} {   \cite{beane,Liebig} 
 }\\
& & \\[-2mm]
& Effect of nucleon recoil in LO diagrams  &
\hspace*{-1.5cm} {   \cite{recoil,recoil_BER}
 }\\[2mm]
\hline  
&  &  \\[-2mm]
{$\text{N}^{3/2}\text{LO}=\Order(p^{3/2})$} &   \parbox[c]{7.cm}{\includegraphics[width=\linewidth]{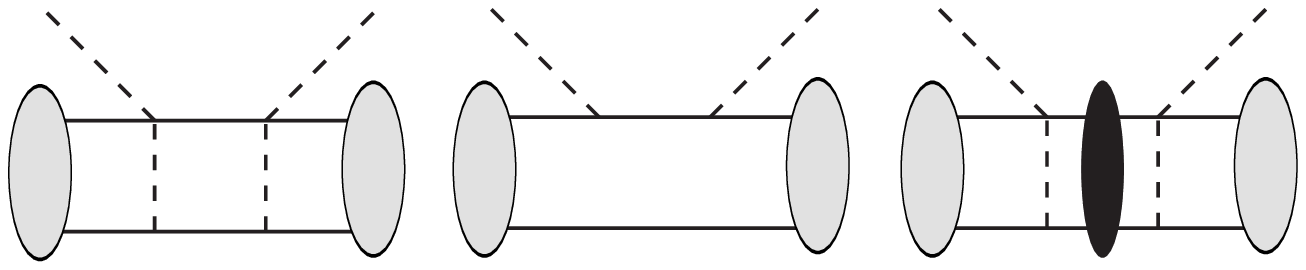}} 
&   \hspace*{-1.5cm} {    \cite{disp}
}\\[8mm]
&     \parbox[c]{7cm}{\includegraphics[width=\linewidth]{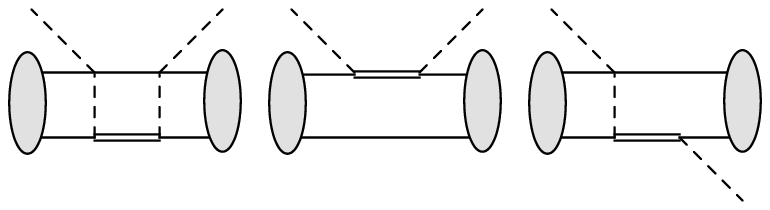}}  
&  \hspace*{-1.5cm} {  \cite{delta}
}
\\
& & \\[-2mm]
&   Effect of nucleon recoil in LO diagrams &
  \hspace*{-1.5cm} {  \cite{recoil,recoil_BER}   
}\\[2mm]
\hline\hline\\[-2mm]
{$\text{N}^2\text{LO}=\Order(p^2)$} &  \hspace*{2.3cm}\parbox[c]{2.3cm}{\includegraphics[width=\linewidth]{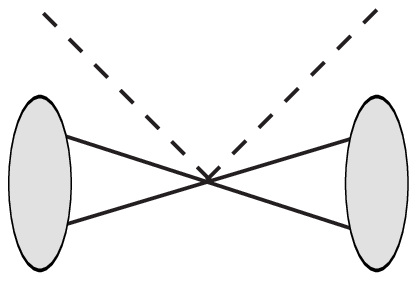}}\ 
 \parbox[c]{2cm}{\vspace*{0.5cm}\quad\Large $+\,\cdots$}  & \\
\end{tabular}
\caption{Hierarchy of isospin-conserving three-body operators 
within Weinberg power counting. Solid (open) circles correspond to
leading (sub-leading)  vertices, grey blobs indicate the deuteron wave functions,
and the black ellipse corresponds to $NN$ interactions in the intermediate state.
Solid single, solid double, and dashed lines correspond to nucleons, $\Delta(1232)$-isobars, and pions, respectively. }
 \label{threebody}
\end{center}
\end{table}
However,
it has been  known for years (see e.g.~\cite{ELT,BK} and references therein) 
that the  double-scattering diagram (the first diagram in the first row of Table~\ref{threebody})
alone is close to the experimental
value of the scattering length,  whereas the term proportional to $a^+$ is significantly smaller. 
This drawback  of the  counting scheme may bring into question the theoretical uncertainty estimate
obtained within ChPT and 
thus the reliability of a ChPT extraction of the $\pi N$ scattering lengths from pionic deuterium.
On the other hand, practical calculations demonstrate that  Weinberg's power counting  still works quite well 
once it is applied to two- and three-body operators independently---in spite of the difficulties that this power counting has in accounting for the relative size of these two classes of contribution. 
In particular, it was shown in~\cite{Liebig} that an application of the Weinberg scheme allows one to systematically account for
strong three-body contributions to $a_{\pi^- d}$  to very high accuracy. In this work we demonstrate (see Sect.~\ref{sec:virt})
that Weinberg power counting is also fully in line with the actual size of the isospin-violating three-body contributions. 
Since isospin breaking in the two-body sector is also well under control~\cite{HKM}, this permits
a precise extraction of $\tilde a^+$.

Therefore, in what follows we consider a power counting within the class of  three-body 
contributions. All such diagrams are ordered using the Weinberg scheme, with their order quoted as the predicted size in that counting relative to the leading, double-scattering term. The isospin-conserving three-body diagrams are illustrated in Table~\ref{threebody}. The table also quantifies their relative contribution according to the small expansion parameter $p$. We note that the ordering in Table~\ref{threebody} goes beyond naive dimensional analysis, since known enhancements and contributions at fractional orders in the expansion parameter are already taken into account (see below).  
The goal is to  include all three-body operators  up to one order lower than the contribution 
of the leading unknown $(N^\dagger N)^2\pi\pi$ contact term, which appears 
at  next-to-next-to-leading order (N$^2$LO).  Its contribution cannot easily
be determined from data, and is a key source of uncertainty in our result.
Given that $\Order(p) \sim \chi=\mpi/\mpp$, we anticipate an accuracy of a few percent for
threshold $\pi^- d$ scattering. 

This expectation is substantiated by the
sensitivity of our integrals to the choice of the deuteron wave function (see Sect.~\ref{subsec:numerics}). 
Convolving the operators of Table~\ref{threebody} with different wave functions derived from chiral 
and phenomenological $NN$ potentials  we find a variation in the results of about $5\,\%$: an
independent estimate of the contact term's effect. 
The explicit cutoff dependence of the individual diagrams at LO and NLO has been 
extensively discussed in the literature~\cite{Platter,PavonValderrama,Nogga,Liebig}
and it is now well established that, for deuteron wave functions based on the one-pion-exchange
interaction, the results become cutoff-independent in the limit of a large cutoff.

The hierarchy of three-body operators contributing up to N$^2$LO is shown in Table~\ref{threebody}. 
In addition to the double-scattering diagram there 
are also LO diagrams involving $3\pi N^\dagger N$ and $4\pi$ vertices, which individually depend on the parametrization of the pion fields, while only the sum is parametrization independent. The effect of these diagrams
 is numerically irrelevant~\cite{weinberg,beane}. The reason for that 
 was understood in~\cite{Liebig} to be due to an accidental 
suppression of the spin-isospin matrix element, which appears to be  more than one order of magnitude smaller 
in these diagrams than that in double scattering, although the momenta in the diagrams are in line with Weinberg power counting.
  
The operators at NLO involve sub-leading vertices and were shown to cancel amongst themselves in~\cite{beane}. 
In addition, at NLO there is a triple-scattering term that requires some care. The problem is that the actual size of this 
diagram is enhanced as compared to the estimate based on dimensional analysis, which predicts that this diagram 
contributes only at N$^2$LO.  Specifically, it was shown  in~\cite{Liebig} that the long-ranged (infrared) part 
of this diagram is enhanced numerically by a factor of $\pi^2$ whereas the rest behaves in accord with Weinberg counting.
The origin of this enhancement was associated in~\cite{Liebig} with the special topology of the diagram  consisting of two 
consecutive pion exchanges with Coulombic-type pion propagators. 
It is interesting to note that enhancements by factors of $\pi$ were already observed to emerge
also in pion-loop contributions to the $NN$ potential~\cite{Friar:2003yv}, the scalar nucleon form factor~\cite{Becher:1999he},
the $\pi^0$ photoproduction amplitude~\cite{Bernard:1991rt}, and even isospin violation in $\pi N$ scattering~\cite{GR02,HKM} itself, cf.~\eqref{IV_piN}, from similar topologies as those discussed here.
A deeper understanding of when these dimensionless factors appear would be very desirable. 
For $\pi^-d$ scattering this sort of numerical (not parametric) 
enhancement may also appear in topologically analogous diagrams 
that belong to the so-called multiple-scattering series. This leads to concerns regarding the quadruple-scattering term: although it formally appears only at N$^4$LO,
which is far beyond the edge of the theoretical accuracy, a potential numerical enhancement needs to be carefully studied since 
it may affect the uncertainty estimate.
In fact, the whole class of multiple-scattering diagrams can be summed up to all orders, and we find the effect from 
quadruple-scattering and higher diagrams to be negligible, see Sect.~\ref{mult} for more details. 
Thus, this class of potentially dangerous diagrams does not affect our uncertainty estimate.

In addition, starting from NLO, nucleon-recoil effects to the leading double-scattering operator 
have to be taken into account. The nucleon recoil has been widely studied in the literature both phenomenologically, 
see e.g.~\cite{kolyb,Faeldt}, and within effective field theory (EFT)~\cite{recoil,recoil_BER,gammad}. 
In~\cite{recoil_BER} it was demonstrated how one can calculate 
the recoil effect to all orders in a systematic expansion within EFT. 
At leading order nucleons are considered as infinitely heavy, such that the pion is scattered
off two fixed centers. At NLO the nucleon kinetic energies enter and the static pion propagator 
needs to be replaced by the full propagator corresponding to the three-body $\pi NN$ intermediate state. In the regime where all momenta
in the diagram are of order of $M_{\pi}$  the nucleon-recoil effect is purely perturbative, and thus it can be calculated
by expanding the nucleon kinetic energies using standard   heavy-baryon techniques. In this regime the recoil effect contributes
at integer powers in the expansion, i.e.~at $\Order(p)$, $\Order(p^2)$, etc., relative to the leading, double-scattering effect. In addition, there is also a non-perturbative regime 
in which the three-body propagator goes to zero---the regime of the three-body singularity. In this regime the pion kinetic energy 
is of the order of the nucleon recoil, such that pion momenta appear to be suppressed by $\sqrt{M_{\pi}/\mpp}$~\footnote{Here and below we identify the nucleon mass with the mass of the proton, apart from instances where we explicitly discuss the impact of the nucleon mass difference $\Delta_{\rm N}$ on our results.}
compared to the characteristic momenta in the deuteron.
Thus, the expansion of the double-scattering diagram contains half-integer powers of ${M_{\pi}/\mpp}$ due to the recoil effect.  
Note that the potentially largest isovector recoil correction at order 
$\Order(p^{1/2})$ fully determined by the small scales
 vanishes exactly as a consequence of the Pauli principle, which prohibits 
the $NN$ intermediate state to be in an $S$-wave in this case~\cite{recoil,recoil_BER}. 
Furthermore, it was demonstrated in~\cite{recoil_BER} that the recoil effect for 
$\pi^-d$ scattering is relevant only at orders  $\Order(p)$ and  $\Order(p^{3/2})$, whereas the contributions at higher 
orders are already negligible. This can be considered 
further support for the Weinberg counting scheme and our resulting uncertainty estimate.

In order to achieve the desired accuracy we must, in addition, compute the dispersive corrections as well as the contributions that emerge due to 
the explicit treatment of the $\Delta(1232)$ resonance~\cite{disp,delta}, which enter at order $\Order(p^{3/2})$, 
see Sect.~\ref{dispdel} for more details. Both effects involve new scales. 
The dispersive corrections due to the process $\pi d\to NN\to \pi d$  are linked closely 
to pion production in $NN$ collisions, where, as a consequence of  the large   momentum between the $NN$  pair, the expansion parameter is   
$\sqrt{M_{\pi}/\mpp}$.  
The $\Delta$-resonance in the $\Delta$-less theory is hidden in the low-energy constants $c_i$ of $\pi N$ scattering
and it  contributes to $\pi^-d$ scattering through the so-called boost (Fermi motion) 
correction~\cite{beane,Liebig}.\footnote{Note that
the LEC $c_1$ and the linear combination of LECs $c_2+c_3$ contribute to 
 the $\pi N$ scattering length $a^+$  and through that also to $\pi^-d$ scattering. However, neither  $c_1$
nor $c_2+c_3$ are affected by the $\Delta$-isobar up to order $\Order(p^2)$~\cite{BKM_NPA615}, 
although the values of $c_2$ and $c_3$ individually are strongly saturated by the $\Delta$ and thus change significantly when considering the $\Delta$ as an explicit degree of freedom~\cite{BKM_NPA615,epelbaum}.}  
This two-body correction was shown to be significant  but strongly model-dependent~\cite{beane}, although in a 
more refined treatment~\cite{Liebig} the model dependence was shown to be smaller. 
In any case, the $\Delta$--nucleon mass difference is just about twice as large as the pion
mass and it is profitable to  include the $\Delta(1232)$ dynamically and in this way increase the breakdown scale of the theory. 
When this is done  the value of the relevant LEC  $c_2$ is reduced by almost an order of magnitude. Therefore the inclusion of the $\Delta$ as an explicit  degree of freedom  
allows one to reduce the model dependence and  to achieve a faster converging series through the explicit 
calculation of a certain class of  diagrams~\cite{delta,Liebig}. The residual 
boost correction is negligible in the $\Delta$-full theory~\cite{Liebig}.

\subsection{Isospin-violating operators}
\label{IVO}

\begin{sloppypar}
Two-body isospin-violating corrections were already discussed in Sect.~\ref{sec:IV_scatt}. 
Here we will follow the logic of the previous subsection and apply the Weinberg power counting to 
the isospin-violating three-body corrections. Thus, we are going to discuss 
the hierarchy of isospin-violating three-body operators relative to each other, and their relative
suppression compared to the
leading-order isospin-conserving operators. This is sufficient to perform
a high-accuracy  calculation of the $\pi^- d$ scattering length.
\end{sloppypar} 

\begin{table}
\begin{center}
\begin{tabular}{llr}
Chiral order &  \hspace*{2cm}Three-body operator & \hspace*{-1cm}Reference\\
\hline\\[-2mm]
{$\text{LO}=\Order(e^2)$} & \hspace{-1.2cm}
\parbox[c]{9.3cm}{\includegraphics[width=\linewidth]{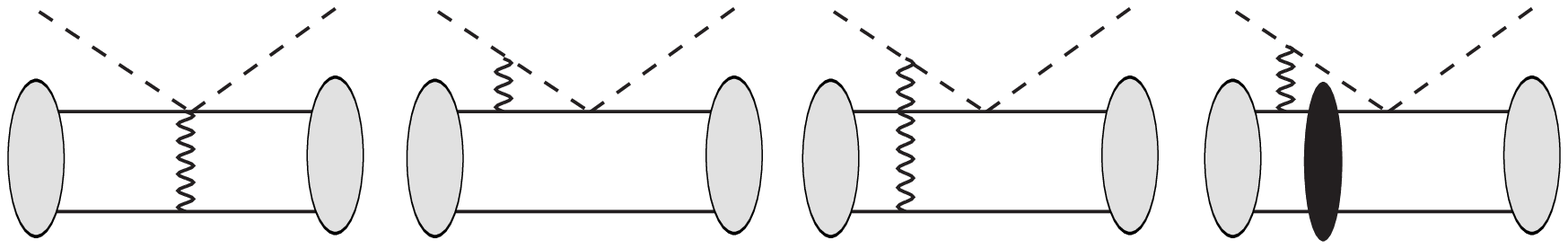}} &\\[8mm]
& \parbox[c]{7cm}{\includegraphics[width=\linewidth]{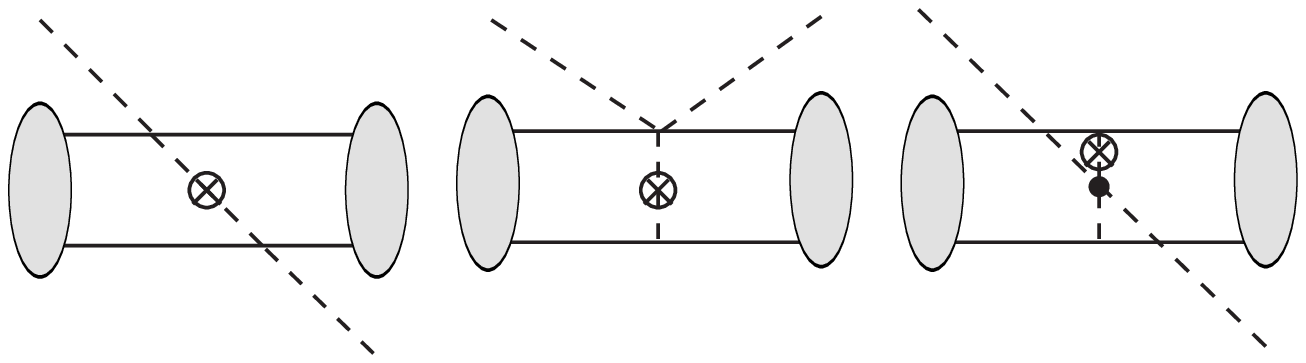}} & \\[10mm]
& \hspace{2.4cm}\parbox[c]{2.3cm}{\includegraphics[width=\linewidth]{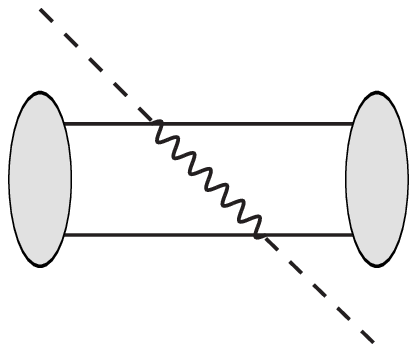}} & \hspace*{-5mm}\cite{disp}
\\[10mm]
\hline\\[-2mm]
{$\text{N}^{1/2}\text{LO}= \Order(e^2p^{1/2})$} & \hspace*{2.35cm} \parbox[c]{2.3cm}{\includegraphics[width=\linewidth]{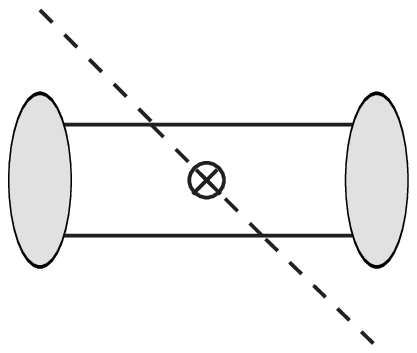}} &
\\[10mm]
\hline\hline\\[-2mm]
{$\text{NLO}= \Order(e^2p)$} & \hspace*{1cm} \parbox[c]{4.6cm}{\includegraphics[width=\linewidth]{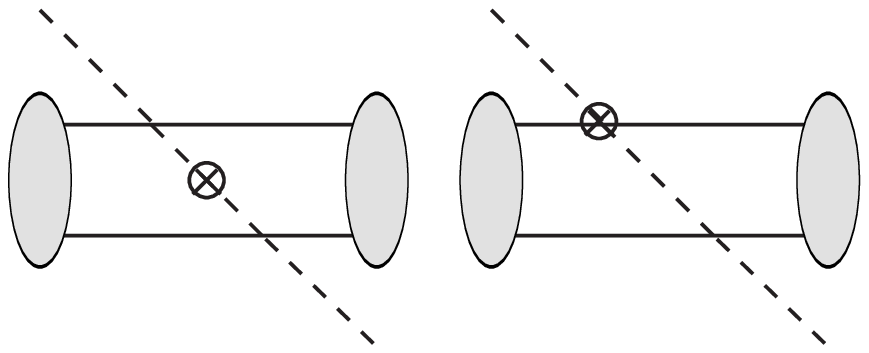}}  \ 
 \parbox[c]{2cm}{\quad\Large $+\,\cdots$} & \\[10mm]
\hline\\[-2mm]
{$\text{N}^{2}\text{LO}= \Order(e^2p^2)$} &  \hspace{-0.3cm} \parbox[c]{7cm}{\includegraphics[width=\linewidth]{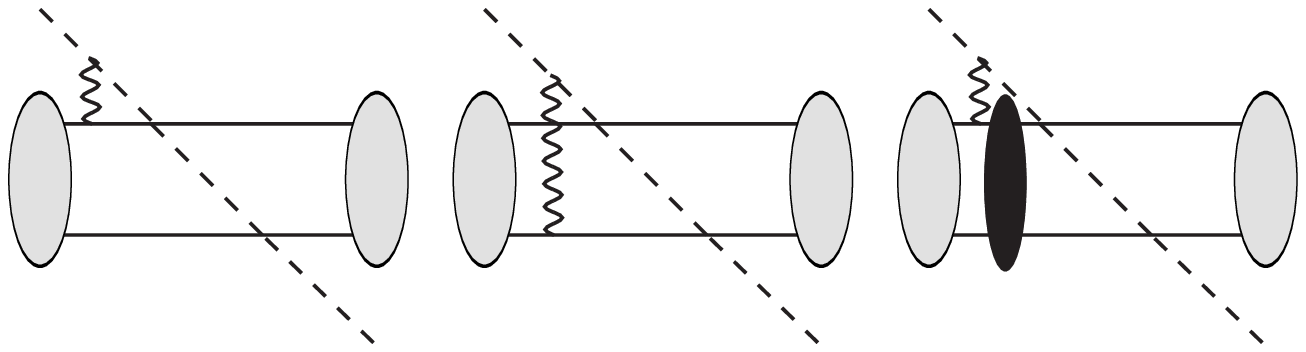}}\ \parbox[c]{2cm}{\quad\Large $+\,\cdots$}
   &\\
\end{tabular}
\caption{Hierarchy of isospin-violating three-body operators 
within Weinberg power counting. The suppression of these operators is given with respect to the 
isospin-conserving diagrams at LO (cf.~Table~\ref{threebody}).
Isospin violation appears either due to the inclusion of virtual 
photons or due to mass differences and electromagnetic effects marked by crossed circles. Note that diagrams with intermediate $NN\gamma$ states were already considered in~\cite{disp}. The contribution of these diagrams is included in the result for the dispersive corrections which we will adopt from~\cite{disp}.
}
  \label{threebodyIV}
\end{center}
\end{table}

\begin{sloppypar}
There are three classes of isospin-violating three-body contributions:  
first, isospin-breaking corrections appear in $\pi NN$ propagators due to pion and nucleon mass 
differences; second, isospin-violating  $\pi N$ interactions can occur in the diagrams introduced in  
Sect.~\ref{ICO}. The operators corresponding to these two classes of isospin-violating mechanisms are marked as crossed circles in Table~\ref{threebodyIV}.
Since isospin violation in hadron masses and $\pi N$ interactions can occur due to both electromagnetic and quark-mass effects both these classes could be either $\sim \md - \muu$ or $\sim e^2$. The third class is a purely electromagnetic effect: a new set of diagrams involving (low-energy) virtual photons (see Table~\ref{threebodyIV} and 
Sect.~\ref{sec:virt} for more details).
\end{sloppypar}

At leading order in isospin violation diagrams of this third class that involve a virtual photon and one insertion of the isospin-conserving $\pi N$ vertex occur. These are represented by the first 
row of diagrams in Table~\ref{threebodyIV}.
At the same order effects due to the pion mass difference in the leading-order
 isospin-conserving diagrams enter (the second row of diagrams in Table~\ref{threebodyIV}). These effects can be computed by working on the particle basis for the pion intermediate states in the leading-order three-body diagrams (see first row of Table~\ref{threebody}), and  explicitly keeping track of charged and neutral pion masses there. However, when this is done the double-scattering diagram must be treated in a special way, because its $\pi NN$ intermediate state, and associated three-body cut, means that the pion-mass-difference effect in this graph generates contributions not just at $\Order(e^2)$, but also at order $\Order(e^2 p^{1/2})$, $\Order(e^2 p)$, etc. The double-scattering graph with one insertion of the pion mass difference is therefore shown in Table~\ref{threebodyIV} at LO, N$^{1/2}$LO, and NLO. The piece of this graph which is LO in isospin violation is suppressed by $e^2 \Fpi^2/\mpi^2$ compared
to the corresponding  isospin-conserving diagrams at LO, as are the other diagrams listed in the first two rows of Table~\ref{threebodyIV}. Diagrams involving $NN\gamma$ intermediate states are also of this size (see third line of Table~\ref{threebodyIV}), but these are included in the calculation of the dispersive corrections in~\cite{disp}, and so will be accounted for in Sect.~\ref{dispdel}.

Next in importance are the non-analytic terms which result from the inclusion of the pion mass difference in the $\pi NN$ propagator of the double-scattering diagram. These yield the N$^{1/2}$LO contribution of Table~\ref{threebodyIV}, in full analogy with the effect of nucleon recoil in the isospin-conserving case.

\begin{sloppypar}
The operators at NLO are 
 suppressed by $\Order(e^2p)$ compared to the three-body isospin-conserving 
operators at LO, and, given the smallness of the expansion parameter,  
are irrelevant for our present purposes. (Furthermore, for full consistency with the power counting,
the inclusion of NLO corrections to the three-body isospin-violating operators would require the calculation of N$^2$LO isospin-violating two-body corrections, something that has not yet been achieved.) Therefore, for this study, it is necessary only to calculate all isospin-violating corrections to the $\pi^- d$ scattering length up to N$^{1/2}$LO. 
\end{sloppypar}

However, due to the appearance of new scales in the three-body problem,
there might be some higher-order operators that are enhanced, even though 
formally they only appear beyond $\Order(e^2 p^{1/2})$. We identify and investigate these contributions explicitly:
 \begin{itemize}
\item[a)] To account 
for all effects related to the three-body cut in the double-scattering diagram we keep all 
terms proportional to the pion and nucleon mass differences in the $\pi NN$ propagator unexpanded 
(see Sect.~\ref{sec:strong}). In particular, we include the nucleon mass difference in the propagators
to have the $\pi^0 nn$ and $\pi^- pn$  thresholds
at the proper places, although this is formally an NLO effect.\footnote{At first order in isospin breaking we have $\Delta_{\rm N}=-4B c_5(\md-\muu)+f_2e^2\Fpi^2$ and $\Delta_\pi=2Ze^2\Fpi^2$, with $c_5\sim f_2 \sim 1/\mpp$ and $Z = \Order(1)$. These quantities then enter the $\pi NN$ propagator in the combination $\rho=2\mpi\Delta_{\rm N}-\Delta_\pi$, cf.~\eqref{integrals}.  Assuming that the electromagnetic and quark-mass-induced  contributions to $\Delta_{\rm N}$ are of the same size, one finds $\mpi\Delta_{\rm N}/ \Delta_\pi \sim \mpi/\mpp \sim \Order(p)$. Therefore the nucleon-mass-difference contribution to $\rho$ is  suppressed by one chiral order compared to that coming from $\Delta_\pi$. After accounting for the modification of the chiral counting due to the presence of the $\pi NN$ cut we find that $\Delta_N$ contributes to $a^{\rm cut}$ at $\Order(e^2 p)$, whereas $\Delta_\pi$ affects the scattering length already at $\Order(e^2)$.}
\item[b)] Due to the large size of the  double-scattering diagram in the isospin-conserving case 
we include isospin violation in the $\pi N$ vertices in this diagram. This effect also starts at NLO.
\item[c)] We study certain virtual-photon corrections to the double-scattering process (formally appearing at N$^2$LO). 
 The presence of virtual photons enhances 
the region of small momenta in these diagrams, such that the integrals become
infrared divergent in the limit of
vanishing deuteron binding energy. 
The finite binding energy of the deuteron renders these 
diagrams finite, but the resulting contribution  is  potentially enhanced. In view of the fact that double scattering is numerically by far the dominant contribution to $\pi^-d$ scattering, these virtual-photon corrections could become relevant for our study.
This effect is discussed in detail in Sect.~\ref{sec:virt}.
\end{itemize}

Note that the isospin-violating three-body mechanisms up to N$^{1/2}$LO are of purely electromagnetic origin, while isospin violation due to the quark mass difference  appears only in higher-order corrections, e.g.~a) and b). 
Recall that isospin violation in the pion mass difference is predominantly an electromagnetic, not a quark-mass effect. There is no piece of this quantity $\sim \md - \muu$ at LO in isospin violation. Nucleon-mass-difference effects enter contribution a) only at $\Order(e^2 p)$. In the case of b), the isospin-violating
$\pi N$ interactions which appear there include terms proportional to the quark mass difference---as reviewed in Sect.~\ref{sec:IV_scatt}. But this whole class of diagrams involving isospin-breaking pion--nucleon vertices does not start until NLO.

We demonstrate in the subsequent sections that it is indeed the case that the additional corrections a)--c) beyond N$^{1/2}$LO 
are significantly smaller than the estimated theoretical uncertainty of the full analysis.  
The explicit calculation of these 
higher-order corrections provides an additional test of our counting scheme and uncertainty estimate.

\section{Strong contributions to $\pi^-d$ scattering}
\label{sec:strong}

\subsection{Double scattering}
\label{sec:double_scat}

\begin{figure}
\centering
\includegraphics[width=0.485\linewidth]{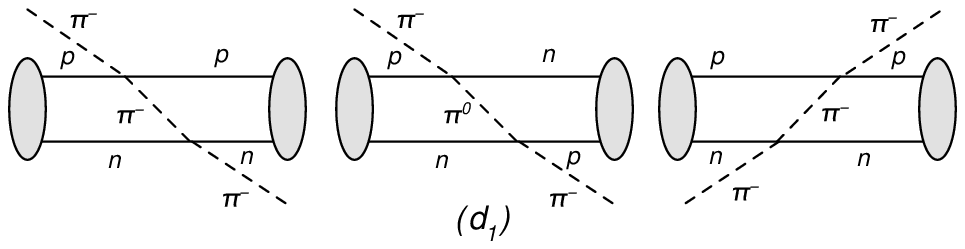}\hfill
\includegraphics[width=0.485\linewidth]{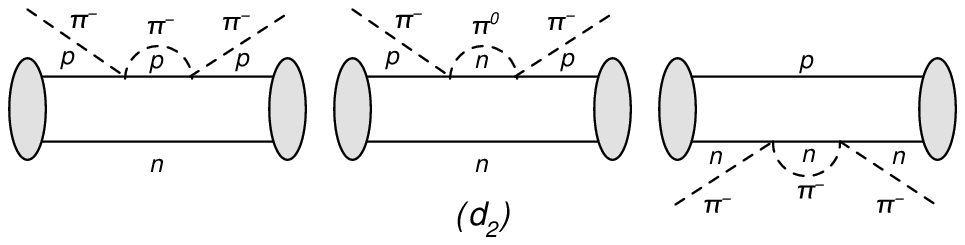}
\caption{Double-scattering contributions to $\pi^-d$ scattering.}
\label{fig:d1d2}
\end{figure}

We start our discussion with the double-scattering diagrams $(d_1)$ and $(d_2)$ (cf.~Fig.~\ref{fig:d1d2}). 
The diagram $(d_1)$  appears already at LO and  contributes to all higher  orders due to the effect of nucleon recoil,
 as was discussed in the previous section. 
The diagram $(d_2)$, however, gives rise to a three-body contribution  only if   nucleon recoil is included. 
In the limit of static (infinitely heavy) nucleons the pion loops in diagram  $(d_2)$ 
are already subsumed in the $\pi N$ scattering lengths, since in this limit $(d_2)$ is nothing but an ordinary loop correction in the chiral expansion of the $\pi N$ scattering lengths. 
In this way, the contribution to $\pi^-d$
scattering from the part of $(d_2)$ corresponding to static nucleons  is always included in the two-body term $a_{\pi^-d}^{(2)}$  proportional to $a^+$, 
see~\eqref{a23} and \eqref{a2}.  
Thus, to obtain an additional three-body correction
we need to investigate the effect of embedding the $\pi N$ amplitude into the $\pi NN$ system. 
In this way we explicitly see one of the recoil effects:
in order to treat the three-body dynamics properly we must 
replace the contribution of the two-body ($\pi N$) cut by that of the three-body
($\pi NN$) cut. For the isospin-conserving case the procedure for this was established 
in~\cite{recoil,recoil_BER}. 
The goal of this section is to extend it to the isospin-violating case.

The isospin-violating corrections occur in the diagrams of Fig.~\ref{fig:d1d2} through the different masses 
of particles in the intermediate states 
and through isospin-violating corrections to the $\pi N$ scattering lengths. 
Note that in the  calculation of the double-scattering diagrams one can safely omit all isoscalar terms, since the term proportional to  $(\tilde a^+)^2$ is tiny  compared to  $(a^-)^2$, while the term 
proportional to the combination   $\tilde a^+ a^-$  cancels. Therefore we calculate the diagrams of Fig.~\ref{fig:d1d2} keeping
only the isovector $\pi N$ scattering amplitude, retaining the isospin-violating correction $\Delta a^-$. 

In this way, we can use the form 
of $a^{\rm cut}$ that is correct in the presence of  a pure isovector $\pi N$ interaction. (The inclusion of recoil effects in the isoscalar case is discussed in~\cite{recoil_BER}.)
Following the procedure described in~\cite{recoil}, we obtain:
{\allowdisplaybreaks
\begin{align}
 a^{(d_1)+(d_2)}&=a^{\rm static} + a^{\rm static}_{\rm NLO}+ a^{\rm cut}+ \Delta a^{(2)},\notag\\
a^{\rm static} &= - \bar a ^2 \bigg\langle\frac{1}{\qq^2}\bigg\rangle, \quad
a^{\rm static}_{\rm NLO} = \bar a^2\bigg\langle\frac{1}{\qq^2}\Big(\frac{\omega_\qq}{\omega_{\qq}+\mpp}\Big)\bigg\rangle, \notag \\
a^{\rm cut}&=  \int \diff^3\pp\,\diff^3\qq 
\big(\Psi^\dagger(\pp)-\Psi^\dagger(\pp-\qq)\big)\Psi(\pp)\notag\\ 
&\qquad\times\Bigg\{\bar a^2\bigg(\frac{1}{\qq^2+\delta}-\frac{1}{\qq^2+\tilde\delta}\bigg)
-\bar a_{\rm cex}^2 \bigg(\frac{1}{\qq^2+\delta}-\frac{1}{\qq^2+\delta+\rho}\bigg)\Bigg\}, \notag\\
\Delta a^{(2)} &= \bar a_{\rm cex}^2\int \diff^3\qq \ \bigg(\frac{1}{\qq^2+\tilde\delta}-\frac{1}{\qq^2+\tilde\delta + \rho}\bigg),\label{integrals}
\end{align}}\noindent
where
\begin{align}
\label{deltadef}
\delta_{\pp_1,\pp_2}&=2\omega_{\pp_1-\pp_2}\bigg(\epsilon+\frac{\pp_1^2+\pp_2^2}{2\mpp}\bigg),\quad \delta=\delta_{\pp,\pp-\qq},\quad \tilde\delta=\frac{\omega_\qq\qq^2}{\mpp},\quad
\rho=2\mpi \Delta_{\rm N}-\Delta_\pi,\notag\\
\omega_{\qq}&=\sqrt{\mpi^2+\qq^2},\quad
 \langle f(\qq)\rangle = \int \diff^3\pp\,\diff^3\qq\Psi^\dagger(\pp-\qq)f(\qq)\Psi(\pp),
\end{align}
and $\eps$ is the deuteron binding energy. 
The $\pi N$ scattering lengths in~\eqref{integrals} are defined as
\begin{align}
 \bar a^2 &= \frac{\xip^2}{\pi^2\xid}
\bigg((a^-+\Delta a^-)^2+\frac{1}{2}\left(a_{\pi^- p}^{\rm cex}\right)^2\bigg)=
\frac{\xip^2}{\pi^2\xid}
\Big(2(a^-)^2+2a^-\Delta a^- -\sqrt{2}\,a^-\Delta a_{\pi^- p}^{\rm cex}+\cdots\Big),\notag\\
\bar a_{\rm cex}^2 &= \frac{\xip^2}{2\pi^2\xid}\Big(a_{\pi^- p}^{\rm cex}\Big)^2=\frac{\xip^2}{\pi^2\xid}
\Big((a^-)^2-\sqrt{2}\,a^-\Delta a_{\pi^- p}^{\rm cex}+\cdots\Big),\label{bar_a}
\end{align}
where the ellipses contain higher orders in isospin violation. 
We use a normalization of the deuteron wave functions $\Psi(\pp)$ where
\beq
\int \diff^3\pp\Psi^\dagger(\pp)\Psi(\pp)=1\label{deut_norm}.
\eeq
The individual terms in~\eqref{integrals} can be interpreted as follows: $a^{\rm static}$ corresponds to $(d_1)$ evaluated with a static pion propagator, and is numerically by far the dominant contribution. Recoil corrections to the static pion propagator are incorporated in $a^{\rm static}_{\rm NLO}$, while $a^{\rm cut}$ comprises effects due to the three-body $\pi^0 nn$ and $\pi^- pn$ cuts. $\Delta a^{(2)}$ emerges as an isospin-violating correction in this rearrangement, which in the end does not constitute a true three-body effect---as indicated by the absence of the deuteron wave function. In the isospin limit  $\rho = \Delta_{\pi}=\Delta_{N}=\Delta a^-=\Delta a_{\pi^- p}^{\rm cex}=0$
 \eqref{integrals} reduces to the result derived in~\cite{recoil}.

\begin{sloppypar}
Our power counting is based on dimensional analysis assuming all integrals
scale only with~$M_\pi$. In fact, the integrals in~\eqref{integrals} involve
other scales too: $\sqrt{M_\pi \epsilon}$, $\sqrt{M_\pi/\mpp}\, M_\pi$---due to the three-body cut---, and
$\sqrt{\mpp \epsilon}$, thanks to the deuterium wave functions. 
The appearance of the first two scales becomes apparent if one realizes that in the regime of the cut the
pion kinetic energy is of the order of the nucleon recoil so that pion momenta are suppressed
in this regime.
\end{sloppypar}

At first glance, the presence of a three-body cut in the integral for $a^{\rm cut}$
makes it appear to be enhanced over its naive ChPT order by $\sqrt{\mpp/M_\pi}$~\cite{gammad}.
Indeed,  it was shown in~\cite{recoil_BER}  that the full result for the 
double-scattering diagrams can be expanded  in half-integer powers of $\chi =M_\pi/\mpp$
\beq
a^{(d1)+(d2)}=a^{\rm static}+ {{\chi}^{1/2}}\, a_1 +{\chi}\, a_2 +
{{\chi}^{3/2}}\, a_3 + \cdots,
\label{doubleexpan}
\eeq
where non-integer powers appear due to the presence of the  three-body cut. 
However, the leading non-integer recoil correction at order ${\chi}^{1/2}$ dominated by the small scales 
vanishes, because the Pauli principle and the isovector 
character of the leading $\pi N$ scattering operator ensure that the intermediate $NN$ state is projected
onto a $P$-wave~\cite{recoil}, and thus $a_1=0$.  Specifically, to account for the leading correction  at order ${\chi}^{1/2}$  
in the expression of $a^{\rm cut}$  one has to drop small pion momenta $\bf q$ with respect to $\bf p$ in the wave functions, which immediately gives zero, as per the second line of~\eqref{integrals}. In consequence the scales 
$\sqrt{M_\pi  \epsilon}$,  $\sqrt{M_\pi/\mpp}\, M_\pi$,
 and $\sqrt{\mpp \epsilon}$ do not enter 
at this order: all enhanced contributions cancel due to a subtle interplay between the 
two diagrams $(d_1)$ and $(d_2)$ that is dictated by the Pauli principle. 
The combined integral is,
as originally assumed in establishing the ChPT ordering of diagrams, then dominated by
momenta of order $M_\pi$. 
Half-integer corrections at order $\chi^{3/2}$ and above can contain the small scales 
as well, but still momenta of order of $\mpi$ will have the largest impact on $a_3$ in~\eqref{doubleexpan}.

The explicit calculation of terms of order $\chi^2$ and above in this expansion 
shows that the relevant corrections to the static term appear only at orders $\chi$ and $\chi^{3/2}$~\cite{recoil_BER}.
 The net result for the recoil correction stemming from these orders is of natural size. 
In Sect.~\ref{subsec:numerics} we present the results of explicit evaluations of the integrals in~\eqref{integrals}.

In principle, there are also contributions with $P$-wave interactions 
between nucleons in the
intermediate state. Examination of the integrand for $a_{\pi^- d}$ in this case shows that
its dominant contribution comes from pion momenta $|\qq| \sim \chi^{1/2} |\pp|$, 
where $\pp$ is the momentum of the nucleons in the intermediate state. 
Taking $|\pp| \sim \mpi$,  
we find a contribution from this graph of chiral order $\chi^{3/2}$ relative to leading.
It would appear, then, that this contribution must be calculated explicitly, since we need to compute effects at $\Order(\chi^{3/2})$ in order to achieve our accuracy goal. 

However, this $P$-wave intermediate-state-interaction graph also includes
a factor of the $NN$ amplitude, evaluated at an energy $\sim -\eps - \qq^2/2\mpi$---something which was not factored into the above assessment of its chiral order. 
The $NN$
interaction that appears here must be evaluated sub-threshold, but the energy  involved is small enough that we can still estimate its effect
using, e.g., the effective-range expansion. When this is done we find that the
$NN$ amplitude will be (perturbatively) small at the energies where it is needed for this graph, essentially because the $P$-wave phase shifts are small throughout the domain of validity of the effective-range expansion. 
Hence, we estimate that the contribution to the $\pi^- d$ scattering length of the $NN$ interaction in $P$-waves will be of the order of
\beq
\chi^{3/2} \big|\delta_{^3P_J}\big(E=\mpi^2/\mpp\big)\big| a^{\rm static},
\eeq
where $J\in\{0,1,2\}$, and, if the phase shifts $\delta_{^3P_J}$ are small, then the effective-range-expansion amplitude has approximately the same size at positive and negative energies of equal magnitude. 
The $NN$ amplitude thus produces 
a further reduction beyond the straightforward chiral-EFT counting of $\sim 0.15$, so we find that $P$-wave intermediate-state $NN$ interactions affect our final result by only about $0.2\cdot 10^{-3} \mpi^{-1}$, which is significantly below the accuracy that we seek.

\subsection{Further leading-order diagrams}
\label{secd3d4}

\begin{figure}
\centering
\includegraphics[width=0.8\linewidth]{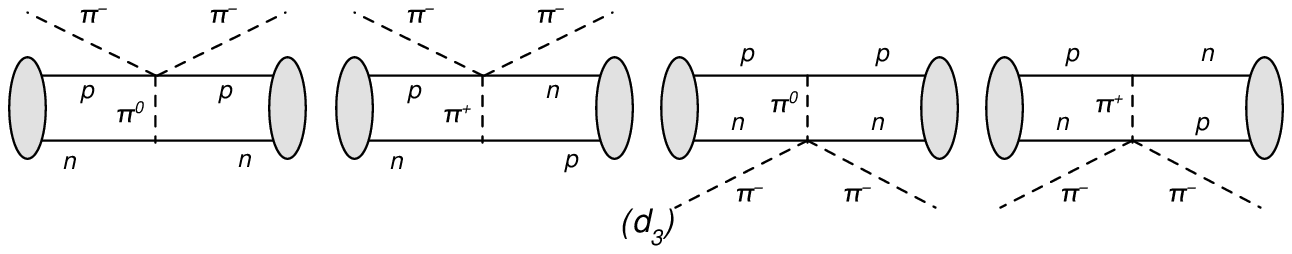}\\[2mm]
\includegraphics[width=0.4\linewidth]{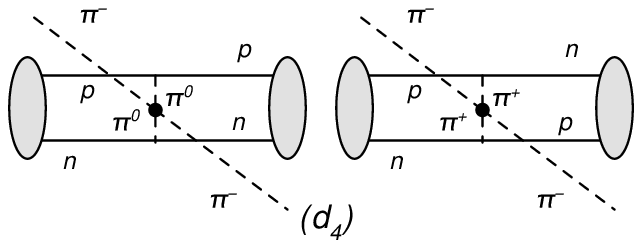}
\caption{Isospin violation in $(d_3)$ and $(d_4)$.}
\label{fig:d3d4}
\end{figure}

According to the power counting, the diagrams $(d_3)$ and $(d_4)$ (cf.~Fig.~\ref{fig:d3d4}) should be of the same order as double scattering. However, these diagrams are suppressed by accidentally small spin-isospin factors~\cite{Liebig}. As the Clebsch--Gordan coefficients are always smaller than $1$, it is guaranteed that this mechanism can only lead to a suppression and not to an enhancement, such that our accuracy estimate remains unaffected. Nevertheless, we have calculated the full isospin-violating corrections to this class of diagrams, as required by the power counting, 
\beq
a^{\pi\pi} =\frac{\ga^2\mpii^2}{128\pi^4\xid\Fpi^4}\bigg\{\bigg\langle\frac{\qq\cdot\boldsymbol{\sigma}_1\qq\cdot\boldsymbol{\sigma}_2}{(\qq^2+\mpii^2)^2}\bigg\rangle
-\frac{4\Delta_\pi}{\mpii^2}\bigg\langle\frac{\qq\cdot\boldsymbol{\sigma}_1\qq\cdot\boldsymbol{\sigma}_2}{(\qq^2+\mpi^2)^2}\bigg\rangle\bigg\}.
\eeq
We find that isospin-breaking corrections due to the pion mass difference amount to $4\Delta_\pi / \mpii^2\approx 28\,\%$. This is a large isospin-violating effect, which is, however, of little practical relevance given the overall suppression of this contribution.

As was shown in~\cite{beane}, NLO contributions to the leading-order diagrams vanish in the isospin limit. As isospin-breaking corrections to this are suppressed by another two orders, sub-leading corrections to $(d_1)$--$(d_4)$ may therefore be safely ignored.

\subsection{Triple scattering and the multiple-scattering series}
\label{mult}

Although the triple-scattering diagram  (see Table~\ref{threebody}) is formally suppressed by $p^2$ 
relative to the double-scattering operator, it was shown in~\cite{beane,Liebig} that this diagram is enhanced by a factor of $\pi^2$ over 
its power-counting estimate and hence has to be included in order to achieve the desired accuracy. Neglecting 
isoscalar contributions as well as isospin-breaking corrections, one finds
\beq
\label{atriple}
a^{\rm triple} = \frac{(\xip a^-)^3}{\pi \xid}\bigg \langle\frac{1}{|\qq|}\bigg\rangle.
\eeq
The enhancement can be traced back to the occurrence of two successive Coulombic propagators. This special topology produces a dimensionless integral that is not $\Order(1)$, as assumed in dimensional analysis, but instead yields a factor $\pi^2$. This enhancement of the ``triangle topology'' is not accounted for by Weinberg's application of ChPT counting to the irreducible $\pi NN$ graphs~\cite{Liebig}. Similar enhancements occur at higher orders in the multiple-scattering series, too, and one might worry that this spoils the convergence of the perturbative expansion. However, we find from explicit numerical evaluation that the result for the full multiple-scattering series resummed in configuration space~\cite{MRR_Kd,Kamalov,Kolybasov72,Brueckner53}
\beq
a^{\rm ms}=-\frac{4}{\xid}\Bigg\langle\frac{\frac{(\xip a^-)^2}{r}-\frac{(\xip a^-)^3}{r^2}}
{1 +  \frac{(\xip a^-)^2}{r^2} -  2 \frac{(\xi_p a^-)^3}{r^3}}\Bigg\rangle
=-\frac{4(\xip a^-)^2}{\xid}\left\langle\frac{r}{r^2+r\xip a^-+2(\xip a^-)^2}\right\rangle
\eeq
differs from the first two terms (double and triple scattering) by only $0.1\cdot 10^{-3}\mpi^{-1}$. This is 
significantly below the estimated uncertainty due to the contact term of about $1\cdot 10^{-3}\mpi^{-1}$. Hence, the 
multiple-scattering series converges sufficiently quickly that quadruple scattering and higher orders can be 
neglected in the calculation of $a_{\pi^- d}$. In particular, we stress that this result 
indicates that the enhancement of triple scattering with respect to its Weinberg-counting estimate does not lead to an enhancement of the $\pi NN$ contact operators contributing to $\pi^-d$ scattering. These still occur at $\Order(p^2)$ relative to leading order, preserving the estimates of the contact term given above.
 For a more detailed discussion of the multiple-scattering series in meson--nucleus scattering we refer to~\cite{MSS}.

\subsection{Evaluation of the matrix elements}
\label{subsec:numerics}

\begin{table}
\centering
\begin{tabular}{cccccccc}
\hline\hline
 model   &  $\Lambda / \tilde \Lambda$  &    $\langle 1/\qq^2\rangle$ &  $I^{\rm NLO}$    &   $I^{\rm cut}$   &    $\langle 1/|\qq|\rangle$    &   $a^{\pi\pi}$ & $I^{\rm EM}$\\\hline
AV 18& --- & $ 12.7$ &  $1.94$ & $-2.68$ &  $7.4$ &    $-0.00050$ & $9.92$\\	
CD-Bonn & --- & $12.8$ & $2.04$ & $-2.47$ & $8.7$ &  $-0.00037$ & $9.85$\\ 
ChPT N$^2$LO & $450/500$ & $13.0$ & $2.12$ & $-2.29$ & $9.7$ & $-0.00007$ & $9.85$\\ 
ChPT N$^2$LO & $600/500$ & $12.8$ & $2.04$ & $-2.43$ & $8.8$ & $-0.00025$ & $9.94$\\ 
ChPT N$^2$LO & $550/600$ & $12.9$ & $2.09$ & $-2.39$ & $9.5$ & $-0.00020$ & $9.91$\\
ChPT N$^2$LO & $450/700$ & $13.2$ & $2.15$ & $-2.30$ & $9.9$ & $-0.00007$ & $9.95$\\ 
ChPT N$^2$LO & $600/700$ & $12.9$ & $2.09$ & $-2.43$ & $9.4$ & $-0.00025$ & $9.85$\\ 
\hline\hline
\end{tabular}
\caption{Matrix elements for phenomenological as well as N$^2$LO chiral wave functions. The cutoffs $\Lambda/\tilde \Lambda$ are given in~MeV and specify the version of the chiral interaction as given in~\cite{NNLO}. All integrals are given in appropriate powers of~$\mpi$. $I^{\rm EM}$ is defined in Sect.~\ref{sec:numerics} and needed for diagrams involving virtual photons.}
\label{table:str_integrals}
\end{table}

Our results for the wave-function averages are shown in Table~\ref{table:str_integrals}. We give results for two different modern phenomenological interactions, AV18~\cite{AV18} and CD-Bonn~\cite{CDBonn}, and the implementations of~\cite{NNLO} of chiral interactions at order N$^2$LO. For the calculations, we used wave functions obtained from numerical solutions of the Schr\"o\-dinger equation in momentum space. In order to facilitate the calculation, we used a recently developed Monte Carlo scheme to evaluate the integrals. For details on the numerical procedure, we refer to~\cite{Liebig}. The calculation was also cross-checked with the standard method of 
Gaussian numerical integration.

As the isospin-breaking corrections to the $\pi N$ scattering lengths in~\eqref{bar_a} are relevant only for $a^{\rm static}$, to which they contribute about $1\%$, we may write the integrals for the contribution of the cut term and the NLO correction to the static $\pi NN$ propagator as 
\beq
a^{\rm cut} =\frac{2\xip^2}{\pi^2\xid}(a^-)^2 I^{\rm cut}, \quad 
a^{\rm static}_{\rm NLO} =\frac{2\xip^2}{\pi^2\xid}(a^-)^2I^{\rm NLO},
\eeq
with
\begin{align}
I^{\rm cut}&=\int \diff^3\pp\,\diff^3\qq 
\big(\Psi^\dagger(\pp)-\Psi^\dagger(\pp-\qq)\big)\Psi(\pp) 
\Bigg\{\frac{1}{\qq^2+\delta}-\frac{1}{\qq^2+\tilde\delta}
-\frac{1}{2}\bigg(\frac{1}{\qq^2+\delta}-\frac{1}{\qq^2+\delta+\rho}\bigg)\Bigg\},\notag\\
 I^{\rm NLO}&=\bigg\langle\frac{1}{\qq^2}\Big(\frac{\omega_\qq}{\omega_{\qq}+\mpp}\Big)\bigg\rangle,
\end{align}
and $\delta$, $\tilde\delta$, and $\rho$ defined in~\eqref{deltadef}.

The statistical uncertainties of the numerical results are not significant and are therefore not given in the table.  An appreciably larger uncertainty is introduced by the different short-distance ($r \ll 1/M_\pi$) physics of the $NN$ wave functions used.  To combine the results for the different deuteron wave functions, we take the average of all seven potentials as our mean value, while the uncertainty is taken to be the maximum deviation from this average. In this way, we obtain the individual contributions to $a^{\rm str}$ given in Table~\ref{table:strong} (note that $\Delta a^{(2)}$ is independent of the deuteron wave function). They produce a total
\beq  
a^{\rm str} = (-22.6\pm 1.1\pm 0.4)\cdot10^{-3}\mpi^{-1}, 
\eeq
where the first error refers to the model dependence of the matrix elements and the second to the uncertainty in the isospin-breaking shifts in the $\pi N$ scattering lengths. (Here, and in Table~\ref{table:str_integrals}, results are quoted for $a^-=86.1\cdot 10^{-3}\mpi^{-1}$. When our correlation band is obtained in Fig.~\ref{fig:bands} below, the full $a^-$ dependence of $a^{\rm str}$ is taken
into account.)

\begin{table}
\centering
\begin{tabular}{cccccc}
\hline\hline
$a^{\rm static}$ & $-24.1\pm 0.7$ & $a^{\rm static}_{\rm NLO}$ & $3.8\pm 0.2$ & $a^{\rm cut}$ & $-4.8\pm 0.5$  \\\hline
$a^{\rm triple}$  & $2.6\pm 0.5$ & $a^{\pi\pi}$ & $-0.2\pm 0.3$ & $\Delta a^{(2)}$ & $0.2$\\
\hline\hline
\end{tabular}
\caption{Strong contributions to $a^{(3)}_{\pi^- d}$ in units of $10^{-3} \mpi^{-1}$ for $a^-=86.1\cdot 10^{-3}\mpi^{-1}$.}
\label{table:strong}
\end{table}

\section{Virtual photons}
\label{sec:virt}

The improved Deser formula~\eqref{pid_Deser} is derived in an EFT that resums the effects of the photon ladder in pionic deuterium. This calculation includes effects due to virtual-photon momenta in the ``hadronic-atom regime'' $|\kk| \sim r_{\rm B}^{-1}$ 
with the Bohr radius $ r_{\rm B} = (\alpha M_\pi)^{-1}$. 
Our calculation of $a_{\pi^- d}$ should therefore include all contributions above this scale. In the ChPT counting where momenta are assumed to be of order $\mpi$ the leading contributions  due to the exchange of (Coulomb) photons  between the $\pi^-$ and the proton are given by the diagrams shown in Fig.~\ref{fig:structureless}: $(d_6)$, $(d_7)$, and $(d_8)$.  
Photon exchange is perturbative at $|\kk| \sim M_\pi$, and the pertinent pieces of these graphs enter at $\Order(e^2)$ 
relative to $(d_1)$, cf.~Table~\ref{threebodyIV}. 
Such effects in the other diagrams are of a higher ChPT order than considered here. 

However, diagrams $(d_6)$ and $(d_8)$--$(d_{10})$ are reducible in the sense originally defined by
Weinberg~\cite{weinberg}, with the $\pi NN$ intermediate state involving relative momenta $\sim \sqrt{M_\pi \epsilon} \ll M_\pi$. 
In fact, these diagrams are ``would-be infrared singular'' in the sense that in the limit 
$\epsilon\rightarrow 0$ (and for static nucleons) they involve the (singular) matrix elements $\langle 1/\qq^4\rangle$. 
This leads to a potential enhancement $\sim\sqrt{\mpi/\epsilon}$ for physical values of $\epsilon$.
Furthermore, the intermediate $NN$ pair can be in an $S$-wave, such that we must allow for the possibility of $NN$ interactions while the pion is ``in flight''. In the isoscalar partial wave the intermediate $NN$ Green's function will, in particular, include the deuteron pole. In that contribution we must separate the low-momentum part of these contributions that exactly corresponds to the combined quantum-mechanical effect of the Coulomb potential and the $\pi^- d$ scattering length accounted for in the improved Deser formula, where the infrared divergence is regulated by the presence of the hadronic-atom binding energy.  

We will now discuss all these effects in some detail. As the appearance of additional scales might require modifications of the ChPT counting rules, we will also consider the double-scattering  diagrams $(d_9)$ and $(d_{10})$, which formally enter at higher order.

\subsection{Diagrams without intermediate-state $NN$ rescattering}
\label{sec:structureless}

\begin{figure}
\centering
\includegraphics[width=0.4\linewidth]{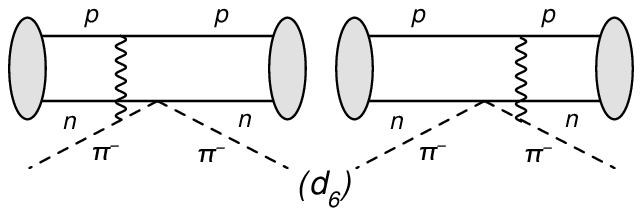}\quad
\includegraphics[width=0.2\linewidth]{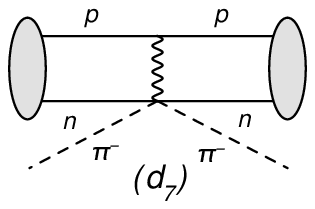}\\[2mm]
\includegraphics[width=0.4\linewidth]{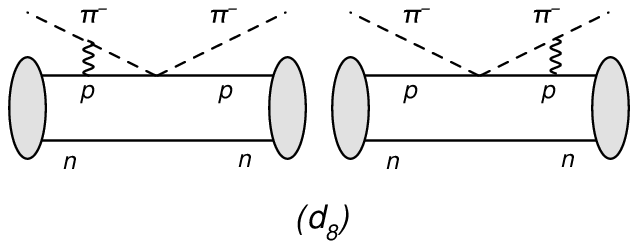}\\[2mm]
\includegraphics[width=0.8\linewidth]{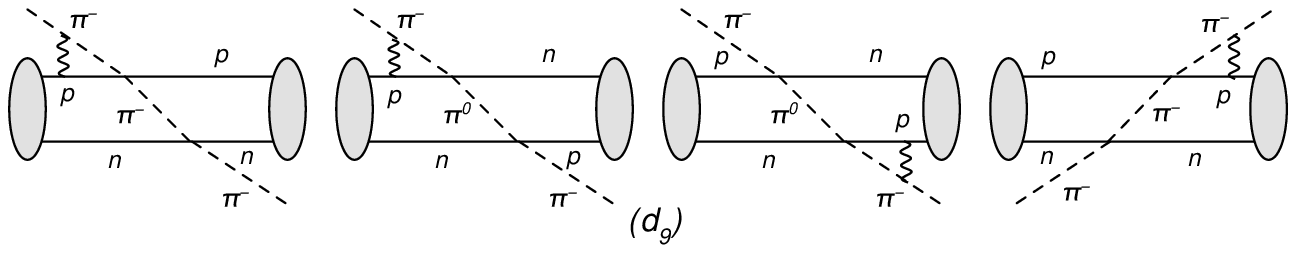}\\[2mm]
\includegraphics[width=0.4\linewidth]{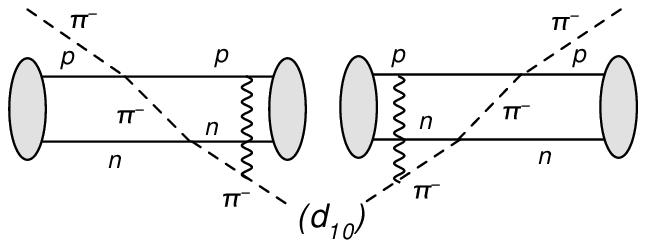}
\caption{Virtual-photon diagrams in $\pi^- d$ scattering. Note that the charges of the particles are shown explicitly to illustrate how the photon couples to the charged particles. The effect of mass differences between charged and neutral particles is not considered, since it is higher order in isospin breaking.}
\label{fig:structureless}
\end{figure}

As a first step, we consider the diagrams depicted in Fig.~\ref{fig:structureless} where there is no intermediate-state $NN$ interaction. For $(d_6)$ and $(d_8)$ we obtain
\beq
a^{(d_6)+(d_8)}=\frac{2\alpha\mpi\xip}{\pi^2\xid}\int \diff^3\mathbf{p}\,\diff^3\mathbf{q} 
\big(a_{\pi^-n}\Psi^\dagger(\pp-\qq)\Psi(\pp)+ a_{\pi^-p}\Psi^\dagger(\pp)\Psi(\pp)\big)\frac{1}{|\qq|\big(|\qq|+\delta/2\omega_\qq\big)\big(\qq^2+\delta\big)},
\label{ad6plusd8}
\eeq
where we have used time-ordered perturbation theory to include the nucleon recoil both in the photon and the pion propagator. 
It is now convenient to split this expression into isovector and isoscalar $\pi N$ interactions. Dropping isospin violation in the scattering lengths, the isospin $T=1$ part becomes
\beq
a^{(d_6)+(d_8)}_{T=1}=\frac{2\alpha\mpi\xip a^-}{\pi^2\xid}\int \diff^3\mathbf{p}\,\diff^3\mathbf{q} 
\big(\Psi^\dagger(\pp)-\Psi^\dagger(\pp-\qq)\big)\Psi(\pp)\frac{1}{|\qq|\big(|\qq|+\delta/2\omega_\qq\big)\big(\qq^2+\delta\big)}
\label{isoscalar}.
\eeq 
Similar to the double-scattering process the expansion of $(d_6)$ and $(d_8)$   in $\mpi/\mpp$ induces non-analytic terms due to nucleon recoil. However, in contradistinction to $a^{\rm cut}$ in~\eqref{integrals} the region of small momenta in~\eqref{isoscalar}
is significantly enhanced due to the presence of the photon. As a consequence, the expansion\footnote{Note that this concerns only the aforementioned expansion in $\mpi/\mpp$, whereas the scale of the individual contributions is hidden in the coefficients $b_i$. For example, $b_0\sim 1/\gamma$ is dominated by momenta $\sim\gamma$, which together with $1/\sqrt{\chi}$ from~\eqref{ad6d8_exp} produces $1/\sqrt{\mpi\eps}$ in~\eqref{ad6d8LO}.} of $(d_6)$ and $(d_8)$  
starts from $1/\sqrt{\chi}$
\beq
a^{(d_6)+(d_8)}=\frac{b_0}{\sqrt{\chi}} +b_{\rm static} +{b_1}{\sqrt{\chi}}+\cdots.
\label{ad6d8_exp}
\eeq
Indeed, at leading order in $M_\pi/\mpp$, which appears from the non-perturbative regime 
of the three-body cut ($|\qq|\sim \sqrt{M_\pi/\mpp} |\pp| \ll |\pp|$), the contributions of $(d_6)$ and $(d_8)$ 
are  equal and both involve the integral
\beq
\int \diff^3\pp\,\diff^3\qq\frac{\Psi^\dagger(\pp)\Psi(\pp)}{\qq^2\big(\qq^2+2\mpi(\epsilon+\pp^2/\mpp)\big)}=
\frac{\sqrt{2}\,\pi^2}{\sqrt{\chi}} \int  
\diff^3\pp\,\frac{\Psi^\dagger(\pp)\Psi(\pp)}{\sqrt{\pp^2+\gamma^2}}
\eeq
 with $\gamma=\sqrt{\mpp\eps}$.
If we use asymptotic deuteron wave functions
\beq
\Psi(\pp)= \frac{\sqrt{\gamma}}{\pi(\pp^2+\gamma^2)} \label{asymp}
\eeq
 to perform the integral, we find at leading order
\beq
a^{(d_8)}_{T=1}=-a^{(d_6)}_{T=1}=\frac{2\alpha\mpi\xip a^-}{\pi^2\xid}\frac{8\pi}{3\sqrt{2}}\frac{1}{\sqrt{\mpi\epsilon}}.
\label{ad6d8LO}
\eeq
In this way, we see that indeed the individual diagrams are enhanced by $\sqrt{\mpi/\epsilon}$, but these contributions exactly cancel in the sum of $(d_6)$ and $(d_8)$, such that 
$b_0=0$ in~\eqref{ad6d8_exp} (in close analogy to $a_1=0$ in~\eqref{doubleexpan}). 
The physical explanation for this cancellation is provided by the Pauli principle: as the $\pi N$ interaction does not change the spin, it implies that for the isovector case the intermediate $NN$ pair must be in a $P$-wave, which is reflected by the relative sign between the wave functions in~\eqref{isoscalar}. As $P$-wave $NN$ interactions are small (cf.~discussion in Sect.~\ref{sec:double_scat}), the only non-vanishing isovector contribution is therefore generated by the residual effects in~\eqref{isoscalar}.

The calculation of the gauged Weinberg--Tomozawa diagram $(d_7)$ including the nucleon recoil in the photon propagator proceeds along the same lines as the decomposition of $(d_1)$ and $(d_2)$. Subtracting the appropriate two-body diagram, we obtain:
\begin{align}
 a^{(d_7)}&=-\frac{\alpha }{16\pi ^3\Fpi^2\xid}\bigg\langle\frac{1}{\qq^2}\bigg\rangle+\frac{\alpha }{16\pi ^3\Fpi^2\xid}\bigg\langle\frac{1}{|\qq|(2\mpp+|\qq|)}\bigg\rangle\notag\\
&+\frac{\alpha }{16\pi ^3\Fpi^2\xid}\int \text{d}^3\pp\,\text{d}^3\qq\big(\Psi^\dagger(\pp)-\Psi^\dagger(\pp-\qq)\big)\Psi(\pp)\bigg\{\frac{1}{|\qq|(|\qq|+\delta/2\omega_\qq)}-\frac{1}{|\qq|(|\qq|+\qq^2/2\mpp)}\bigg\},
\label{d7}
\end{align}
where the second and third terms are analogs of $a^{\rm static}_{\rm NLO}$ and $a^{\rm cut}$, respectively. 
Note that when momenta are of order $M_{\pi}$ the recoil correction in the photon  propagator is, in principle, a 
higher-order effect. Indeed, 
we find that  the corrections to the static photon propagator are numerically very small, only 
about $-0.045\cdot 10^{-3}\mpi^{-1}$, and may therefore be safely neglected. 

In contrast, the Pauli principle allows for $S$-wave $NN$ interactions in the isoscalar part of $(d_6)$ and $(d_8)$. These will be discussed in Sect.~\ref{sec:single}, while the numerical results are summarized in Sect.~\ref{sec:numerics}. However, if the $T=0$ part of these diagrams were significant, then one would also be concerned about virtual-photon exchange within the more sizeable, double-scattering, diagram. For this reason,
we also give the expressions for $(d_9)$ and $(d_{10})$. Neglecting the nucleon recoil in the photon 
propagator but keeping it in the relevant $\pi NN$ intermediate state,  the result reads
\beq
 a^{(d_9)+(d_{10})}=-\frac{3\alpha\mpi(\xip a^-)^2}{\pi^4\xid}\Bigg\langle\frac{1}{\qq^2}\int\frac{\diff^3\lbf}{\lbf^2\big(\lbf^2+\delta_{\pp,\pp+\lbf}\big)}\Bigg\rangle-\frac{\alpha\mpi(\xip a^-)^2}{\pi^4\xid}\Bigg\langle\int\frac{\diff^3\lbf}{\left(\qq+\lbf\right)^2\lbf^2\big(\lbf^2+\delta_{\pp,\pp+\lbf}\big)}\Bigg\rangle,
\eeq
which can be separated into its isoscalar and isovector pieces as follows
\begin{align}
 a^{(d_9)+(d_{10})}_{T=0}&=-\frac{2\alpha\mpi(\xip a^-)^2}{\pi^4\xid}\Bigg\langle\frac{1}{\qq^2}\int\frac{\diff^3\lbf}{\lbf^2\big(\lbf^2+\delta_{\pp,\pp+\lbf}\big)}+\int\frac{\diff^3\lbf}{\left(\qq+\lbf\right)^2\lbf^2\big(\lbf^2+\delta_{\pp,\pp+\lbf}\big)}\Bigg\rangle,\notag\\
 a^{(d_9)+(d_{10})}_{T=1}&=-\frac{\alpha\mpi(\xip a^-)^2}{\pi^4\xid}\Bigg\langle\frac{1}{\qq^2}\int\frac{\diff^3\lbf}{\lbf^2\big(\lbf^2+\delta_{\pp,\pp+\lbf}\big)}-\int\frac{\diff^3\lbf}{\left(\qq+\lbf\right)^2\lbf^2\big(\lbf^2+\delta_{\pp,\pp+\lbf}\big)}\Bigg\rangle\label{d9d10_structureless}.
\end{align}
Again, the leading, potentially enhanced contribution in the isovector case cancels in accordance with the Pauli principle. The isoscalar case, including intermediate-state $NN$ interactions, will be addressed in Sect.~\ref{sec:double}.

\subsection{The role of rescattering I: single scattering with photon exchange}
\label{sec:single}

\begin{figure}
\centering
\includegraphics[width=0.7\linewidth]{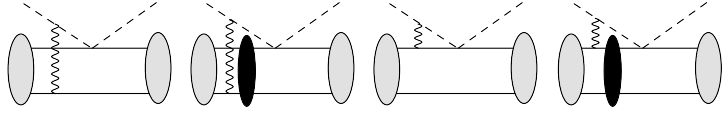}
\caption{Leading virtual-photon diagrams  with the black blobs indicating intermediate $NN$ interactions. The time-reversed diagrams are not shown explicitly, but are included in the calculation.}
\label{fig:d6d8NN}
\end{figure}

\begin{figure}
\centering
\includegraphics[width=0.5\linewidth]{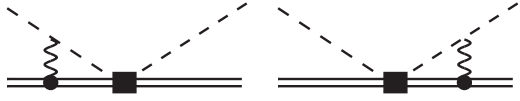}
\caption{Contribution to the modified Deser formula in the $\pi D$ atom due to the part of the Coulomb-photon ladder that is subtracted in~\eqref{NN_isoscalar} and \eqref{NN_isoscalar_double}.
The double line labels the deuteron, the box corresponds to $\pi^- d$ scattering, and the circle to the photon coupling of the deuteron. }
\label{dpole}
\end{figure}

The isoscalar contribution of $(d_6)$ and $(d_8)$ including intermediate-state $NN$ interactions (ISI) (see Fig.~\ref{fig:d6d8NN}) is given by
\begin{align}
a^{(d_6)+(d_8)}_{T=0}+a^{(d_6)+(d_8)}_{T=0,\,{\rm ISI}}&=-\frac{8\pi\alpha\xip a^+}{(2\pi)^6\xid} \int \frac{\diff^3 \kk}{\kk^2} \int \diff^3 \qq\, \diff^3\qq' \Psi^\dagger(\qq') 
\bigg\{G_s\Big(\qq'-\frac{\kk}{2},\qq-\frac{\kk}{2};-\eps - \frac{\kk^2}{2\mpi},-\kk\Big) \notag\\
 &+ G_s\Big(\qq'-\frac{\kk}{2},-\qq+\frac{\kk}{2};-\eps - \frac{\kk^2}{2\mpi},-\kk\Big)
-  \frac{2(2\pi)^3 \Psi(\qq')\Psi^\dagger(\qq)}{-\kk^2/2 \mu_D + i \eta}\bigg\} \Psi(\qq)\label{NN_isoscalar},
\end{align}
where $G_s(\pp',\pp;E,\PP)$ is the isoscalar $NN$ Green's function for initial (final) relative momentum $\pp$ ($\pp'$) and a state of total energy $E$ and momentum $\PP$. Note that to  obtain this result we have neglected the recoil effect in the photon propagator, and  treated the pion as a non-relativistic particle. In the regime which is the focus of this section, where the pion momentum is much less than its mass, both of these are perturbative corrections to the main result. Therefore, we neglect them for the time being, and focus our efforts  on investigating the infrared enhancement of these graphs. The two Green's functions may be interpreted as ``direct'' and ``exchange'' contributions, i.e.\ the photon can couple either to the nucleon that undergoes the $\pi N$ interaction or to the other one. The last term subtracts the 
deuteron pole ($\eta\rightarrow 0^+$), as this part is already accounted for in the modified Deser formula,
cf.~Fig.~\ref{dpole} for a graphical illustration of this piece. The shift of the 1s level in pionic deuterium is proportional 
to the convolution of the $\pi^-d$ scattering length with the Coulombic  wave function of the atom, which diagrammatically 
correspond to an infinite ladder of Coulomb photons. The simplest example shown in Fig.~\ref{dpole} thus needs to be subtracted in~\eqref{NN_isoscalar} 
in order to avoid double counting. 
The details of the derivation of~\eqref{NN_isoscalar} are relegated to \ref{app:single_ChET}.  

The isoscalar propagator $G_s$ is real for energies below the $NN$ threshold and constructed out of continuum states normalized as\footnote{Note that this means that our continuum $NN$ wave functions are normalized differently to our deuterium wave function.}
\beq
\int \frac{\diff^3 \qq}{(2 \pi)^3} \Psi_\pp^{s\,\dagger}(\qq) \Psi^s_{\pp'}(\qq)=(2 \pi)^3 \delta^{(3)}(\pp'-\pp),
\eeq
with $\Psi_\pp^s(\qq)$ obeying
\beq
\bigg(\frac{\pp^2}{\mpp} - \frac{\qq^2}{\mpp}\bigg)\Psi^s_\pp(\qq)=\int \frac{\diff^3\qq'}{(2 \pi)^3} V_s(\qq,\qq') \Psi^s_\pp(\qq'),
\label{schr}
\eeq
where $V_s$ is the projection of the $NN$ potential onto the isoscalar part.
The free part of $G_s$ is therefore
\beq
G^{(0)}(\qq',\qq;E,\PP)= \frac{(2 \pi)^3 \delta^{(3)}(\qq'-\qq)}{E + i \eta - \PP^2/4 \mpp - \qq^2/\mpp}\label{free_Greens},
\eeq
while the total Green's function can be related to the $NN$ scattering $T$-matrix  as
\beq
G_s(\qq',\qq;E,\PP)=G^{(0)}(\qq',\qq;E,\PP) +\frac{T(\qq',\qq;E,\PP)}{(E + i \eta - \PP^2/4 \mpp- \qq^2/\mpp)
(E + i \eta - \PP^2/4 \mpp- \qq'^2/\mpp)},
\label{Gs}
\eeq
where $T$ is connected to the $NN$ phase shifts via 
\beq
T(k,k;E,\PP)=-\frac{4\pi}{\mpp} \frac{1}{k \cot\delta(k)-ik} 
\eeq
 with $k=\sqrt{\mpp(E-\PP^2/4\mpp)}$.
Alternatively,  $G_s$ can be rewritten in terms of the bound and continuum state wave functions 
\beq
G_s(\qq',\qq;E,\PP)=\frac{(2\pi)^3\Psi(\qq')\Psi^\dagger(\qq)}{E + i \eta + \eps - \PP^2/4 \mpp} + \int \frac{\diff^3\pp}{(2 \pi)^3} \frac{\Psi^s_\pp(\qq')\Psi^{s\,\dagger}_\pp(\qq)}{E + i \eta - \pp^2/\mpp - \PP^2/4\mpp}\label{decomp_Gs}.
\eeq
The additional factor $(2\pi)^3$ for the deuteron-pole part is due to our conventions for the deuteron wave functions~\eqref{deut_norm}. Inserting the free part of the Green's function~\eqref{free_Greens} into~\eqref{NN_isoscalar} reproduces the expressions for the structureless diagrams discussed in Sect.~\ref{sec:structureless} up to higher-order terms neglected in the derivation of~\eqref{NN_isoscalar}. Using the decomposition~\eqref{decomp_Gs}, the isoscalar contributions can be cast into the form
\begin{align}
a^{(d_6)+(d_8)}_{T=0}+a^{(d_6)+(d_8)}_{T=0,\,{\rm ISI}}&=\frac{16\pi\alpha\xip a^+}{\xid}\int\frac{\diff^3\kk}{(2\pi)^3}\frac{1}{\kk^2}
\bigg\{\frac{|F(\kk)|^2-1}{\kk^2/2\mu_D-i\eta}\notag\\
&+\int \frac{\diff^3\pp}{(2\pi)^6}
\frac{1}{\epsilon+\pp^2/\mpp+\kk^2/2\mu_D-i\eta}G_\pp^s(\kk)\frac{1}{2}(G_\pp^{s\,\dagger}(\kk)+G^{s\,\dagger}_\pp(-\kk))\bigg\},
\label{cancd6d8}
\end{align}
where
\beq
F(\kk)=\int \diff^3\qq\Psi^\dagger(\qq)\Psi(\qq-\kk/2),\qquad G_\pp^s(\kk)=\int \diff^3\qq\Psi^\dagger(\qq)\Psi_\pp^s(\qq-\kk/2).
\label{cancd6d8_1}
\eeq
Now, from the normalization condition of $\Psi(\qq)$, and the orthogonality of
$\Psi(\qq)$ and $\Psi_\pp^s(\qq)$ for vanishing momentum transfer, it follows that 
\beq
|F(\kk)|^2-1=\Order(\kk^2),\qquad G_\pp^s(\kk)=\Order(\kk).
\label{cancd6d8_2}
\eeq
In this way, we conclude that also in the isoscalar case no terms enhanced by $\sqrt{\mpi/\eps}$ remain. Due to the chiral suppression of $a^+$ the sub-leading corrections can simply be dropped. 
This reasoning used here---based on exploiting orthogonality of bound state and continuum wave functions---follows the calculation of recoil corrections to $\bar K d$ scattering in~\cite{recoil_BER}.

Alternatively, the cancellation can be derived within H$\pi$EFT. In this case it is also convenient to split the total effect into parts without and parts with an intermediate-state $NN$ interaction. The portion without intermediate-state $NN$ interaction gives (cf.~\eqref{ad6d8_exp} and \eqref{ad6d8LO} in Sect.~\ref{sec:structureless})
\begin{align}
a_{T=0}^{(d_8)}&=a^+ \frac{16 \alpha}{3 \pi} \sqrt{\frac{M_\pi}{2 \epsilon}} \left(1 + \Order\left(\frac{M_\pi}{\mpp}\right)\right),\notag\\
a_{T=0}^{(d_6)}&=a^+ \frac{16 \alpha}{3 \pi} \sqrt{\frac{M_\pi}{2 \epsilon}} \left(1 -\frac{3 \pi}{8 \sqrt{2}} \sqrt{\frac{M_\pi}{\mpp}}+ \Order\left(\frac{M_\pi}{\mpp}\right)\right).
\label{eq:d6pionless}
\end{align} 
Again, these expressions show the anticipated infrared enhancement.

Using the leading-order form of the $NN$ scattering amplitude in pionless EFT~\cite{vK99,Ka98A,Ka98B,Ge98,Bi99}
\begin{equation}
T_{s,NN}({\bf q}',{\bf q};E,{\bf P})=\frac{4 \pi}{\mpp} \frac{1}{\gamma + i \sqrt{\mpp \left(E-\frac{{\bf P}^2}{4 \mpp}\right)}},
\end{equation}
we evaluate the diagrams with ISI and find, after removing the deuteron-pole piece already accounted for in the modified Deser formula,
\begin{equation}
a_{T=0,\,{\rm ISI}}^{(d_6)+(d_8)}=-\frac{32 \alpha}{3\pi} a^+ \sqrt{\frac{M_\pi}{2 \epsilon}},
\end{equation}
which precisely cancels the leading piece of the free part of the diagrams in~\eqref{eq:d6pionless}. Furthermore, any contributions of momenta $|{\bf k}| \sim \sqrt{M_\pi \epsilon}$ 
may only appear in the non-integer terms in the expansion~\eqref{ad6d8_exp}
and thus are suppressed by an additional power of $M_\pi/\mpp$ beyond this (i.e.~they correspond to the third term in~\eqref{ad6d8_exp}). These contributions thus have an overall size
\begin{equation}
\frac{32 \alpha}{3\pi} \frac{M_\pi}{\mpp}  \sqrt{\frac{M_\pi}{2 \epsilon}} a_+ \approx 2.8\, \alpha a^+,
\end{equation}
and are therefore well beyond the accuracy we claim for our calculation. 

There is still a possible contribution from ISI diagrams with momenta of order $\gamma$, though. This would be enhanced by $M_\pi/\gamma$ compared
to its naive ChPT order, and so could be relevant for our analysis. Decomposing the Green's function according to
\begin{align}
G_s(\qq',\qq;E,\PP)&=G^{(0)} (\qq',\qq;E,\PP) + \frac{(2\pi)^3\Psi(\qq')\Psi^\dagger(\qq)}{E+i\eta+\eps-\PP^2/4\mpp}\notag\\
&+ \frac{1}{E + i \eta - \PP^2/4\mpp - \qq'^2/\mpp} T_{s,NN}^{\rm np}(\qq',\qq;E,\PP) \frac{1}{E + i \eta - \PP^2/4\mpp - \qq^2/\mpp},
\end{align}
i.e.~into its free part, the deuteron pole, and the non-pole isoscalar $NN$ $T$-matrix $T_{s,NN}^{\rm np}$, the additional contributions due to intermediate $NN$ interactions are given by
\begin{align}
\label{formfactor}
a^{(d_6)+(d_8)}_{T=0,\,{\rm ISI}}&= -\frac{8\pi\alpha \xip a^+}{\xid} \int \frac{\diff^3 \kk}{(2 \pi)^3} \frac{1}{\kk^2} \bigg\{  \int \frac{\diff^3 \qq\,\diff^3 \qq'}{(2 \pi)^3}  \Psi^\dagger(\qq') \frac{1}{-\eps - \kk^2/2\mu_D - (\qq' - \kk/2)^2/\mpp}\notag\\
&\quad \times T_{s,NN}^{\rm np} \frac{1}{-\eps - \kk^2/2 \mu_D - (\qq - \kk/2)^2/\mpp}  \Psi(\qq) 
 + 2\frac{|F(\kk)|^2 - 1}{-\kk^2/2 \mu_D + i \eta}\bigg\},\\
T_{s,NN}^{\rm np}&= T_{s,NN}^{\rm np}\Big(\qq'-\frac{\kk}{2},\qq-\frac{\kk}{2};-\eps - \frac{\kk^2}{2\mpi},-\kk\Big)
+T_{s,NN}^{\rm np}\Big(\qq'-\frac{\kk}{2},-\qq+\frac{\kk}{2};-\eps - \frac{\kk^2}{2\mpi},-\kk\Big)\notag.
\end{align}
This equation will undergo an explicit numerical evaluation in Sect.~\ref{sec:numerics}.
The part of the integral involving momenta $|{\bf k}| \sim M_\pi$ only contributes at the naive chiral order of the pertinent diagrams. For that effect we only need to evaluate $(d_6)$--$(d_8)$ without any intermediate-state interaction, and it is furthermore sufficient to keep only the $T=1$ pieces of the free parts of these graphs, see~\eqref{isoscalar} and \eqref{d7} above. 

\subsection{The role of rescattering II: double scattering with photon exchange}
\label{sec:double}

If only momenta $|{\bf k}|\sim M_\pi$ were important, there would be no need to consider the diagrams $(d_9)$ and $(d_{10})$. But, the full isoscalar contribution to $(d_9)$ and $(d_{10})$ reads (for the derivation see~\ref{app:double_ChET})
\begin{align}
\label{NN_isoscalar_double}
a^{(d_9)+(d_{10})}_{T=0}+a^{(d_9)+(d_{10})}_{T=0,\,{\rm ISI}}&=\frac{32\pi^2\alpha(\xip a^-)^2}{\xid} \int \frac{\diff^3 \kk}{(2 \pi)^3}\int \frac{\diff^3 \lbf}{(2 \pi)^3} \frac{1}{\kk^2\lbf^2} \int \frac{\diff^3 \qq\,\diff^3 \qq'}{(2 \pi)^3}\notag\\
&\times\Psi^\dagger(\qq')\bigg\{2G_s\Big(\qq'-\frac{\kk}{2},\qq-\frac{\kk}{2}+\lbf;-\eps - \frac{\kk^2}{2\mpi},-\kk\Big)\\
&+ 2G_s\Big(\qq'-\frac{\kk}{2},-\qq+\frac{\kk}{2}-\lbf;-\eps - \frac{\kk^2}{2\mpi},-\kk\Big)
- \frac{4(2\pi)^3\Psi(\qq')\Psi^\dagger(\qq-\lbf)}{-\kk^2/2 \mu_D + i \eta}\bigg\} \Psi(\qq)\notag.
\end{align}
These diagrams, too, are reducible, i.e.\ would-be infrared divergent. 
The same arguments as for single scattering yield that the leading contributions from momenta $|{\bf k}| \sim \sqrt{M_\pi \epsilon}$ cancel. This leaves an effect from these momenta that has a numerical size $\approx 1.4\,  \alpha a_{\pi^- d}$, 
which is significantly below the few percent accuracy that we seek.

Again, however, we are concerned about momenta of order $\gamma$, which could yield contributions enhanced by $\mpi/\gamma$, and so compromise the accuracy of our calculation. To evaluate this contribution explicitly note that the free part of the Green's function reproduces~\eqref{d9d10_structureless}, while the intermediate-state $NN$ interactions lead to 
\begin{align}
a^{(d_9)+(d_{10})}_{T=0,\, {\rm ISI}}&=\frac{32\pi^2\alpha(\xip a^-)^2}{\xid} \int \frac{\diff^3 \kk}{(2 \pi)^3}\int \frac{\diff^3 \lbf}{(2 \pi)^3} \frac{1}{\kk^2\lbf^2} \bigg\{\int \frac{\diff^3 \qq\,\diff^3 \qq'}{(2 \pi)^3} \frac{\Psi^\dagger(\qq')}{-\eps - \kk^2/2 \mu_D - (\qq' - \kk/2)^2/\mpp}\notag\\
&\quad\times\tilde T_{s,NN}^{\rm np} \frac{\Psi(\qq) }{-\eps - \kk^2/2 \mu_D - (\qq - \kk/2+\lbf)^2/\mpp}  
 + 4\frac{F(\kk)F(\kk-2\lbf) - F(2\lbf)}{-\kk^2/2 \mu_D + i \eta}\bigg\},\notag\\
\tilde T_{s,NN}^{\rm np}&= 2T_{s,NN}^{\rm np}\Big(\qq'-\frac{\kk}{2},\qq-\frac{\kk}{2}+\lbf;-\eps - \frac{\kk^2}{2\mpi},-\kk\Big)\notag\\
&+2T_{s,NN}^{\rm np}\Big(\qq'-\frac{\kk}{2},-\qq+\frac{\kk}{2}-\lbf;-\eps - \frac{\kk^2}{2\mpi},-\kk\Big)\label{NN_isoscalar_double1},
\end{align}
where we have used repeatedly that $\Psi(\pp)=\Psi(-\pp)$.

\subsection{Numerical evaluations}
\label{sec:numerics}

We explicitly evaluate the isoscalar contributions to single and double scattering for which expressions were derived in the previous two subsections. 
For this purpose we use a separable $NN$ interaction, since we anticipate that the integral is dominated 
by low-momentum modes, so details of the potential are not important. Specifically, we use the effective potential
\beq 
V({\bf p}, {\bf p'} )=\lambda g(\pp)g(\pp'),\quad  g(\pp)=\frac{1}{\pp^2+\beta^2}, 
\eeq
where $\lambda$ is a constant tuned to reproduce the binding momentum $\gamma=\sqrt{m_p \epsilon}$,
and   ${\beta=1.4\,{\rm fm}^{-1}}$ is introduced to parameterize the effective range of $pn$ scattering, which 
enters in realistic potentials through  the one-pion exchange. Solving the 
Schr\"odinger equation~\eqref{schr} for the bound state, one obtains the deuteron wave function of the Hulth\'{e}n type 
\beq
 \Psi(\pp)=N \frac{g(\pp)}{\pp^2+\gamma^2}, \quad
N=\frac{1}{\pi}\sqrt{\gamma\beta(\gamma+\beta)^3}. 
\eeq 
Using this wave function and evaluating~\eqref{NN_isoscalar} with  $G_s$ from~\eqref{Gs},
we find
\beq
a^{(d_6)+(d_8)}_{T=0}+a^{(d_6)+(d_8)}_{T=0,\,{\rm ISI}}=-0.034\,a^+.
\label{eq:pionfuld6plusd8}
\eeq
We note that the individual contributions $a^{(d_6)+(d_8)}_{T=0}$ and $a^{(d_6)+(d_8)}_{T=0,\,{\rm ISI}}$ are 5--6 
times larger than their sum, which attests to the cancellation derived in Sect.~\ref{sec:single}:
at leading order both $a^{(d_6)+(d_8)}_{T=0}$ and $a^{(d_6)+(d_8)}_{T=0,\,{\rm ISI}}$
acquire large contributions from momenta of order $\sqrt{\mpi \epsilon}$ which, however, cancel completely in the sum. 
 The deviation from 
zero in this result is mainly provided by higher-order corrections in the expansion of $a^{(d_6)+(d_8)}_{T=0}$, 
according to~\eqref{ad6d8_exp}, whereas higher-order corrections to $a^{(d_6)+(d_8)}_{T=0,\,{\rm ISI}}$ appear to be 
numerically negligible.  

The dominant effect in this result is ultimately due to momenta of order $\gamma$, and so should also be 
accessible in heavy-pion effective field theory (H$\pi$EFT). 
From~\eqref{eq:d6pionless} we see that, in H$\pi$EFT, the free piece of the isoscalar contribution to $(d_6)$
is
\begin{equation}
a^{(d_6)+(d_8)}_{T=0}=- \alpha a^+ \frac{M_\pi}{2 \gamma}\left[1 + \Order\left(\sqrt{\frac{M_\pi}{\mpp}}\right)\right] \approx -0.022 a^+,
\label{d6d8gamma}
\end{equation}
where the explicitly evaluated contribution arises from momenta of order $\gamma$. 
This number is actually quite close to~\eqref{eq:pionfuld6plusd8}, which suggests that indeed momenta of order $\gamma$ are largely responsible for this contribution.\footnote{Taking into account 
the  wave-function-renormalization factor $Z=1.690$ from~\cite{PRS}, which is necessary to ensure 
the correct asymptotic $S$-state  normalization of the deuteron wave function, 
\eqref{d6d8gamma} changes to $-0.037a^+$, which is even closer to~\eqref{eq:pionfuld6plusd8}.}

In addition,   the ISI part generated by the ``form factor'' in~\eqref{formfactor} can also be 
calculated explicitly within (H$\pi$EFT)
\beq
a^{(d_6)+(d_8)}_{\rm FF}=\frac{32\pi\alpha\mpi \xip a^+}{(2\pi)^3\xid^2} \int \frac{\diff^3 \kk}{\kk^4} \big(|F(\kk)|^2 - 1\big).
\eeq
For momenta $|{\bf k}| \sim \gamma$ the deuteron wave function can be replaced by its H$\pi$EFT approximation~\eqref{asymp}, which leads to
\beq
F(\kk)=\frac{4\gamma}{|\kk|}\arctan\frac{|\kk|}{4\gamma}.
\eeq
Performing the last integration then yields\footnote{Strictly speaking, $a^+$ should be replaced by $\tilde a^++\Delta \tilde a^+$ in order to account for isospin violation, however this does not change the prefactor. Moreover, in the experimental analysis of the level shift in $\pi D$ also corrections due to the electromagnetic radius of the deuteron are taken into account. We show in~\ref{app:hpiEFT_radius} that this amounts only to a tiny modification of the prefactor.}
\beq
a^{(d_6)+(d_8)}_{\rm FF}=-\frac{2}{3}\alpha(1+\log 4)\frac{\mpi}{\gamma}\frac{\xip}{\xid^2}a^+=-0.035\,a^+\label{FF_single},
\eeq
which is even larger than~\eqref{d6d8gamma}. However, the numerical 
smallness of the   full contribution from intermediate 
$NN$ interaction apart from the scale $\sqrt{\mpi \epsilon}$ suggests that the contribution of 
 the form factor should  be canceled by the non-pole part of the $NN$ amplitude.

The numerical evaluation of~\eqref{NN_isoscalar_double}  with the pionful wave functions
based on the separable $NN$ interaction described above yields
\beq
a^{(d_9)+(d_{10})}_{T=0}+a^{(d_9)+(d_{10})}_{T=0,\,{\rm ISI}}=0.3\cdot 10^{-3}\mpi^{-1}.
\label{pionfuld9plusd10}
\eeq
Again, this result basically stems from residual contributions of the free diagrams from scales above
$\sqrt{M_\pi\epsilon}$. Contrary to  $(d_6)$ and $(d_8)$, which are dominated by momenta of order  
$\gamma$, in this case also momenta of order $\mpi$ significantly contribute to the diagrams. 
On the other hand, the ``form-factor'' contribution  $a^{(d_9)+(d_{10})}_{T=0}$ can still be addressed in H$\pi$EFT, 
along the same lines as employed for single 
scattering (see~\ref{app:hpiEFT_double}). The corresponding result for the form-factor contribution is large,
\beq
a^{(d_9)+(d_{10})}_{\rm FF}=\frac{8\alpha\mpi(\xip a^-)^2}{\pi^2\xid^2}\Big(\frac{\pi^2}{2}-2.916\Big)
-\frac{\alpha\mpi a^{\rm static}}{2\gamma\xid}=0.35\cdot10^{-3}\mpi^{-1},\label{FF_double}
\eeq
however, it should again be compensated by the non-pole part of the diagrams, since the residual contribution 
of the $NN$ ISI is numerically negligible.

These full evaluations show that, despite their ostensible infrared enhancement, the isoscalar parts of $(d_6)+(d_8)$ and $(d_9) + (d_{10})$ have contributions from momenta of order $\gamma$ that yield parts of $a_{\pi^- d}$ which are still significantly smaller than our theoretical uncertainty. We will therefore simply drop the isoscalar contributions in what follows.

Similarly, full evaluation shows that---after the Pauli-principle-enforced cancellation of contributions of order $\sqrt{M_\pi/\epsilon}$---isovector contributions of $(d_9) + (d_{10})$ are very small, only about $-0.1\cdot 10^{-3}\mpi^{-1}$. Thus the same sort of cancellations that preclude the existence of a $\chi^{1/2}$ contribution due to recoil in the double-scattering diagrams
also enforces the smallness of this $T=1$ part of $(d_9) + (d_{10})$. We therefore conclude that $(d_9)$ and $(d_{10})$ may be omitted altogether from our analysis. Indeed, had we found that $(d_9)$ and $(d_{10})$ were necessary for a precision evaluation of $a_{\pi^- d}$, we would have been forced to consider all photon diagrams at this order, since $(d_9)$ and $(d_{10})$ do not, on their own, form a gauge-invariant set of diagrams.  

In fact, the results found here as regards ``would-be infrared-divergent'' diagrams are very important, as the cancellations we have identified guarantee that the original ChPT power counting, which assesses the impact of momenta $\sim M_\pi$ on the integrals, provides a reasonable estimate of diagrams involving virtual photons, since the remaining infrared enhancement is too weak to severely violate the ChPT estimates. In this way, we are left with the diagrams $(d_6)$\bis$(d_8)$ 
\begin{align}
\label{aem}
 a^{\rm EM}&=\frac{2\alpha\mpi\xip a^-}{\pi^2\xid}I^{\rm EM}-\frac{\alpha }{16\pi ^3\Fpi^2\xid}\bigg\langle\frac{1}{\qq^2}\bigg\rangle,\notag\\
I^{\rm EM}&=\int \diff^3\mathbf{p}\,\diff^3\mathbf{q} 
\big(\Psi^\dagger(\pp)-\Psi^\dagger(\pp-\qq)\big)\Psi(\pp)\frac{1}{|\qq|\big(|\qq|+\delta/2\omega_\qq\big)\big(\qq^2+\delta\big)}.
\end{align}
With the numerical results from Sect.~\ref{subsec:numerics} and $a^-=86.1\cdot 10^{-3}\mpi^{-1}$, we obtain
\beq
a^{\rm EM} = (0.94\pm 0.01)\cdot 10^{-3}\mpi^{-1},
\eeq
where the error again reflects the wave-function dependence
as follows from Table~\ref{table:str_integrals}. Thus, virtual photons increase $\Re a_{\pi^- d}$ by about $4\,\%$. 

\begin{sloppypar}
In summary, we have shown that there are no infrared-enhanced photon contributions from momenta $\sim\sqrt{\eps\mpi}$---due to subtle cancellations both for isoscalar and isovector $\pi N$ interactions---and that the infrared enhancement provided by momenta $\sim\gamma$ is too weak to generate effects that significantly exceed the estimates for momenta $\sim \mpi$. The size of the virtual-photon corrections is thus roughly in line with the original ChPT power counting.
\end{sloppypar}

\section{Dispersive and \boldmath{$\Delta$} corrections}
\label{dispdel}

There are two additional contributions to the $\pi^-d$ scattering length that have not been mentioned so far. First, diagrams with pure $NN$ or $NN\gamma$ intermediate states yield so-called dispersive corrections. It is this class of diagrams that produces the imaginary part of $a_{\pi^- d}$, although here their leading contribution to the real part of $a_{\pi^- d}$ is suppressed by $p^{3/2}$ relative to $(d_1)$~\cite{disp}. Diagrams with explicit $\Delta$ degrees of freedom enter at the same order and provide the second class of contributions we consider in this section~\cite{delta}. The $\Delta(1232)$ contribution is a true three-body effect, since the nucleon recoil is needed if this $P$-wave resonance is to contribute to $S$-wave $\pi^-d$ scattering. Typical examples for each of these two effects are depicted in Table~\ref{threebody}.
Both classes were computed in~\cite{disp,delta} using a calculation for $NN\to d\pi$ up to NLO in ChPT~\cite{pp2dpi} in which all integrals were cut off  at $1\,{\rm GeV}$. We have checked that varying this cutoff does not introduce additional uncertainty, and
the effect in $a_{\pi^- d}$ is then
\beq
a^{{\rm disp}+\Delta}=(-0.6\pm 1.5)\cdot 10^{-3}\mpi^{-1}.
\eeq
Since this contribution is only $\Order(p^{1/2})$ larger than the contact term, we need not include isospin-violating 
corrections to $a^{{\rm disp}+\Delta}$, which, counting $e \sim p$, would be suppressed by another two orders.

\section{Summary of  three-body  contributions to the pion--deuteron scattering length}
\label{sec:summary}

We now summarize the content of Sects.~\ref{counting}--\ref{dispdel} by listing the three-body corrections that need to be taken into account in the actual calculation. We recapitulate the numerical results for these contributions, and analyze 
the different sources of uncertainty in the calculation.

The essence of the discussion in  Sect.~\ref{counting} was to demonstrate that all isospin-conserving 
three-body corrections can be reliably calculated up to $\Order(p^{3/2})$, i.e.\ a relative accuracy of $(\mpi/\mpp)^{3/2}$. This is half an order lower than the contribution of the leading unknown $(N^\dagger N)^2\pi\pi$ contact term, which is  $\Order(p^{2})$.
The uncertainty anticipated due to the truncation of higher-order terms is a few percent, as follows from 
a naive dimensional analysis. However, to achieve this accuracy one also has to account for three-body isospin-violating corrections, which are suppressed by $e^2$ compared to the leading isospin-conserving terms. Therefore we also presented a complete calculation of the isospin-violating corrections up to $\Order(e^2 p^{1/2})$.
To this order the three-body isospin-violating corrections are of electromagnetic nature, that is they appear due to virtual photons and the pion 
mass difference. The quark mass difference starts to contribute only at $\Order(e^2p)$. 

In order to reach our accuracy goal,  in Sect.~\ref{sec:strong} we explicitly evaluated the diagrams of a multiple-scattering 
topology  as well as those that involve 3$\pi N^{\dagger}N$ and 4$\pi$ vertices (cf.~Fig.~\ref{threebody}).
In particular, in Sect.~\ref{sec:double_scat} we derived the expressions for the most relevant double-scattering 
operator including isospin-violating corrections.  To better understand the relevant scales  contributing to the process the full double-scattering expression was divided,
without making any  approximations, into three parts: the static term at LO, 
its correction at NLO, and the contribution that contains the three-body cut. 
As stated above, the leading isospin-violating correction is due to the pion mass difference and thus it appears explicitly in the three-body-cut
contribution with charge exchange, see~\eqref{integrals}.
In order to have the $\pi^0 nn$ and $\pi^- pn$  thresholds
at the proper places we also kept the nucleon mass difference in the propagators, 
although this is formally a higher-order effect ($\Order(e^2 p)$).  
Due to the presence of the cut the effect of the nucleon mass difference  could 
be enhanced. In fact, it proves to be negligibly small: the direct 
evaluation gives just a $0.5\,\%$ correction to the static term compared, e.g., to a $2\,\%$ correction 
from the pion mass difference. 
In this way, we conclude that the effect of the quark mass difference in three-body operators 
is negligible.

Other diagrams of a multiple-scattering topology appear at $\Order(p)$ relative to the double-scattering diagram,
according to Table~\ref{threebody}.
The effect of the nucleon recoil in these diagrams can thus be neglected there, with, e.g.,
cancellations enforced by  the Pauli principle guaranteeing that
the nucleon recoil in the triple-scattering term starts to contribute only at 
$\Order(p^2)$.
 Thus, it suffices to evaluate the triple-scattering diagram in the static approximation, see~\eqref{atriple}. The effect of all multiple-scattering diagrams 
beyond  triple-scattering  was also evaluated in the static limit and shown to be negligible, cf.~Sect.~\ref{mult}. Similarly, isospin-violating corrections are irrelevant for all terms in the multiple-scattering series beyond double scattering.

The amplitudes $(d_3)$  and $(d_4)$, which involve 3$\pi N^{\dagger}N$ and 4$\pi$ vertices, were calculated  in Sect.~\ref{secd3d4}
including the leading isospin-violating corrections due to the pion mass difference. The contribution of these diagrams appears to 
be suppressed numerically due to accidentally small spin-isospin factors~\cite{Liebig}. 

The combination of all effects discussed thus far in this section defines the ``strong'' contribution to the $\pi^- d$ scattering length $a^{\rm str}$. The numerical result for $a^{\rm str}$ was presented in Sect.~\ref{subsec:numerics}, via evaluations of the wave-function averages 
using modern phenomenological $NN$ potentials, AV18~\cite{AV18} 
and CD-Bonn~\cite{CDBonn}, as well as chiral $NN$ interactions 
at order N$^2$LO~\cite{NNLO}. We found (with $a^-=86.1\cdot 10^{-3}\mpi^{-1}$) 
\beq  
a^{\rm str} = (-22.6\pm 1.1\pm 0.4)\cdot10^{-3}\mpi^{-1}, 
\eeq
where the first uncertainty arises from the different short-distance 
($r \ll 1/M_\pi$) physics of the $NN$ wave functions, and the second from the isospin-breaking shifts in the $\pi N$ scattering lengths. The variation 
in the results due to the use of different wave functions is  about $5\,\%$. This provides 
an independent confirmation of the contact term's effect.
 
The combined effect of the dispersive corrections and the $\Delta(1232)$ contributions 
at  $\Order(p^{3/2})$  was discussed in Sect.~\ref{dispdel} and taken from~\cite{disp,delta}
\beq
 a^{{\rm disp}+\Delta} = (-0.6\pm 1.5)\cdot 10^{-3} \mpi^{-1}.
 \eeq
Isospin-violating corrections to these diagrams are irrelevant at the order to which we work. 

Finally, Sect.~\ref{sec:virt} was devoted to a thorough  investigation of the effects related to virtual photons. 
Due to the presence of photon and pion
propagators these diagrams are infrared enhanced. Therefore,  keeping the full dynamical structure 
of the $\pi NN$ propagator (including the nucleon recoil)  is mandatory to avoid infrared-divergent integrals. 
To the order we are working, it is sufficient to consider the diagrams $(d_6)$--$(d_8)$, which form a gauge-invariant set 
of diagrams at order $\Order(e^2)$. The explicit expressions for these diagrams
were derived using time-ordered perturbation theory, cf.~\eqref{aem}. 
Note that for the gauged Weinberg--Tomozawa diagram $(d_7)$ no   $\pi NN$ propagators
emerge. As a consequence, that contribution is infrared finite even in the static limit, cf.~the 
second term in the first line of \eqref{aem}. The inclusion of the recoil in the photon propagator was also considered, 
cf.~\eqref{d7}. It is, however, a negligible effect. The ultimate result for diagrams $(d_6)$--$(d_8)$ is then
\beq 
a^{\rm EM} =(0.94\pm 0.01)\cdot  10^{-3} \mpi^{-1},
\eeq
with only a $\sim 1\,\%$ wave-function dependence (see last column of Table~\ref{table:str_integrals}). 
Motivated by the large magnitude of the double-scattering term,
 we also investigated associated virtual-photon corrections, cf.~diagrams $(d_9)$ and $(d_{10})$,
 although formally they  contribute only at $\Order(e^2p^{2})$ in the  power counting. The explicit computation of these diagrams showed
that the magnitude of these corrections is far beyond the accuracy we seek.

The three pieces $a^{\rm str}$, $a^{{\rm disp} + \Delta}$, and $a^{\rm EM}$, when added together,
constitute the three-body contribution to the $\pi^-d$ scattering length. 
In fact, to a large extent, the novel three-body effects computed in this study
accidentally cancel
\beq
\Delta a^{(2)} + a^{\rm static}_{\rm NLO} + a^{\rm cut} + a^{\rm EM} = (0.1\pm 0.7)\cdot 10^{-3} \mpi^{-1}.
\eeq
This cancellation is, in itself, somewhat remarkable, since, e.g.\ $a^{\rm static}_{\rm NLO}$ is $\sim 35$ 
times larger than the final central value. The effect of the cancellation is that the main impact of our 
analysis on the extraction of pion--nucleon scattering lengths is our consideration of NLO isospin-breaking 
corrections---in particular the large shift $\Delta \tilde a^+=(-3.3\pm 0.3)\cdot 10^{-3}\mpi^{-1}$---in the $\pi N$ amplitude~\cite{HKM_CD09}.

\section{Pion--nucleon scattering lengths}
\label{sec:results}

\begin{figure}
\centering
\includegraphics[width=0.8\linewidth,clip]{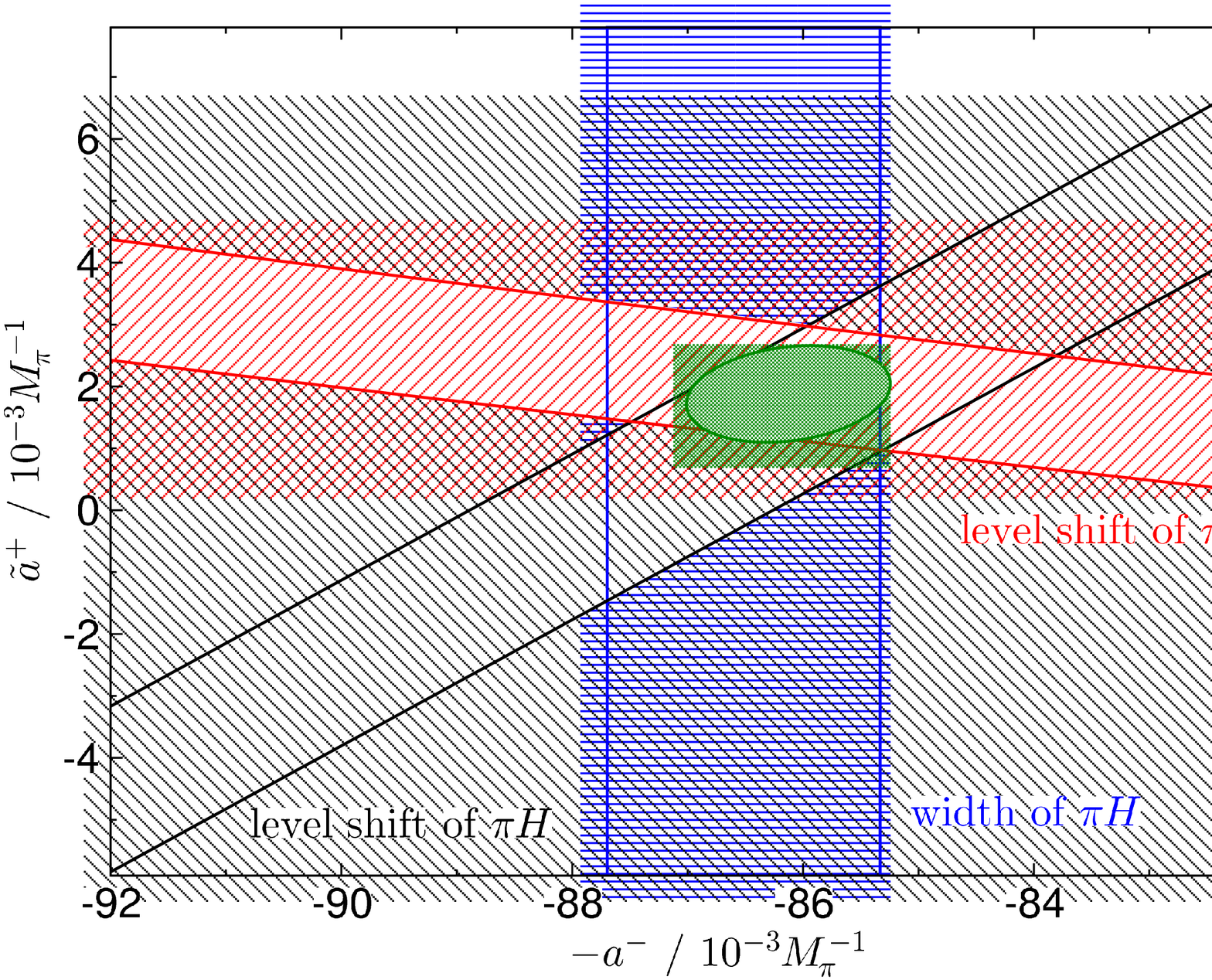}
\caption{Combined constraints in the $\tilde a^+$--$a^-$ plane from data on the width and energy
shift of $\pi H$, as well as the $\pi D$ energy shift. Figure from~\cite{JOB}.}
\label{fig:bands}
\end{figure}

Combining the dependence of the $\pi^- d$ scattering length on $\tilde a^+$ and
$a^-$ and the results for $\pi H$ discussed above, we find the constraints depicted in 
Fig.~\ref{fig:bands}. 
The combined $1\sigma$ error ellipse yields 
\beq
\tilde{a}^+=(1.9\pm 0.8)\cdot 10^{-3} \mpi^{-1},\quad a^-=(86.1\pm 0.9)\cdot 10^{-3}\mpi^{-1},
\label{atilde}
\eeq
with a correlation coefficient $\rho_{a^-\tilde a^+}=-0.21$.  
We find that the inclusion of the $\pi D$ energy shift reduces the uncertainty of $\tilde{a}^+$
by more than a factor of 2 and the correlation between $\tilde{a}^+$ and $a^-$ by more than a factor of 3.
Note that in the case of the $\pi H$ level shift the width of the band is dominated
by the theoretical uncertainty in $\Delta \tilde a_{\pi^- p}$,
whereas for the $\pi H$ width the experimental error is about $50\,\%$ larger than the theoretical one.

Table~\ref{table:piDlevel} shows the individual contributions to the $\pi D$ error band: as with the $\pi H$ level shift, the experimental error is much smaller than the combined theoretical uncertainty, whose largest individual contribution is produced by the uncertainty in $a^{{\rm disp} + \Delta}$.
The wave-function averages contribute about $0.5 \cdot 10^{-3} M_\pi^{-1}$ to the overall uncertainty in $\tilde a^+$, 
which is in line with the estimated impact on $a_{\pi^- d}$ of the $\Order(p^2)$---relative
to $(d_1)$---contact term.
\begin{table}
\centering
\begin{tabular}{ccccc}
\hline\hline
$\epsilon_{1s}^D$ & $\Delta a^-, \Delta a_{\pi^- p}^{\rm cex}$ & $a^{{\rm disp}+\Delta}$ & $\Delta \tilde{a}^+$ & Wave-function averages \\\hline
$16\, \%$ & $21\, \%$  & $75\, \%$ & $30\, \%$ & $53\, \%$\\
\hline\hline
\end{tabular}
\caption{Individual contributions to the error on $\tilde{a}^+$ are added in quadrature to obtain the uncertainty depicted in the bands of Fig.~\ref{fig:bands}. The impact of each source of error is given as a percentage of the total (where the second column gives the uncertainty in the isospin-breaking shifts of $\pi N$ scattering lengths that occur in $a^{\rm str}$, cf.~\eqref{bar_a}). Table from~\cite{JOB}.}
\label{table:piDlevel}
\end{table}

To deduce a value for $a^+$ itself, further input on $c_1$ and $f_1$ is required according to~\eqref{atilde_def}. $c_1$ is related to the $\pi N$ $\sigma$-term: $\sigma_{\pi N}=(45\pm 8)\,{\rm MeV}$ as quoted in~\cite{sigmaterm}\footnote{This value is consistent with recent determinations of $\sigma_{\pi N}$ from the lattice, $\sigma_{\pi N}=(50\pm 10)\,{\rm MeV}$~\cite{thomas_sigma}
and $\sigma_{\pi N}=(50\pm 10\pm 10)\,{\rm MeV}$~\cite{Durr}.} corresponds to $c_1=(-0.9\pm 0.1)\,{\rm GeV}^{-1}$ and $c_1=(-1.0\pm 0.2)\,{\rm GeV}^{-1}$ at third and forth chiral order, respectively~\cite{BL01}. Recent determinations based on $\pi N$ threshold parameters yield $c_1=(-0.93\pm 0.07)\,{\rm GeV}^{-1}$~\cite{GR02} and $c_1=(-1.2\pm 0.3)\,{\rm GeV}^{-1}$~\cite{hadatoms}, while an investigation of $\pi N$ scattering inside the Mandelstam triangle led to ${c_1=(-0.81\pm 0.12)\,{\rm GeV}^{-1}}$~\cite{BM99}. Finally, fits of ChPT amplitudes to phase-shift analyses provide values in the range $c_1=-(1.2\ldots 1.4)\,{\rm GeV}^{-1}$~\cite{FM00}.
In conclusion, we consider
\beq
c_1=(-1.0\pm 0.3)\,{\rm GeV}^{-1}
\eeq
as a reasonable average of the present knowledge on this LEC. Taken together with the rough estimate $|f_1|\leq 1.4 \,{\rm GeV}^{-1}$~\cite{f1,f1b}, this value for $c_1$ and \eqref{atilde} yield a non-zero $a^+$ at better than the $95\,\%$ confidence level
\beq
a^+=(7.6\pm 3.1)\cdot 10^{-3}\mpi^{-1}.
\eeq
The final result for $a^+$ is only a little larger than several of the contributions considered in this work. 
This emphasizes the importance of a systematic ordering scheme, and a careful treatment of
isospin violation and three-body dynamics. 
A reduction of the theoretical uncertainty beyond that of the present analysis will be hard to achieve without 
additional QCD input that helps pin down the unknown contact-term contributions 
in both the $\pi N$ and $\pi NN$ sectors.

Finally, we can combine our values for the scattering lengths in the isospin limit with the isospin-breaking corrections~\cite{HKM} to arrive at the $\pi N$ scattering lengths for the physical channels summarized in Table~\ref{table:physical_channels}. Note that the difference between scattering lengths in the same isospin channel is better known than the scattering lengths individually, since the scattering lengths in the isospin limit and the associated uncertainties drop out. For example, the difference between $a_{\pi^0p}$ and $a_{\pi^0 n}$ at NLO is given by~\cite{HKM}
\begin{align}
a_{\pi^0p}-a_{\pi^0n}&=\frac{1}{4\pi\xip} \biggl\{\frac{4c_5 B(\md-\muu)}{\Fpi^2}-\frac{\mpi^2}{8\pi\Fpi^4}\Big(\sqrt{\Delta_\pi+2\mpi\Delta_{\rm N}}-\sqrt{\Delta_\pi-2\mpi\Delta_{\rm N}}\Big)\biggr\}\notag\\
&=(-3.4\pm 0.4)\cdot 10^{-3}\mpi^{-1}, \label{scattlengthdiff}
\end{align}
with $c_5$ being related to the strong contribution to the proton--neutron mass difference. Although it is formally of higher chiral order, here the contribution from the cusp due to $\pi^+n$ and $\pi^-p$ intermediate states has been kept, since it is enhanced by half an order in the isospin-breaking parameter $\delta=\{e^2,\md-\muu\}$. And indeed, it ultimately contributes about $30\,\%$ to the number quoted in~\eqref{scattlengthdiff}.

\begin{table}
\centering
\begin{tabular}{ccccc}
\hline\hline
isospin limit & channel & scattering length & channel & scattering length \\\hline
$a^++a^-$ & $\pi^-p\rightarrow \pi^-p$ & $86.1\pm 1.8$ & $\pi^+n\rightarrow \pi^+n$ & $85.2\pm 1.8$ \\
$a^+-a^-$ & $\pi^+p\rightarrow \pi^+p$ & $-88.1\pm 1.8$ & 	$\pi^-n\rightarrow \pi^-n$  & $-89.0\pm 1.8$ \\
$-\sqrt{2}\,a^-$ & $\pi^-p\rightarrow \pi^0n$ & $-121.4\pm 1.6$ &  	$\pi^+n\rightarrow \pi^0 p$ & $-119.5\pm 1.6$\\
$a^+$ & $\pi^0p\rightarrow \pi^0p$ & $2.1\pm 3.1$ & 		$\pi^0n\rightarrow \pi^0 n$ &  $5.5\pm 3.1$\\\hline\hline
\end{tabular}
\caption{$\pi N$ scattering lengths for the physical channels in units of $10^{-3}\mpi^{-1}$.}
\label{table:physical_channels}
\end{table}

\section{Goldberger--Miyazawa--Oehme sum rule}
\label{sec:GMO}

The Goldberger--Miyazawa--Oehme (GMO) sum rule~\cite{GMO} relates the charged-pion--nucleon coupling constant $g_c^2/4\pi$ to ${a_{\pi^-p}-a_{\pi^+p}}$ and the integral over the cross sections $\sigma_{\pi^-p}^{\rm tot}(k)-\sigma_{\pi^+p}^{\rm tot}(k)$ measured in the laboratory frame for pion momenta $k$ ranging from zero to infinity
\begin{align}
 \frac{g_c^2}{4\pi}&=\bigg(\bigg(\frac{\mpp+\mn}{\mpi}\bigg)^2-1\bigg)\bigg\{\bigg(1+\frac{\mpi}{\mpp}\bigg)
\frac{\mpi}{4}(a_{\pi^-p}-a_{\pi^+ p})-\frac{\mpi^2}{2}J^-\bigg\},\notag\\
 J^-&=\frac{1}{4\pi^2}\int_0^\infty d k\frac{\sigma_{\pi^-p}^{\rm tot}(k)-\sigma_{\pi^+p}^{\rm tot}(k)}{\sqrt{\mpi^2+k^2}}.\label{GMO}
\end{align}
This result is derived by writing down dispersion relations for 
$\pi^\pm p\to \pi^\pm p$ for fixed $t$, assuming that the amplitudes are analytic functions of $s$ with a right-hand cut starting at $s_{\rm thr}=(\mpp+\mpi)^2$ and a left-hand cut starting at $s=(\mpp-\mpi)^2-t$. These dispersion relations are then evaluated at threshold. 
The scattering lengths enter as the $\pi N$ amplitude at threshold, while the coupling constant $g_c$ is related to the residue of the nucleon pole. Finally, one employs the optical theorem in the laboratory frame to replace the imaginary part of the amplitude by the total cross section.

\subsection{Isospin violation}
\label{subsec:IV}

There are two ways in which isospin violation affects the derivation of the GMO sum rule: mass effects and virtual photons. 

The proton--neutron mass difference enters through the intermediate neutron in the nucleon-pole diagram, which is already taken into account in~\eqref{GMO}. Additionally, the threshold for $\pi^-p\to\pi^0n$ lies below $s_{\rm thr}$, such that the right-hand cut for $\pi^-p$ already starts at $(\mn+\mpii)^2$. Thus, the total cross section for $\pi^- p$ diverges at threshold due to the lower threshold of $\pi^0n$. However, this divergence corresponds just to the right half of a principal-value integral:  the dispersion integral for the reaction $\pi^-p$ really starts at $s=(\mn+\mpii)^2$, and the resulting pole at $s_{\rm thr}$ can be taken care of in the usual way by the principal-value prescription. To estimate the remaining effect, one may use the fact that the imaginary part of the amplitude for $s < (\mpp + M_{\pi^-})^2$ can be well approximated by the imaginary part of the $\pi^-p$ scattering length due to the $\pi^0 n$ intermediate state~\cite{HKM}. In this way, we find a shift in $J^-$ by about $-0.005\,{\rm mb}$, which we will take into account in our uncertainty estimate for $J^-$ below. In conclusion, mass effects do not invalidate the GMO sum rule and the necessary modifications are quite well under control.

In contrast, we cannot write down the GMO sum rule in the presence of virtual photons, as e.g.~the nucleon pole is not separated any more from the $\gamma N$ cut. Therefore, \eqref{GMO} is only applicable if all ingredients are purified from virtual-photon effects to ensure that the analytic structure of the corresponding amplitude coincides with what was assumed in the original  derivation. For this reason, we will adopt the following point of view: we assume that the removal of electromagnetic effects in the cross sections using the Tromborg procedure~\cite{Tromborg} works sufficiently well that the resulting value for $J^-$ is compatible with the above analyticity assumptions. Moreover, we subtract virtual-photon effects in the scattering lengths based on~\cite{HKM}, but keep isospin-violating effects due to the nucleon and pion mass difference (we will dwell on this procedure in Sect.~\ref{subsec:resGMO}). In this way, our final result for $g_c$ consistently refers to the scenario where all particle masses are fixed at their physical value, but virtual photons are switched off.   

Finally, we comment on the definition of scattering lengths and coupling constants in the presence of electromagnetic interactions. 
Even in principle, the calculation of electromagnetic corrections is a scale-dependent procedure~\cite{GRS}, which, however, can only be systematically addressed if the underlying theory is known. Within an effective theory, ChPT in our case, a consistent treatment of electromagnetic corrections is possible, apart from the fact that the ambiguities in the separation of photon effects present in full QCD should be reflected in additional uncertainties in the LECs. To the best of our knowledge, the practical consequences of~\cite{GRS} for ChPT calculations have yet to be explored. However, the study of the linear $\sigma$ model in~\cite{GRS} suggests that such effects are not relevant at the level of accuracy at which the LECs can be usually pinned down. 
In addition, the definition of a scattering length for charged particles is a subtle matter~\cite{bethe49}, and recent attempts to define a strong proton--proton scattering length yield only scale-dependent quantities~\cite{Sauer,Kong98,Gegelia03,Pavon09}. An effect analogous to that discussed in these works
is also present in the calculation of the $\pi N$ scattering lengths, but, due to the perturbative nature of $\pi N$ dynamics, this effect is negligible as we will show in the following. Removing the Coulomb phase $\theta_{\rm C}(|\pp|)$, the behavior of the $\pi^-p$ scattering amplitude $T_{\pi^-p}$ at threshold is given by~\cite{LR00}
\beq
e^{-2i\alpha \theta_{\rm C}(|\pp|)}T_{\pi^-p}=\frac{B_1}{|\pp|}+B_2\log\frac{|\pp|}{\mu_H}+T_{\pi^-p}^{\rm thr}+\Order(|\pp|),
\label{eq:pimpamp}
\eeq
where 
\beq
4 \pi \xip a_{\pi^-p}=T_{\pi^-p}^{\rm thr},\quad B_1=4\pi^2\alpha\mpi a_{\pi^-p},\quad B_2=-8\pi\alpha\mpi\big(a_{\pi^-p}\big)^2.
\eeq
The scale ambiguity represented by the presence of $\log \mu_H$ is not induced by an ultraviolet divergence, but by a kinematic singularity at threshold. While the $1/|\pp|$ term (the leading approximation to the Gamow--Sommerfeld factor~\cite{Gamow28,Sommerfeld39}) can be unambiguously separated, the $\log |\pp|$ requires the choice of a scale in order to define the strong threshold amplitude $T_{\pi^-p}^{\rm thr}$. In~\eqref{eq:pimpamp} that scale has been chosen to be $\mu_H$. However, $B_2$ differs from zero only at two-loop level, i.e.\ it is suppressed by two chiral orders compared to the accuracy at which the isospin-breaking corrections~\cite{HKM} are known. Thus, choosing the mass of the $\rho$ meson, rather than the reduced mass $\mu_H$, shifts $a_{\pi^-p}$ by
\beq
\frac{B_2}{4\pi\xip}\log\frac{M_\rho}{\mu_H}=-2\alpha\mu_H \big(a_{\pi^-p}\big)^2\log\frac{M_\rho}{\mu_H}=-0.2\cdot 10^{-3}\mpi^{-1},
\eeq 
an effect fully in line with its two-loop estimate that can therefore safely be neglected.

One might worry that our definition of $g_c$ is not exactly what is measured in experiment, because, despite the application of electromagnetic corrections, the full range of virtual-photon effects is not captured by present-day analyses. For this reason, one could try to add a certain class of virtual-photon diagrams in order to obtain a quantity that corresponds better to the experimentally accessible one. In fact, this is quite a difficult enterprise: to extract the coupling constant, we need the amplitude at $s=\mn^2$, where threshold ambiguities as in the case of the scattering lengths do not occur. However, the cancellation of infrared divergences that is ensured at threshold by phase-space arguments no longer works, which makes the inclusion of bremsstrahlung inevitable. In order to estimate the size of such effects, one may in a first rough approximation consider only the leading bremsstrahlung contribution that involves logarithms of the detector resolution $E_{\rm max}$. In this naive approach---described in more detail in~\ref{app:gc}---we find a shift of about $0.07$ in $g_c^2/4\pi$ for $E_{\rm max}=10\,{\rm MeV}$, which is thus significantly below the accuracy we claim for our final result below. We conclude that, in order to address virtual-photon effects systematically, one is forced to perform the full radiative corrections for a given process, which is beyond the scope of this work.

\subsection{Evaluation of $J^-$}
\label{subsec:jminus}

The evaluation of $J^-$ has recently been discussed in great detail in~\cite{ELT} and~\cite{AMS}, hereafter referred to as ELT and AMS. The main difference between both analyses is that the former relies on phase-shift solutions to determine the cross sections, while the latter uses data directly. Both investigations apply the Tromborg procedure~\cite{Tromborg} to remove electromagnetic effects.\footnote{Above the energy range where the Tromborg corrections are available the effect due to the Coulomb barrier is accounted for following the potential-model calculation~\cite{Bugg}.} The quoted results
\begin{align}
J^-_{\rm ELT}&=(-1.083 \pm 0.032)\,{\rm mb},\qquad J^-_{\rm AMS}=(-1.060\pm 0.030)\,{\rm mb},
\end{align}
are consistent within the uncertainties and we take the discrepancy between the mean values as an indication of the final accuracy one can hope to achieve in an evaluation of $J^-$. In order to obtain an average value of $J^-$ that combines ELT and AMS, we now compare both analyses in the low-momentum ($k\leq 2\,{\rm GeV}/c$), the high-momentum ($2\,{\rm GeV}/c\leq k\leq 240\,{\rm GeV}/c$), and the Regge regime ($k\geq 240\,{\rm GeV}/c$) separately. In general, we employ the uncertainties quoted by ELT, whose error estimates tend to be more conservative than those of AMS. 

In the low-momentum region, we average the mean of the results for the SM95~\cite{ELT,SM95}, SM99~\cite{ELT,SM99}, and FA02~\cite{sainio,FA02} phase-shift solutions with AMS. In these determinations, the threshold region constitutes an additional source of uncertainty due to a lack of very low-energy data. Therefore, for the interval $(0\bis80)\,{\rm MeV}/c$ an interpolation between the cross sections at threshold
\beq
\Delta\sigma(0)=\sigma_{\pi^-p}^{\rm tot}(0)-\sigma_{\pi^+p}^{\rm tot}(0)=4\pi\Big((a_{\pi^-p})^2+(a_{\pi^-p}^{\rm cex})^2-(a_{\pi^+p})^2\Big)
\eeq 
and the available scattering data is needed. To estimate the impact of the scattering length on $J^-$, we consider the $S$-wave part of this interpolation, which changes by about $0.009\,{\rm mb}$ if one varies $\Delta\sigma(0)$ by $30\,\%$. In view of the fact that the difference between our value for $\Delta\sigma(0)$ and those of AMS and ELT lies below $20\,\%$, this should provide a conservative estimate of the additional uncertainties to be expected in the threshold region. Adding in quadrature this estimation, the uncertainty quoted by ELT, and the effect due to the lower $\pi^0 n$ threshold discussed in Sect.~\ref{subsec:IV}, yields the $\pm 0.023$ error given in Table~\ref{table:jminus}. In the high-momentum region we use the mean of AMS and ELT with the ELT error, while the contributions from the Regge regime above $240\,{\rm GeV}/c$ are determined as the average of the five models employed in AMS and ELT with an error chosen generously to encompass all models (cf.~Table~\ref{table:jminus}). In this way, we finally obtain for our average 
\beq
J^-=(-1.073\pm 0.034)\,{\rm mb}\label{jminus}.
\eeq

\begin{table}
\centering
\begin{tabular}{cccc}
\hline\hline
$k \ \text{in} \ {\rm GeV}/c$ &&&\\\hline
$0\bis2$ & SM95 \cite{ELT,SM95} & SM99 \cite{ELT,SM99} & FA02 \cite{sainio,FA02}  \\
 & $-1.302$ & $-1.314$ & $-1.3043$ \\
& AMS & average &\\
& $-1.3003$ & $-1.304\pm 0.023$ & \\\hline
$2\bis240$ & ELT \cite{ELT} & AMS \cite{AMS} & average \\
& $0.197$ & $0.2149$ & $0.206 \pm 0.024$\\\hline
$240\bis\infty$ & H\"ohler \cite{hoehler} & Donnachie--Landshoff \cite{Donnachie_Landshoff} & Gauron--Nicolescu \cite{Gauron_Nicolescu} \\
& $0.0222$ & $0.0294$ & $0.0244$ \\ 
& Regge94 \cite{Regge94} & Regge98 \cite{Regge98}& average\\
& $0.030$ & $0.018$ & $0.025\pm 0.007$\\
\hline\hline
\end{tabular}
\caption{Contributions to $J^-$ from different momentum regions in mb. For the detailed comparison, we use that the regions $(2\bis2.03)\,{\rm GeV}/c$ and $(240\bis350)\,{\rm GeV}/c$ yield a contribution of $0.0027\,{\rm mb}$ and $0.0043\,{\rm mb}$ to $J^-$, respectively~\cite{sainio}.}
\label{table:jminus}
\end{table}

\subsection{Results for the pion--nucleon coupling constant}
\label{subsec:resGMO}

Inspired by~\cite{ELT}, we adopt the following strategy to determine $a_{\pi^-p}-a_{\pi^+p}$. Writing
\beq
a_{\pi^-p}-a_{\pi^+p}=2a_{\pi^-p}-\big(a_{\pi^-p}+a_{\pi^-n}\big)+a_{\pi^-n}-a_{\pi^+p}=2a_{\pi^-p}-2\big(\tilde a^++\Delta\tilde a^+\big)+a_{\pi^-n}-a_{\pi^+p},
\label{rewrite}
\eeq
we can take
\beq
a_{\pi^-p}=(85.66\pm 0.14)\cdot 10^{-3}\mpi^{-1}
\label{apimp}
\eeq
directly from the level shift in $\pi H$, $\tilde a^+$ from~\eqref{atilde}, and 
\beq
a_{\pi^-n}-a_{\pi^+p}=\frac{e^2}{4\pi\xip}\Big\{f_2+2\mpi\big(2g^{\rm r}_6+g^{\rm r}_8\big)\Big\}
\label{diffa}
\eeq
from~\cite{HKM}. As the $g^{\rm r}_i$ cancel between $a_{\pi^-n}-a_{\pi^+p}$ and $2\Delta\tilde a^+$
  (cf.~\eqref{IV_piN} and \eqref{diffa}), 
such that only the rather well-determined LEC $f_2$ remains, this procedure is particularly stable with respect to unknown LECs. We find
\beq
a_{\pi^-p}-a_{\pi^+p}=(173.2\pm 1.6)\cdot 10^{-3}\mpi^{-1}.
\eeq
However, these scattering lengths still contain virtual-photon effects, such that we need to subtract the corresponding contribution (cf.~\ref{app:virt_phot})
\beq
a_{\pi^-p}-a_{\pi^+p}\Big|_{e^2}=(2.1\pm 1.8)\cdot 10^{-3}\mpi^{-1}.\label{virt_phot_scatt}
\eeq
The GMO sum rule~\eqref{GMO} with the input~\eqref{jminus},  \eqref{rewrite}, and \eqref{apimp} then
yields
\beq
\frac{g_c^2}{4\pi}=13.69\pm 0.12\pm0.15,\label{gc}
\eeq
where the first error gives the uncertainty due to the scattering lengths and the second that due to $J^-$. This value is in agreement with determinations from $NN$ ($g_c^2/4\pi=13.54\pm 0.05$~\cite{deSwart}) and $\pi N$ ($g_c^2/4\pi=13.75\pm 0.10$~\cite{FA02}, $g_c^2/4\pi=13.76\pm 0.01$~\cite{arndt06}) scattering data.
We stress that the errors quoted in~\cite{deSwart,FA02,arndt06} mainly reflect statistical uncertainties. The systematic subtleties associated with isospin violation that were discussed above in Sect.~\ref{subsec:IV} were not quantified in these previous studies.\footnote{In the nucleon--nucleon case electromagnetic corrections to the one-pion-exchange potential were calculated in~\cite{vanKolck97,Kaiser06}. The renormalization procedure chosen in~\cite{vanKolck97} implies that electromagnetic corrections to $g_c$ are small. However, it is unclear to us how these conventions should be translated to $\pi N$ scattering.}

\section{Conclusions}

\label{sec:conc}

We have presented a calculation of the $\pi^-d$ scattering length based on ChPT including full isospin-violating corrections in both the two- and three-body sectors. In the isospin-conserving part we included all contributions below the order 
at which an unknown $(N^\dagger N)^2\pi\pi$ counter term enters. The remaining cutoff dependence of our results is consistent with dimensional analysis of the counter term, and thus provides an estimate of the theoretical uncertainty of the calculation. We have also considered isospin-violating three-body contributions 
below the order $\Order(e^2 p)$ relative to the leading isospin-conserving operator.
Moreover, we considered several higher-order diagrams that could potentially be enhanced due to the dominance of double scattering in $a_{\pi^- d}$ or because of small scales associated with the deuteron binding energy. In the end, we find no significantly enhanced virtual-photon effects. Ostensibly enhanced contributions from momenta of order $\sqrt{\mpi\eps}$ vanish for both isovector and isoscalar $\pi N$ scattering, where the cancellation can be traced back to the Pauli principle and the orthogonality of deuteron and continuum wave functions, respectively. We conclude that a higher accuracy in both the two-body isospin-violating corrections and the three-body part of the $\pi^-d$ scattering length requires additional information on $\pi N$ and $\pi NN$ contact terms, which are not constrained by chiral symmetry. Finally, we presented the consequences of the calculation for the $\pi N$ scattering lengths and---after carefully revisiting the GMO sum rule in the presence of isospin violation---for the charged-pion--nucleon coupling constant.

\section*{Acknowledgments}

\begin{sloppypar}
We thank D.~Gotta, A.~Kudryavtsev, U.-G.~Mei\ss ner, A.~Rusetsky, M.~Sainio, and A.~W.~Thomas for useful discussions.
This research was supported by the DFG (SFB/TR 16, ``Subnuclear Structure of Matter''), DFG-RFBR grant (436 RUS 113/991/0-1), the Mercator Programme of the DFG, the Helmholtz Association through funds provided
to the virtual institute ``Spin and strong QCD'' (VH-VI-231) and the young investigator group ``Few-Nucleon Systems in Chiral Effective Field Theory'' (grant VH-NG-222), the Bonn-Cologne Graduate School of Physics and Astronomy, the DAAD,
the project ``Study of Strongly Interacting Matter''
(HadronPhysics2, grant No.~227431) under the 7th Framework Programme of the EU, 
the US Department of Energy (Office of Nuclear Physics, under contract No.~DE-FG02-93ER40756 with Ohio University), and the Federal Agency of Atomic Research of the Russian Federation (``Rosatom'').
Computing resources were provided by the JSC, J\"ulich, Germany.
\end{sloppypar}

\appendix

\renewcommand{\thefigure}{\arabic{figure}}
\renewcommand{\thetable}{\arabic{table}}

\section{Effective Lagrangians}
\label{app:lagr}

For the sake of completeness, we review here the effective Lagrangian for nucleons, pions, and virtual photons, as constructed in~\cite{GR02}. The following terms are needed in the context of the present study
\begin{align}
\Lagr_{\rm eff}&=\Lagr_\pi^{(p^{2})}+\Lagr_\pi^{(e^2)}+\Lagr_\pi^{(e^2p^2)}
+\Lagr_{\rm N}^{(p)}+\Lagr_{\rm N}^{(p^2)}+\Lagr_{\rm N}^{(p^3)}+\Lagr_{\rm N}^{(e^2)}+\Lagr_{\rm N}^{(e^2p)}+\Lagr_\gamma,\notag\\
\Lagr_\pi^{(p^2)}&+\Lagr_\pi^{(e^2)}+\Lagr_\gamma=\frac{F^2}{4}\langle d^\mu U^\dagger d_\mu U+\chi^\dagger U+U^\dagger \chi\rangle
+Z F^4\langle \mathcal{Q}U\mathcal{Q}U^\dagger\rangle-\frac{1}{4}F_{\mu\nu}F^{\mu\nu}-\frac{1}{2}\big(\partial_\mu A^\mu\big)^2,\notag\\
\Lagr_\pi^{(e^2p^2)}&=F^2 \Bigl\{\langle d^\mu U^\dagger d_\mu U \rangle \bigl(k_1\langle \mathcal{Q}^2 \rangle
+k_2 \langle \mathcal{Q}U\mathcal{Q}U^\dagger \rangle\bigr)
+k_4\langle d^\mu U^\dagger \mathcal{Q} U \rangle \langle d_\mu U \mathcal{Q} U^\dagger\rangle \notag\\
&+k_3\bigl(\langle d^\mu U^\dagger \mathcal{Q} U \rangle\langle d_\mu U^\dagger \mathcal{Q} U \rangle+
 \langle d^\mu U \mathcal{Q} U^\dagger\rangle\langle d_\mu U \mathcal{Q} U^\dagger\rangle\bigr)\Bigr\} ,\notag\\
\Lagr_{\rm N}^{(p)}&=\bar{\Psi}\Big\{i\slashed{D}-m+\frac{1}{2}g \slashed{u}\gamma_5\Big\}\Psi, \notag\\
\Lagr_{\rm N}^{(p^2)}&=\bar{\Psi}\Big\{c_1 \langle\chi_+\rangle -\frac{c_2}{4m^2}\langle u_\mu u_\nu\rangle D^\mu D^\nu + {\rm h.c.}
+\frac{c_3}{2}\langle u_\mu u^\mu\rangle+\frac{i}{4}c_4 \sigma^{\mu\nu}[u_\mu,u_\nu]+c_5 \hat{\chi}_+\Big\}\Psi\notag,\\
\Lagr_{\rm N}^{(p^3)}&=\frac{i}{2m}\bar{\Psi}\Big\{d_5[\chi_-,u_\mu]D^\mu\Big\}\Psi+{\rm h.c.},\notag\\
\Lagr_{\rm N}^{(e^2)}&=F^2\bar{\Psi}\Big\{f_{1} \langle \hat{Q}^2_+ - Q_-^2\rangle+ f_2\langle Q_+\rangle \hat{Q}_+\Big\}\Psi,\notag\\
\Lagr_{\rm N}^{(e^2p)}&=\frac{F^2}{2}\bar\Psi \Big\{g_1\langle Q_+^2-Q_-^2\rangle\gamma^\mu\gamma_5u_\mu+g_2\langle Q_+\rangle^2\gamma^\mu\gamma_5u_\mu\Big\} \Psi\notag\\
&+\frac{iF^2}{2m}\bar{\Psi}\Big\{g_6\langle Q_+\rangle \langle Q_-u_\mu\rangle D^\mu 
+g_{7} \langle Q_+ u_\mu\rangle Q_- D^\mu+g_{8} \langle Q_- u_\mu\rangle Q_+ D^\mu\Big\}\Psi+{\rm h.c.},
\end{align}
where $\langle A\rangle$ denotes the trace of a matrix $A$, $\hat{A}=A-\langle A\rangle/2$ its traceless part,
$
\bar{\Psi}(\mathcal{O}+{\rm h.c.})\Psi\equiv \bar{\Psi}\mathcal{O}\Psi+{\rm h.c.}
$
for an operator $\mathcal{O}$ and
\begin{align}
 d_\mu U&=\partial_\mu U-i A_\mu[\mathcal{Q},U], \quad \chi=2 B\,\text{diag}(\muu,\md), \quad U=u^2, \quad \mathcal{Q}=\frac{e}{3}\,\text{diag}(2,-1),\notag\\
F_{\mu\nu}&=\partial_\mu A_\nu-\partial_\nu A_\mu,\quad  Q=e\,\text{diag}(1,0),\quad Q_\pm=\frac{1}{2}(u Q u^\dagger \pm u^\dagger Q u),\notag\\
D_\mu &=\partial_\mu + \Gamma_\mu, \quad  \Gamma_\mu= \frac{1}{2}\Big(u^\dagger(\partial_\mu-i Q A_\mu)u+u(\partial_\mu-i Q A_\mu)u^\dagger\Big),
\quad \chi_\pm=u^\dagger \chi u^\dagger\pm u \chi^\dagger u,\notag\\
u_\mu&= i\Big(u^\dagger(\partial_\mu-i Q A_\mu)u-u(\partial_\mu-i Q A_\mu)u^\dagger\Big),\quad [D_\mu,u_\nu]=\partial_\mu u_\nu+[\Gamma_\mu,u_\nu].
\end{align}
$\Psi=(p,n)^T$ contains the nucleon fields and the matrix $U$ 
collects the pion fields.
$F$, $g$, and $m$ are the pion decay constant, the axial charge, and the mass of the nucleon in the chiral limit, respectively. The renormalized LECs are denoted by a superscript r.

\section{Photon diagrams in chiral effective theory}

\subsection{Single scattering with photon exchange}
\label{app:single_ChET}

To begin with we discuss the case that the $\pi NN \rightarrow \pi NN$ operator contains only the (nominally) leading isoscalar contribution. Writing the (isoscalar) threshold  $\pi N$ amplitude as $T^+=4\pi\xip a^+$, we have
\begin{align}
i \M^{(d_6)+(d_8)}&=2\int \frac{\diff^4 k}{(2 \pi)^4} \int \frac{\diff^3 \qq\, \diff^3 \qq'}{(2 \pi)^3} \Psi^\dagger(\qq') i 2 T^+
\frac{i}{(M_\pi v + k)^2 - M_\pi^2 + i \eta} (-ie\, 2 M_\pi) \Big(i \frac{e}{2}\Big) \frac{i}{\kk^2} \notag\\ 
&\times\bigg\{ i G_s\Big(\qq'-\frac{\kk}{2},\qq-\frac{\kk}{2};-\eps - k_0,-\kk\Big)  + i G_s\Big(\qq'-\frac{\kk}{2},-\qq+\frac{\kk}{2};-\eps - k_0,-\kk\Big)\bigg\}
 \Psi(\qq),
\label{MChiET}
\end{align}
where $v=(1,\mathbf{0})$ and we have already used the fact that $k_0 \sim {\bf k}^2/M_\pi \ll {\bf k}$ and therefore can be neglected in the photon propagator. 
The factor of 2 multiplying $T^+$ is present because the pion can interact with either the neutron or the proton, while the overall factor of 2 includes the time-reversed diagram. The factor $e/2$ in the first line occurs because we include only one of the two possible interactions of the nucleons inside the nucleus with the photon. The other is accounted for via the exchange term, which is represented by the second Green's function.  In that portion we have replaced $\pp$ by $-\pp$ as compared to the direct piece of the amplitude. More specifically, we can rewrite $\M$ as
\beq
\M=\langle \Psi|Q G(E)\frac{1}{2}(1-P_{12}) T_{\pi^-N}|\Psi\rangle\label{matrix_elem},
\eeq
where 
\beq
T_{\pi^-N}=\left(T^+\unity^{(1)}+T^-\tau_3^{(1)}\right)\otimes\unity^{(2)}+(1\leftrightarrow 2),\quad  Q=\frac{e}{2}\left(\unity^{(1)}+\tau_3^{(1)}\right)\otimes\unity^{(2)}+(1\leftrightarrow 2)\label{pi_N_operator},
\eeq
the superscript referring to nucleon 1 and 2, respectively. In~\eqref{matrix_elem} $G(E)=1/(E+i\eta-H)$
denotes the Green's function describing the propagation of the $NN$ pair from the $\pi NN$ interaction to the photon coupling, and the projector $(1-P_{12})/2$ has been introduced to impose the Pauli principle ($P_{12}$ interchanges nucleons 1 and 2). Note that the operators in round brackets in~\eqref{pi_N_operator} are implicitly understood to be accompanied by ``shift operators'' $\mathcal{S}^{(i)}$ indicating the momentum shift induced by the pion--nucleon or photon--nucleon interaction, which has to be taken into account when the symmetry properties of the individual terms are analyzed.  
Inserting~\eqref{pi_N_operator} into~\eqref{matrix_elem}, we obtain two distinct contributions (all terms with a single $\tau_3^{(i)}$ involve $\langle T=0|\tau_3^{(i)}|T=0 \rangle=0$ and may thus be dropped). First, the isoscalar piece reads
\beq
\langle \Psi|\frac{e}{2}G(1-P_{12})2T^+|\Psi\rangle\equiv \langle \Psi|\frac{e}{2}(G_s+\tilde{G}_s)2T^+|\Psi\rangle,
\label{Teq0formal}
\eeq
and since the isospin wave function of the isospin-zero state is already antisymmetric under particle exchange, this isoscalar ``direct-minus-exchange'' $G(1-P_{12})\equiv G_{\rm D}-G_{\rm E}$ prescription (in~\eqref{Teq0formal} $\tilde{G}_s$ denotes the outcome for the exchange part) produces the sum of two $G_s$ terms in~\eqref{MChiET}. Second, the ``direct-minus-exchange'' contribution to $\M$ that results from the isovector part of the pion's interaction with the $NN$ system discussed below only contributes for odd partial waves due to the Pauli principle. This can already be seen from the isospin structure: ${\tau^{(i)}_3|T=0\rangle\propto|T=1,T_3=0\rangle}$, and since spin is conserved, the Pauli principle in the form
${(-1)^{L+S+T}=-1}$ requires an odd partial wave.

Because of the presence of the deuteron pole, the expression~\eqref{MChiET} is infrared divergent (not just ``would-be infrared divergent''). It blows up because we have neglected effects due to the atomic-binding energy $B_{\rm at}$: if we included these, it would be regulated at scale $\sqrt{M_\pi B_{\rm at}} \sim \alpha M_\pi$. However, the physics associated with this momentum scale was already included in the atomic-physics calculation. In that calculation, which is done using non-relativistic effective field theory (NREFT), we have an expression for these effects that corresponds to a structureless deuteron. Using the fact that the normalization of any deuteron wave function equals 1, that expression can be written as
\begin{align}
i \M_{\rm IR}&=2\int \frac{\diff^4 k}{(2 \pi)^4} \int \diff^3  \qq\, \diff^3 \qq' \Psi^\dagger(\qq') i 2 T^+ \Psi(\qq')
\frac{i}{-k_0 - \kk^2/2\md + i \eta} \frac{i}{k_0 - \kk^2/2\mpi + i \eta} \notag\\
&\qquad\times\Psi^\dagger(\qq) (-ie)(ie) \frac{i}{\kk^2} \Psi(\qq)\label{IR_amplitude}.
\end{align}
From the difference of~\eqref{MChiET} and \eqref{IR_amplitude}, we can obtain an expression that is safe in the infrared and includes only the effects not already accounted for in the NREFT computation
\begin{align}
\M_{\rm IR~safe}^{(d_6)+(d_8)}&=-2e^2 T^+ \int \frac{\diff^3 \kk}{(2 \pi)^3}\frac{1}{\kk^2} \int \frac{\diff^3 \qq\,\diff^3 \qq'}{(2 \pi)^3}
 \Psi^\dagger(\qq')
\bigg\{\int \frac{\diff k_0}{2 \pi}\bigg[ i G_s\Big(\qq'-\frac{\kk}{2},\qq-\frac{\kk}{2};-\eps - k_0,-\kk\Big)  \notag\\
&\!\!+ i G_s\Big(\qq'-\frac{\kk}{2},-\qq+\frac{\kk}{2};-\eps - k_0,-\kk\Big)\bigg]
\frac{2 M_\pi}{(M_\pi v + k)^2 - M_\pi^2 + i \eta} 
- \frac{2(2\pi)^3\Psi(\qq')\Psi^\dagger(\qq)}{-\kk^2/2 \mu_D + i \eta} \bigg\} \Psi(\qq).
\end{align}
Now, since we are prepared to ignore $\pi \pi NN$ cuts (they only lead to higher-order effects), we drop the ``backward-going pion'' contribution. The evaluation of the $k_0$ integral in the first term can then be done by picking up the pion pole. This yields
\begin{align}
\M_{\rm IR~safe}^{(d_6)+(d_8)}&=-2e^2  T^+ \int \frac{\diff^3 \kk}{(2 \pi)^3} \frac{1}{\kk^2} \int \frac{\diff^3 \qq\,\diff^3\qq'}{(2 \pi)^3} 
\Psi^\dagger(\qq') \bigg\{G_s\Big(\qq'-\frac{\kk}{2},\qq-\frac{\kk}{2};-\eps - \frac{\kk^2}{2\mpi},-\kk\Big)  \notag\\
&+ G_s\Big(\qq'-\frac{\kk}{2},-\qq+\frac{\kk}{2};-\eps -  \frac{\kk^2}{2\mpi},-\kk\Big)
-  \frac{2(2\pi)^3\Psi(\qq')\Psi^\dagger(\qq)}{-\kk^2/2 \mu_D + i \eta}\bigg\} \Psi(\qq),
\end{align}
where we have neglected terms that are higher order in $\kk$, and so replaced $\omega_\mathbf{k}$ by $\mpi$ and the kinetic energy of the pion by its non-relativistic form. From this result, we can now read off~\eqref{NN_isoscalar} via
\beq
a^{(d_6)+(d_8)}_{NN}=\frac{\M_{\rm IR~safe}^{(d_6)+(d_8)}}{4\pi\xid}.
\eeq

Finally, the isovector $P$-wave part produces a contribution corresponding to
\begin{align}
a^{(d_6)+(d_8)}_{T=1}&=-\frac{8\pi\alpha\xip a^-}{(2\pi)^6\xid} \int \frac{\diff^3 \kk}{\kk^2} \int \diff^3 \qq\, \diff^3\qq' \Psi^\dagger(\qq') \notag\\
&\times\bigg\{G_v\Big(\qq'-\frac{\kk}{2},\qq-\frac{\kk}{2};-\eps - \frac{\kk^2}{2\mpi},-\kk\Big)
- G_v\Big(\qq'-\frac{\kk}{2},-\qq+\frac{\kk}{2};-\eps - \frac{\kk^2}{2\mpi},-\kk\Big)\bigg\} \Psi(\qq),
\end{align}
where the isovector Green's function $G_v$ is defined analogously to $G_s$ with the deuteron pole removed and $T^{\rm np}_{s,NN}\rightarrow T^{\rm np}_{v,NN}$ (the relative sign is due to the symmetry of the $|T=1,T_3=0\rangle$ state). Although we do not consider isovector $NN$ interactions, it is a valuable check of the calculation that the free part of the Green's function reproduces the expressions of Sect.~\ref{sec:structureless}.

\subsection{Double scattering with photon exchange}
\label{app:double_ChET}

We now generalize the previous discussion to a $\pi NN$ kernel given by double $\pi N$ scattering. Let us first consider the case where the photon is exchanged after the $\pi NN$ interaction. Two distinct processes contribute to the $\pi NN$ interaction: the exchange of a $\pi^-$ corresponds to double elastic $\pi^-N$ scattering, while an intermediate $\pi^0$ requires two charge-exchange reactions. The amplitudes can be written as
\beq
\M^{\pi^-}=\langle \Psi|Q G\frac{1}{2}(1-P_{12}) T_{\pi^-N}T_{\pi^-N}|\Psi\rangle,\quad
\M^{\pi^0}=\langle \Psi|Q G\frac{1}{2}(1-P_{12}) T_{\pi^0N\rightarrow \pi^- N} T_{\pi^-N\rightarrow \pi^0N}|\Psi\rangle\label{double_ampl},
\eeq
where 
\begin{align}
T_{\pi^-N\rightarrow \pi^0N}&=(-\sqrt{2}\,T^-)\tau_-^{(1)}\otimes \unity^{(2)}+(1\leftrightarrow 2),\notag\\
T_{\pi^0N\rightarrow \pi^-N}&=(-\sqrt{2}\,T^-)\tau_+^{(1)}\otimes \unity^{(2)}+(1\leftrightarrow 2),
\end{align}
with raising and lowering operators
\beq
\tau_\pm^{(i)}=\frac{1}{2}\left(\tau_1^{(i)}\pm i\tau_2^{(i)}\right),
\eeq
and $T_{\pi^-N}$ and $Q$ are defined in~\eqref{pi_N_operator}. Again, the momentum shifts are understood to be taken into account implicitly. The evaluation of $\M^{\pi^-}$ proceeds in close analogy to~\eqref{matrix_elem}, 
once the contributions for which both $\pi N$ interactions happen to the same nucleon are excluded. Neglecting isoscalar $\pi N$ scattering terms (which are of higher order), we obtain
\beq
\M^{\pi^-}=-2(T^-)^2\langle \Psi|\frac{e}{2}G\frac{1}{2}(1-P_{12})|\Psi\rangle
=-2(T^-)^2\langle \Psi|\frac{e}{2}(G_s+\tilde{G}_s)|\Psi\rangle.
\eeq
The expression for $\M^{\pi^0}$ is somewhat more involved, the crucial observation being that the states ${\tau_+^{(2)}\tau_-^{(1)}|T=0\rangle}$,  ${\tau_+^{(1)}\tau_-^{(2)}|T=0\rangle}$, $(1+\tau_3^{(1)})|T=0\rangle$, and $(1+\tau_3^{(2)})|T=0\rangle$, are actually a superposition of $|T=0,T_3=0\rangle$ and $|T=1,T_3=0\rangle$, such that both even and odd partial waves contribute.
We find that the former give rise to
\beq
\M^{\pi^0}_{T=0}=-(-\sqrt{2}\,T^-)^2\langle \Psi|(G_s+\tilde{G}_s) \frac{e}{2}|\Psi\rangle,
\eeq
while the latter lead to
\beq
\M^{\pi^0}_{T=1}=-(-\sqrt{2}\,T^-)^2\langle \Psi|(G_v-\tilde{G}_v) \frac{e}{2}|\Psi\rangle.
\eeq
Therefore, the analog of~\eqref{MChiET} becomes
\begin{align}
i \M^{(d_9)+(d_{10})}&=2\int \frac{\diff^4 k}{(2 \pi)^4} \int \frac{\diff^4 l}{(2 \pi)^4} \int \frac{\diff^3 \qq\, \diff^3 \qq'}{(2 \pi)^3} \Psi^\dagger(\qq') 
(-2) (i T^-)^2 (-ie 2 M_\pi) \Big(i \frac{e}{2}\Big) \frac{i}{\kk^2}\notag\\ 
&\times\bigg\{i G_s\Big(\qq'-\frac{\kk}{2},\qq-\frac{\kk}{2}+\lbf;-\eps - k_0,-\kk\Big)  
+ i G_s\Big(\qq'-\frac{\kk}{2},-\qq+\frac{\kk}{2}-\lbf;-\eps - k_0,-\kk\Big)\bigg\}\notag\\
& \times\frac{i}{-v\cdot l+i\eta}\frac{i}{(M_\pi v + l)^2 - M_\pi^2 + i \eta}\frac{i}{(M_\pi v + k)^2 - M_\pi^2 + i \eta} \Psi(\qq)\notag\\
&+2\int \frac{\diff^4 k}{(2 \pi)^4} \int \frac{\diff^4 l}{(2 \pi)^4} \int \frac{\diff^3 \qq\, \diff^3 \qq'}{(2 \pi)^3} \Psi^\dagger(\qq')
 (-1) (-i\sqrt{2}\, T^-)^2 (-ie 2 M_\pi) \Big(i \frac{e}{2}\Big) \frac{i}{\kk^2}\notag\\ 
&\times\bigg\{i G_s\Big(\qq'-\frac{\kk}{2},\qq-\frac{\kk}{2}+\lbf;-\eps - k_0,-\kk\Big)  
+ i G_s\Big(\qq'-\frac{\kk}{2},-\qq+\frac{\kk}{2}-\lbf;-\eps - k_0,-\kk\Big)
\bigg\}\notag\\
&\times\frac{i}{-v\cdot l+i\eta}\frac{i}{(M_\pi v + l)^2 - M_\pi^2 + i \eta}\frac{i}{(M_\pi v + k)^2 - M_\pi^2 + i \eta} \Psi(\qq),
\end{align}
where again the overall factor of $2$ accounts for the time-reversed diagram. Performing the $l^0$ and $k^0$ integrations yields
\begin{align}
\M^{(d_9)+(d_{10})}&=2e^2(T^-)^2\int \frac{\diff^3 \kk}{(2 \pi)^3} \int \frac{\diff^3 \lbf}{(2 \pi)^3} \int \frac{\diff^3 \qq\, \diff^3 \qq'}{(2 \pi)^3} \Psi^\dagger(\qq')  \frac{1}{\kk^2\lbf^2}\notag\\ 
&\times\bigg\{2G_s\Big(\qq'-\frac{\kk}{2},\qq-\frac{\kk}{2}+\lbf;-\eps - \frac{\kk^2}{2\mpi},-\kk\Big)\notag\\
&+
2G_s\Big(\qq'-\frac{\kk}{2},-\qq+\frac{\kk}{2}-\lbf;-\eps - \frac{\kk^2}{2\mpi},-\kk\Big) \bigg\}\Psi(\qq).
\end{align}
To generalize~\eqref{IR_amplitude} we must replace the isoscalar two-body contribution to the $\pi^-d$ scattering length $2T^+$ by the double-scattering analog 
\beq
-4 (T^-)^2\int \frac{\diff^3 \pp\,\diff^3\qq}{(2 \pi)^3} \Psi^\dagger(\pp-\qq)\frac{1}{\qq^2}\Psi(\pp),
\eeq
such that the infrared-safe amplitude is given by
\begin{align}
\M_{\rm IR~safe}^{(d_9)+(d_{10})}&=2e^2  (T^-)^2 \int \frac{\diff^3 \kk}{(2 \pi)^3}\int \frac{\diff^3 \lbf}{(2 \pi)^3} \frac{1}{\kk^2\lbf^2} \int \frac{\diff^3 \qq\,\diff^3 \qq'}{(2 \pi)^3} \notag\\
&\times\Psi^\dagger(\qq')\bigg\{2G_s\Big(\qq'-\frac{\kk}{2},\qq-\frac{\kk}{2}+\lbf;-\eps - \frac{\kk^2}{2\mpi},-\kk\Big)\notag\\
&+2G_s\Big(\qq'-\frac{\kk}{2},-\qq+\frac{\kk}{2}-\lbf;-\eps - \frac{\kk^2}{2\mpi},-\kk\Big)
- \frac{4(2\pi)^3\Psi(\qq')\Psi^\dagger(\qq-\lbf)}{-\kk^2/2 \mu_D + i \eta}\bigg\} \Psi(\qq),
\end{align}
which finally proves~\eqref{NN_isoscalar_double}.

\section{Photon diagrams in heavy-pion EFT}

\subsection{Electromagnetic radius of the deuteron}
\label{app:hpiEFT_radius}

As the correction due to the electromagnetic radius of the deuteron is already included in the experimental value for the level shift, we should subtract the corresponding NREFT contribution as well (in addition to~\eqref{IR_amplitude})
\beq
i \M_{\rm IR}^{\rm radius}=2\int \frac{\diff^4 k}{(2 \pi)^4} i 2 T^+ 
\frac{i}{-k_0 - \kk^2/2\md + i \eta} \frac{i}{k_0-B_{\rm at} - \kk^2/2\mpi + i \eta} (-ie)(ie) \frac{i}{\kk^2}\Big(-\frac{1}{6}\kk^2\big\langle r_d^2\big\rangle\Big),
\eeq
where $B_{\rm at}=\alpha^2\mu_D/2$ includes the atomic binding energy and $\big\langle r_d^2\big\rangle=(8\gamma^2)^{-1}$ for asymptotic wave functions. Performing the $k^0$ integral yields
\beq
\M_{\rm IR}^{\rm radius}=-2e^22T^+\int \frac{\diff^3 \kk}{(2 \pi)^3} \frac{1}{\kk^2} \frac{1}{-B_{\rm at} - \kk^2/2\mu_D} 
\Big(-\frac{1}{6}\kk^2\big\langle r_d^2\big\rangle\Big).
\eeq
In fact, this expression is linearly divergent: the theory is only valid in the hadronic-atom regime. In dimensional regularization we may drop a scaleless integral, which leads to
\beq
\M_{\rm IR}^{\rm radius}=\frac{e^2T^+}{6\pi}\big\langle r_d^2\big\rangle2\mu_D\sqrt{2\mu_D B_{\rm at}},
\eeq  
and subtracting this contribution from~\eqref{FF_single} gives
\beq
a^{(d_6)+(d_8)}_{\rm FF,\,radius}=-\frac{2}{3}\alpha\Big(1+\log 4+\frac{\alpha\mpi}{4\gamma\xid}\Big)\frac{\mpi}{\gamma}\frac{\xip}{\xid^2}a^+.
\eeq
Therefore, the additional shift due to the deuteron charge radius $\sim\alpha\mpi/4\gamma\xid=0.005$ is certainly negligible.

\subsection{Double scattering with photon exchange}
\label{app:hpiEFT_double}

We start from~\eqref{NN_isoscalar_double1} in the form
\begin{align}
a^{(d_9)+(d_{10})}_{\rm FF}&=\frac{32\pi^2\alpha(\xip a^-)^2}{\xid} \int \frac{\diff^3 \kk}{(2 \pi)^3}\int \frac{\diff^3 \lbf}{(2 \pi)^3} \frac{1}{\kk^2\lbf^2}  4\frac{F(\kk)F(\kk-2\lbf) - F(2\lbf)}{-\kk^2/2 \mu_D}\notag\\
&=-\frac{256\pi^2\alpha\mpi(\xip a^-)^2}{\xid^2}\int\frac{\diff^3\kk}{\kk^4}\frac{\gamma}{\pi^2}\Big(I(|\kk|)\frac{4\gamma}{|\kk|}\arctan\frac{|\kk|}{4\gamma}-I(0)\Big)\label{start_double_hpiEFT},
\end{align}
where
\beq
I(|\kk|)=\int\frac{\diff^3\lbf}{(2\pi)^3}\frac{\diff^3\pp}{(2\pi)^3}\frac{1}{\lbf^2}\frac{1}{(\pp-\kk/2+\lbf)^2+\gamma^2}\frac{1}{\pp^2+\gamma^2}.
\eeq
Introducing Feynman parameters in the standard way, this becomes
\begin{align}
I(|\kk|)&=\frac{1}{16\pi}\int_0^1\frac{\diff x}{(1-x)^3}\int_0^{1-x}\diff y\int\frac{\diff^3\lbf}{(2\pi)^3}\bigg\{\frac{y}{1-x}\Big(1-\frac{y}{1-x}\Big)\bigg(\frac{\kk}{2}-\lbf\bigg)^2+\frac{x}{1-x}\lbf^2+\gamma^2\bigg\}^{-3/2}\notag\\
&=\frac{1}{16\pi}\int_0^1\frac{\diff x}{\sqrt{1-x}}\int_0^1\frac{\diff y}{(y(1-y)(1-x)+x)^{3/2}}\int\frac{\diff^3\lbf}{(2\pi)^3}\frac{1}{\big(\lbf^2+h(1-h)\frac{\kk^2}{4}+\tilde\gamma^2\big)^{3/2}},
\end{align}
with
\beq
h=\frac{y(1-y)(1-x)}{y(1-y)(1-x)+x},\qquad \tilde \gamma^2=\frac{1-x}{y(1-y)(1-x)+x}\gamma^2,
\eeq
and hence
\beq
I(|\kk|)-I(0)=-\frac{1}{64\pi^3}\int^1_0\frac{\diff x}{\sqrt{1-x}}\int^1_0\frac{\diff y}{(y(1-y)(1-x)+x)^{3/2}}\log\bigg(1+h(1-h)\frac{\kk^2}{4\tilde\gamma^2}\bigg).
\eeq
The corresponding contribution to~\eqref{start_double_hpiEFT} is given by
\begin{align}
 &\frac{16\alpha\mpi(\xip a^-)^2}{\pi^2\xid^2}\int_0^\infty\frac{\diff z}{z^3}\arctan\frac{z}{2}
\int^1_0\frac{\diff x}{\sqrt{1-x}}\int^1_0\frac{\diff y\log\left(1+\frac{x y(1-y)}{y(1-y)(1-x)+x}z^2\right)}{(y(1-y)(1-x)+x)^{3/2}}\notag\\
&=\frac{8\alpha\mpi(\xip a^-)^2}{\pi^2\xid^2}\Bigg\{\frac{\pi^2}{2}+\int^1_0\frac{\diff x}{\sqrt{1-x}}\int^1_0\frac{\diff y}{(y(1-y)(1-x)+x)^{3/2}}
\int_0^\infty\frac{\diff z}{z^2}\bigg(\frac{2}{z}\arctan\frac{z}{2}-1\bigg)\notag\\
&\qquad\times\log\bigg(1+\frac{x y(1-y)}{y(1-y)(1-x)+x}z^2\bigg)\Bigg\}\notag\\
&=\frac{8\alpha\mpi(\xip a^-)^2}{\pi^2\xid^2}\Big(\frac{\pi^2}{2}-2.916\Big),
\end{align}
such that we are left with
\begin{align}
a^{(d_9)+(d_{10})}_{\rm FF}&=\frac{8\alpha\mpi(\xip a^-)^2}{\pi^2\xid^2}\bigg(\frac{\pi^2}{2}-2.916\bigg)
-\frac{256\alpha\mpi\gamma(\xip a^-)^2}{\xid^2}I(0)\int\frac{\diff^3\kk}{\kk^4}\bigg\{\frac{4\gamma}{|\kk|}\arctan\frac{|\kk|}{4\gamma}-1\bigg\}\notag\\
&=\frac{8\alpha\mpi(\xip a^-)^2}{\pi^2\xid^2}\bigg(\frac{\pi^2}{2}-2.916\bigg)
+\frac{64\pi^2\alpha\mpi(\xip a^-)^2}{\xid^2}I(0)\label{interm_result}.
\end{align}
Now, $I(0)$ does not converge: the reason is that the double-scattering contribution to the $\pi^-d$ scattering length itself diverges in heavy-pion EFT and we need a counterterm $D$ to renormalize this diagram. Demanding that the physical double-scattering contribution in the static approximation $a^{\rm static}$ be reproduced, we have
\beq
a^{\rm static}=\frac{1}{4\pi\xid}\bigg\{D-\frac{8}{\pi}(\xip a^-)^2\bigg\langle\frac{1}{\qq^2}\bigg\rangle\bigg\}
=\frac{1}{4\pi\xid}\Big\{D-512\pi^3\gamma(\xip a^-)^2I(0)\Big\}.
\eeq
However, the same counterterm will contribute to $(d_9)$ and $(d_{10})$ as well and needs to be added to~\eqref{interm_result}. Eliminating $D$ therein in favor of $I(0)$ and $a^{\rm static}$, $I(0)$ cancels and we finally obtain~\eqref{FF_double}.

\section{Subtraction of virtual-photon effects} 

\subsection{$\pi N$ scattering lengths}
\label{app:virt_phot}

The isospin-violating corrections to the scattering lengths calculated in~\cite{HKM} involve both contributions due to the pion mass difference $\sim e^2Z$, and those due to virtual photons $\sim e^2$, where $Z=(\mpi^2-\mpii^2)/2e^2\Fpi^2=0.81$. (The nucleon mass difference does not enter in these channels at third chiral order.) Retaining only the $e^2$ part, we obtain the following correction~\cite{HKM}
\beq
a_{\pi^-p}-a_{\pi^+p}\Big|_{e^2}=-\frac{\mpi}{2\pi\xip}\bigg\{\frac{e^2\ga^2}{16\pi^2\Fpi^2}\bigg(1+4\log 2+3\log\frac{\mpi^2}{\mu^2}\bigg)-2e^2\Big(\tilde g^{\rm r}_6+\tilde g^{\rm r}_8-\frac{5}{9\Fpi^2}\tilde k_1^{\rm r}\Big)\bigg\},
\eeq
where $\tilde g_i^{\rm r}$ and $\tilde k_i^{\rm r}$ denote the $e^2$ piece of $g_i^{\rm r}$ and $k_i^{\rm r}$, respectively. The relation between both sets of LECs can be established by means of their $\beta$-functions $\sigma_i$ and $\eta_i$. It is convenient to define scale-independent LECs $\bar g_i$ and $\bar k_i$ by 
\beq
k_i^{\rm r}=\frac{\sigma_i}{16\pi^2}\bigg(\bar k_i+\log\frac{\mpi}{\mu}\bigg),\quad g_i^{\rm r}=\frac{\eta_i}{16\pi^2\Fpi^2}\bigg(\bar g_i+\log\frac{\mpi}{\mu}\bigg),
\eeq
such that 
\beq
\tilde k_i^{\rm r}=\frac{\sigma_i|_{Z=0}}{16\pi^2}\bigg(\bar k_i+\log\frac{\mpi}{\mu}\bigg),\quad \tilde g_i^{\rm r}=\frac{\eta_i|_{Z=0}}{16\pi^2\Fpi^2}\bigg(\bar g_i+\log\frac{\mpi}{\mu}\bigg),
\eeq
which finally leads to
\beq
\tilde k_i^{\rm r}=\frac{\sigma_i|_{Z=0}}{\sigma_i} k_i^{\rm r},\quad \tilde g_i^{\rm r}=\frac{\eta_i|_{Z=0}}{\eta_i} g_i^{\rm r}.
\eeq
Estimating the LECs as in~\cite{HKM} yields the result quoted in~\eqref{virt_phot_scatt}.

\subsection{$\pi NN$ coupling constant}
\label{app:gc}

\begin{figure}
\begin{center}
\includegraphics[width=\linewidth]{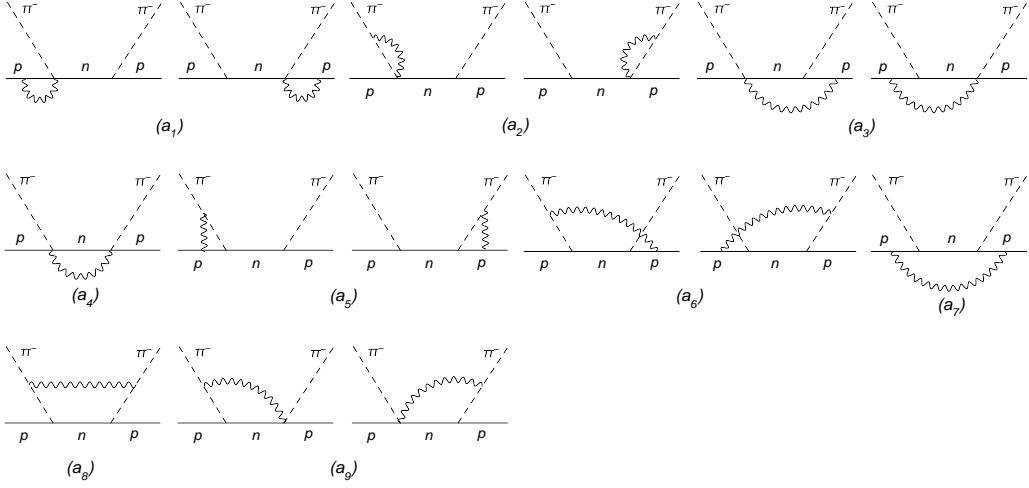}
\end{center}
\caption{Virtual-photon corrections to the nucleon-pole diagram in $\pi^-p$ scattering at $\Order(p^3)$~\cite{HKMlong}.}
\label{fig:axial}
\end{figure}

The full set of virtual-photon corrections to the nucleon-pole diagram in $\pi^-p$ scattering at third chiral order is depicted in Fig.~\ref{fig:axial}. Based on the discussion of these diagrams in~\cite{HKMlong}, the shift of $g_c$ due to virtual photons can be read off from the residue of the scattering amplitude at $s=\mn^2$. To obtain an idea how large these effects are, we consider here the diagrams $(a_1)$, $(a_2)$, and $(a_5)$, which is motivated by the expectation that it ought to be possible to absorb this subset of diagrams into a simple redefinition of $g_c$. And indeed, we find that these diagrams, together with the pertinent contact terms and the wave-function renormalization, represent a scale-independent quantity: all ultraviolet divergences cancel between loops and contact terms. In a strict chiral expansion, we find that $(a_1)$, $(a_2)$, and $(a_5)$ yield (post-renormalization) a
shift in $g_c^2$ of
\beq
\frac{\Delta g_c^2}{4\pi}=\frac{e^2\ga^2\mpp^2}{4\pi\Fpi^2}\bigg\{\frac{2\Fpi^2}{\ga}\big(\tilde g_1^{\rm r}+\tilde g_2^{\rm r}\big)-\frac{20}{9}\tilde k_1^{\rm r}-\frac{1}{8\pi^2}\bigg(3+\log\frac{\mpi^2}{\mu^2}\bigg)+\frac{1}{4\pi^2}\log\frac{\mpi^2}{4E_{\rm max}^2}\bigg\}.
\eeq
The last term is present because infrared divergences only cancel at threshold. To remove these singularities (regulated by a finite photon mass $m_\gamma$ in the actual calculation of the diagrams), we use the leading, logarithmically enhanced part of the bremsstrahlung calculated in~\cite{HKMlong}, which effectively eliminates $m_\gamma$ in favor of twice the detector resolution $E_{\rm max}$. Numerically, this amounts to
\beq
\frac{\Delta g_c^2}{4\pi}=0.07\pm 0.03 \pm 0.04,
\eeq
where we take $E_{\rm max}=10\,{\rm MeV}$ and the errors are, respectively, due to the LECs and a variation of $E_{\rm max}$ by a factor of $2$. Dropping the term due to bremsstrahlung, the shift in $g_c^2/4\pi$ is reduced to $-0.01\pm 0.03$. We stress that these estimates can by no means replace a full analysis of radiative corrections, but we think they can be taken as indicative of the size of virtual-photon effects.


\begin{thebibliography}{999}
\biboptions{sort&compress}


\bibitem{JOB}
  V.~Baru, C.~Hanhart, M.~Hoferichter, B.~Kubis, A.~Nogga, and D.~R.~Phillips,
  Phys.\ Lett.\  B {\bf 694} (2011) 473
  [arXiv:1003.4444 [nucl-th]].

\bibitem{Beane:Lattice}
  S.~R.~Beane, K.~Orginos, and M.~J.~Savage,
  Int.\ J.\ Mod.\ Phys.\  E {\bf 17} (2008) 1157
  [arXiv:0805.4629 [hep-lat]].

\bibitem{Torok}
  A.~Torok {\it et al.},
  Phys.\ Rev.\  D {\bf 81} (2010) 074506
  [arXiv:0907.1913 [hep-lat]].

\bibitem{Weinberg66}
  S.~Weinberg,
  Phys.\ Rev.\ Lett.\  {\bf 17} (1966) 616.

\bibitem{CGL00}
  G.~Colangelo, J.~Gasser, and H.~Leutwyler,
  Phys.\ Lett.\  B {\bf 488} (2000) 261
  [arXiv:hep-ph/0007112].

\bibitem{CGL01}
  G.~Colangelo, J.~Gasser, and H.~Leutwyler,
  Phys.\ Rev.\ Lett.\  {\bf 86} (2001) 5008
  [arXiv:hep-ph/0103063].

\bibitem{BKM93}
  V.~Bernard, N.~Kaiser, and U.-G.~Mei{\ss}ner,
  Phys.\ Lett.\  B {\bf 309} (1993) 421
  [arXiv:hep-ph/9304275].

\bibitem{Bernstein1}
  A.~M.~Bernstein, 
  Phys.\ Lett.\  B {\bf 442} (1998) 20
  [arXiv:hep-ph/9810376].

\bibitem{Bernstein2}
  A.~M.~Bernstein, M.~W.~Ahmed, S.~Stave, Y.~K.~Wu, and H.~R.~Weller,
  Ann.\ Rev.\ Nucl.\ Part.\ Sci.\  {\bf 59} (2009) 115
  [arXiv:0902.3650 [nucl-ex]].

\bibitem{FM00}
  N.~Fettes and U.-G.~Mei{\ss}ner,
  Nucl.\ Phys.\  A {\bf 676} (2000) 311 
  [arXiv:hep-ph/0002162].

\bibitem{sigmaterm}
  J.~Gasser,  H.~Leutwyler, and M.~E.~Sainio,
  Phys.\ Lett.\  B {\bf 253} (1991) 252.

\bibitem{GMO}
  M.~L.~Goldberger, H.~Miyazawa, and R.~Oehme,
  Phys.\ Rev.\  {\bf 99} (1955) 986.

\bibitem{ELT}
  T.~E.~O.~Ericson, B.~Loiseau, and A.~W.~Thomas,
  Phys.\ Rev.\  C {\bf 66} (2002) 014005
  [arXiv:hep-ph/0009312].

\bibitem{AMS}
  V.~V.~Abaev, P.~Mets\"a, and M.~E.~Sainio,
  Eur.\ Phys.\ J.\  A {\bf 32} (2007) 321
  [arXiv:0704.3167 [hep-ph]].

\bibitem{hadatoms}
  J.~Gasser, V.~E.~Lyubovitskij, and A.~Rusetsky,
  Phys.\ Rept.\  {\bf 456} (2008) 167
  [arXiv:0711.3522 [hep-ph]].

\bibitem{Gottawidth}
  D.~Gotta {\it et al.},
  Lect.\ Notes Phys.\  {\bf 745} (2008) 165.

\bibitem{LR00}
  V.~E.~Lyubovitskij and A.~Rusetsky,
  Phys.\ Lett.\  B {\bf 494} (2000) 9
  [arXiv:hep-ph/0009206].

\bibitem{vac_pol}
  D.~Eiras and J.~Soto,
  Phys.\ Lett.\  B {\bf 491} (2000) 101
  [arXiv:hep-ph/0005066].

\bibitem{zemp}
  P.~Zemp, 
  PhD thesis, University of Bern (2004).

\bibitem{Panofsky}
  J.~Spuller {\it et al.},
  Phys.\ Lett.\  B {\bf 67} (1977) 479.

\bibitem{mrr1}
  U.-G.~Mei{\ss}ner, U.~Raha, and A.~Rusetsky,
  Eur.\ Phys.\ J.\  C {\bf 41} (2005) 213
  [arXiv:nucl-th/0501073]. 

\bibitem{Gottapid}
  T.~Strauch {\it et al.},
  Eur.\ Phys.\ J.\  A {\bf 47} (2011) 88
  [arXiv:1011.2415 [nucl-ex]];
%
  EPJ Web Conf.\  {\bf 3} (2010) 03006
  [arXiv:1002.4277 [nucl-ex]].

\bibitem{pid_channel}
  V.~L.~Highland, M.~Salomon, M.~D.~Hasinoff, E.~Mazzucato, D.~F.~Measday, J.~M.~Poutissou, and T.~Suzuki,
  Nucl.\ Phys.\  A {\bf 365} (1981) 333.


\bibitem{HKM}
  M.~Hoferichter, B.~Kubis, and U.-G.~Mei{\ss}ner,
  Phys.\ Lett.\  B {\bf 678} (2009) 65
  [arXiv:0903.3890 [hep-ph]].

\bibitem{HKMlong}
  M.~Hoferichter, B.~Kubis, and U.-G.~Mei{\ss}ner,
  Nucl.\ Phys.\  A {\bf 833} (2010) 18
  [arXiv:0909.4390 [hep-ph]].

\bibitem{GR02}
  J.~Gasser, M.~A.~Ivanov, E.~Lipartia, M.~Moj\v zi\v s, and A.~Rusetsky,
  Eur.\ Phys.\ J.\  C {\bf 26} (2002) 13
  [arXiv:hep-ph/0206068].

\bibitem{GL82}
  J.~Gasser and H.~Leutwyler,
  Phys.\ Rept.\  {\bf 87} (1982) 77.

\bibitem{Beane06}
  S.~R.~Beane, K.~Orginos, and M.~J.~Savage,
  Nucl.\ Phys.\  B {\bf 768} (2007) 38
  [arXiv:hep-lat/0605014].


\bibitem{Blum10}
  T.~Blum, R.~Zhou, T.~Doi, M.~Hayakawa, T.~Izubuchi, S.~Uno, and N.~Yamada,
  Phys.\ Rev.\  D {\bf 82} (2010) 094508
  [arXiv:1006.1311 [hep-lat]].

\bibitem{filin}
  A.~Filin, V.~Baru, E.~Epelbaum, J.~Haidenbauer, C.~Hanhart, A.~E.~Kudryavtsev, and U.-G.~Mei{\ss}ner,
  Phys.\ Lett.\  B {\bf 681} (2009) 423
  [arXiv:0907.4671 [nucl-th]].

\bibitem{FLAG}
  G.~Colangelo {\it et al.},
  Eur.\ Phys.\ J.\  C {\bf 71} (2011) 1695
  [arXiv:1011.4408 [hep-lat]].

\bibitem{BP97}
  J.~Bijnens and J.~Prades,
  Nucl.\ Phys.\  B {\bf 490} (1997) 239
  [arXiv:hep-ph/9610360].

\bibitem{AM04}
  B.~Ananthanarayan and B.~Moussallam,
  JHEP {\bf 0406} (2004) 047
  [arXiv:hep-ph/0405206].

\bibitem{piK_IV}
  B.~Kubis and U.-G.~Mei{\ss}ner,
  Nucl.\ Phys.\  A {\bf 699} (2002) 709
  [arXiv:hep-ph/0107199].

\bibitem{mrr2}
  U.-G.~Mei{\ss}ner, U.~Raha, and A.~Rusetsky,
  Phys.\ Lett.\  B {\bf 639} (2006) 478
  [arXiv:nucl-th/0512035].

\bibitem{weinberg}
  S.~Weinberg,
  Phys.\ Lett.\  B {\bf 295} (1992) 114
  [arXiv:hep-ph/9209257].

\bibitem{beane98}
  S.~R.~Beane, V.~Bernard, T.~S.~H.~Lee, and U.-G.~Mei{\ss}ner,
  Phys.\ Rev.\  C {\bf 57} (1998) 424
  [arXiv:nucl-th/9708035].

\bibitem{beane}
  S.~R.~Beane, V.~Bernard, E.~Epelbaum, U.-G.~Mei{\ss}ner, and D.~R.~Phillips,
  Nucl.\ Phys.\  A {\bf 720} (2003) 399
  [arXiv:hep-ph/0206219].

\bibitem{Liebig} 
  S.~Liebig, V.~Baru, F.~Ballout, C.~Hanhart, and A.~Nogga,
  Eur.\ Phys.\ J.\  A {\bf 47} (2011) 69
  [arXiv:1003.3826 [nucl-th]].

\bibitem{recoil}
  V.~Baru, C.~Hanhart, A.~E.~Kudryavtsev, and U.-G.~Mei{\ss}ner,
  Phys.\ Lett.\  B {\bf 589} (2004)  118
  [arXiv:nucl-th/0402027].

\bibitem{recoil_BER}
  V.~Baru, E.~Epelbaum, and A.~Rusetsky,
  Eur.\ Phys.\ J.\  A {\bf 42} (2009) 111
  [arXiv:0905.4249 [nucl-th]].

\bibitem{disp}
  V.~Lensky, V.~Baru, J.~Haidenbauer, C.~Hanhart, A.~E.~Kudryavtsev, and U.-G.~Mei\ss ner,
  Phys.\ Lett.\  B {\bf 648} (2007) 46
  [arXiv:nucl-th/0608042].

\bibitem{delta}
  V.~Baru, J.~Haidenbauer, C.~Hanhart, A.~E.~Kudryavtsev, V.~Lensky, and U.-G.~Mei\ss ner,
  Phys.\ Lett.\  B {\bf 659} (2008) 184
  [arXiv:0706.4023 [nucl-th]].

\bibitem{BK}
  V.~V.~Baru and A.~E.~Kudryavtsev,
  Phys.\ Atom.\ Nucl.\  {\bf 60} (1997) 1475
  [Yad.\ Fiz.\  {\bf 60} (1997) 1620].

\bibitem{Platter}
  L.~Platter and D.~R.~Phillips,
  Phys.\ Lett.\ B {\bf 641} (2006) 164
  [arXiv:nucl-th/0605024].

\bibitem{PavonValderrama}
  M.~Pavon Valderrama and E.~R.~Arriola,
  arXiv:nucl-th/0605078.

\bibitem{Nogga}
  A.~Nogga and C.~Hanhart,
  Phys.\ Lett.\ B {\bf 634} (2006) 210
  [arXiv:nucl-th/0511011].

\bibitem{Friar:2003yv}
  J.~L.~Friar, U.~van Kolck, G.~L.~Payne, and S.~A.~Coon,
  Phys.\ Rev.\  C {\bf 68} (2003) 024003
  [arXiv:nucl-th/0303058].

\bibitem{Becher:1999he}
  T.~Becher and H.~Leutwyler,
  Eur.\ Phys.\ J.\  C {\bf 9} (1999) 643
  [arXiv:hep-ph/9901384].

\bibitem{Bernard:1991rt}
  V.~Bernard, N.~Kaiser, J.~Gasser, and U.-G.~Mei{\ss}ner,
  Phys.\ Lett.\ B {\bf 268} (1991) 291.

\bibitem{kolyb}
  V.~M.~Kolybasov and V.~G.~Ksenzov,
  Zh.\ Eksp.\ Teor.\ Fiz.\  {\bf 71} (1976) 13. 

\bibitem{Faeldt}
  G.~F\"aldt,
  Phys.\ Scripta {\bf 16} (1977) 81.

\bibitem{gammad}
  V.~Lensky, V.~Baru, J.~Haidenbauer, C.~Hanhart, A.~E.~Kudryavtsev, and U.-G.~Mei{\ss}ner,
  Eur.\ Phys.\ J.\  A {\bf 26} (2005) 107
  [arXiv:nucl-th/0505039].

\bibitem{BKM_NPA615}
  V.~Bernard, N.~Kaiser, and U.-G. Mei{\ss}ner, 
  Nucl.\ Phys.\  A {\bf 615} (1997) 483
  [arXiv:hep-ph/9611253].

\bibitem{epelbaum}
  H.~Krebs, E.~Epelbaum, and U.-G.~Mei{\ss}ner,
  Eur.\ Phys.\ J.\  A {\bf 32} (2007) 127
  [arXiv:nucl-th/0703087].

\bibitem{MRR_Kd}
  U.-G.~Mei{\ss}ner, U.~Raha, and A.~Rusetsky,
  Eur.\ Phys.\ J.\  C {\bf 47} (2006) 473
  [arXiv:nucl-th/0603029].

\bibitem{Kamalov}
  S.~S.~Kamalov, E.~Oset, and A.~Ramos,
  Nucl.\ Phys.\  A {\bf 690} (2001) 494
  [arXiv:nucl-th/0010054].

\bibitem{Kolybasov72}
  V.~M.~Kolybasov and A.~E.~Kudryavtsev,
  Nucl.\ Phys.\  B {\bf 41} (1972) 510.

\bibitem{Brueckner53}
  K.~A.~Brueckner,
  Phys.\ Rev.\  {\bf 89} (1953) 834.

\bibitem{MSS}
  V.~Baru, E.~Epelbaum, C.~Hanhart, M.~Hoferichter, A.~E.~Kudryavtsev, and D.~R.~Phillips, in preparation. 

\bibitem{NNLO}
  E.~Epelbaum, W.~Gl\"{o}ckle, and U.-G.~Mei{\ss}ner,
  Nucl.\ Phys.\  A {\bf 747} (2005) 362
  [arXiv:nucl-th/0405048].

\bibitem{AV18}
  R.~B.~Wiringa, V.~G.~J.~Stoks, and R.~Schiavilla,
  Phys.\ Rev.\  C {\bf 51} (1995) 38
  [arXiv:nucl-th/9408016].

\bibitem{CDBonn}
  R.~Machleidt,
  Phys.\ Rev.\  C {\bf 63} (2001) 024001
  [arXiv:nucl-th/0006014].

\bibitem{vK99}  U.~van Kolck,
  Nucl.\ Phys.\  A {\bf 645}  (1999) 273
  [arXiv:nucl-th/9808007].   

\bibitem{Ka98A}
  D.~B.~Kaplan, M.~J.~Savage, and M.~B.~Wise,
  Phys.\ Lett.\  B {\bf 424}  (1998) 390
  [arXiv:nucl-th/9801034].

\bibitem{Ka98B}
  D.~B.~Kaplan, M.~J.~Savage, and M.~B.~Wise,
  Nucl.\ Phys.\  B {\bf 534} (1998) 329
  [arXiv:nucl-th/9802075].

\bibitem{Ge98}
  J.~Gegelia,
  Phys.\ Lett.\  B {\bf 429}  (1998) 227.

\bibitem{Bi99}
  M.~C.~Birse, J.~A.~McGovern, and K.~G.~Richardson,
  Phys.\ Lett.\  B {\bf 464} (1999) 169
  [arXiv:hep-ph/9807302].

\bibitem{PRS}
  D.~R.~Phillips, G.~Rupak, and M.~J.~Savage,
  Phys.\ Lett.\  B {\bf 473} (2000) 209
  [arXiv:nucl-th/9908054].

\bibitem{pp2dpi}
  V.~Lensky, V.~Baru, J.~Haidenbauer, C.~Hanhart, A.~E.~Kudryavtsev, and U.-G.~Mei\ss ner,
  Eur.\ Phys.\ J.\  A {\bf 27} (2006) 37
  [arXiv:nucl-th/0511054].

\bibitem{HKM_CD09}
  M.~Hoferichter, B.~Kubis, and U.-G.~Mei\ss ner,
  PoS {\bf CD09} (2009) 014
  [arXiv:0910.0736 [hep-ph]].

\bibitem{thomas_sigma}
  R.~D.~Young and A.~W.~Thomas,
  Phys.\ Rev.\  D {\bf 81} (2010) 014503
  [arXiv:0901.3310 [hep-lat]].

\bibitem{Durr}
  S.~D\"urr {\it et al.},
  PoS {\bf LATTICE2010} (2010) 102
  [arXiv:1012.1208 [hep-lat]].

\bibitem{BL01}
  T.~Becher and H.~Leutwyler,
  JHEP {\bf 0106} (2001) 017 
  [arXiv:hep-ph/0103263].

\bibitem{BM99}
  P.~B{\"u}ttiker and U.-G.~Mei{\ss}ner,
  Nucl.\ Phys.\  A {\bf 668} (2000) 97 
  [arXiv:hep-ph/9908247].

\bibitem{f1}
  J.~Gasser, M.~A.~Ivanov, E.~Lipartia, M.~Moj\v zi\v s, and A.~Rusetsky,
  Eur.\ Phys.\ J.\  C {\bf 26} (2002) 13
  [arXiv:hep-ph/0206068].

\bibitem{f1b}
  N.~Fettes and U.-G.~Mei{\ss}ner,
  Phys.\ Rev.\  C {\bf 63} (2001) 045201
  [arXiv:hep-ph/0008181].

\bibitem{Tromborg}
  B.~Tromborg, S.~Waldenstr\o m, and I.~\O verb\o,
  Phys.\ Rev.\  D {\bf 15} (1977) 725.

\bibitem{GRS}
  J.~Gasser, A.~Rusetsky, and I.~Scimemi,
  Eur.\ Phys.\ J.\  C {\bf 32} (2003) 97
  [arXiv:hep-ph/0305260].

\bibitem{bethe49}
  H.~A.~Bethe,
  Phys.\ Rev.\  {\bf 76} (1949) 38.

\bibitem{Sauer}
  P.~U.~Sauer,
  Phys.\ Rev.\ Lett.\  {\bf 32} (1974) 626.

\bibitem{Kong98}
  X.~Kong and F.~Ravndal,
  Phys.\ Lett.\  B {\bf 450} (1999) 320
  [arXiv:nucl-th/9811076].

\bibitem{Gegelia03}
  J.~Gegelia,
  Eur.\ Phys.\ J.\  A {\bf 19} (2004) 355
  [arXiv:nucl-th/0310012].

\bibitem{Pavon09}
  M.~Pavon Valderrama and E.~Ruiz Arriola,
  Phys.\ Rev.\  C {\bf 80} (2009) 024001
  [arXiv:0904.1120 [nucl-th]].

\bibitem{Gamow28}
  G.~Gamow,
  Z.\ Phys.\  {\bf 51} (1928) 204.

\bibitem{Sommerfeld39}
  A.~Sommerfeld, {\em Atombau und Spektrallinien}, Volume 2,
  Vieweg, Braunschweig (1939).

\bibitem{Bugg}
  D.~V.~Bugg and A.~A.~Carter,
  Phys.\ Lett.\  B {\bf 48} (1974) 67.

\bibitem{SM95}
  R.~A.~Arndt, I.~I.~Strakovsky, R.~L.~Workman, and M.~M.~Pavan,
  Phys.\ Rev.\  C {\bf 52} (1995) 2120
  [arXiv:nucl-th/9505040];
Scattering Interactive Dial-Up (SAID), VPI, Blacksburg,
  The VPI/GWU $\pi N$ solution SM95 (1995); http://gwdac.phys.gwu.edu/analysis/pin\_analysis.html.
 
\bibitem{SM99}
  M.~M.~Pavan, R.~A.~Arndt, I.~I.~Strakovsky, and R.~L.~Workman,
  PiN Newslett.\  {\bf 15} (1999) 171
  [Phys.\ Scripta {\bf T87} (2000) 65]
  [arXiv:nucl-th/9910040];
Scattering Interactive Dial-Up (SAID), VPI, Blacksburg,
  The VPI/GWU $\pi N$ solution SM99 (1999); http://gwdac.phys.gwu.edu/analysis/pin\_analysis.html.
 

\bibitem{sainio}
  M.~Sainio, private communication.

\bibitem{FA02}
  R.~A.~Arndt, W.~J.~Briscoe, I.~I.~Strakovsky, R.~L.~Workman, and M.~M.~Pavan,
  Phys.\ Rev.\  C {\bf 69} (2004) 035213
  [arXiv:nucl-th/0311089].

\bibitem{hoehler}
  G.~H\"ohler, {\em Pion--Nucleon Scattering}, Landolt--B\"ornstein, Volume~I/9b2, Springer, Berlin (1983).

\bibitem{Donnachie_Landshoff}
  A.~Donnachie and P.~V.~Landshoff,
  Phys.\ Lett.\  B {\bf 296} (1992) 227
  [arXiv:hep-ph/9209205].

\bibitem{Gauron_Nicolescu}
  P.~Gauron and B.~Nicolescu,
  Phys.\ Lett.\  B {\bf 486} (2000) 71
  [arXiv:hep-ph/0004066].

\bibitem{Regge94}
  L.~Montanet {\it et al.}  [Particle Data Group],
  Phys.\ Rev.\  D {\bf 50} (1994) 
  1335.

\bibitem{Regge98}
  C.~Caso {\it et al.}  [Particle Data Group],
  Eur.\ Phys.\ J.\  C {\bf 3} (1998) 
  205.

\newpage

\bibitem{deSwart}
  J.~J.~de Swart, M.~C.~M.~Rentmeester, and R.~G.~E.~Timmermans,
  PiN Newslett.\  {\bf 13} (1997) 96
  [arXiv:nucl-th/9802084].

\bibitem{arndt06}
  R.~A.~Arndt, W.~J.~Briscoe, I.~I.~Strakovsky, and R.~L.~Workman,
  Phys.\ Rev.\  C {\bf 74} (2006) 045205 
  [arXiv:nucl-th/0605082].

\bibitem{vanKolck97}
  U.~van Kolck, M.~C.~M.~Rentmeester, J.~L.~Friar, J.~T.~Goldman, and J.~J.~de Swart,
  Phys.\ Rev.\ Lett.\  {\bf 80} (1998) 4386
  [arXiv:nucl-th/9710067].

\bibitem{Kaiser06}
  N.~Kaiser,
  Phys.\ Rev.\  C {\bf 73} (2006) 044001
  [arXiv:nucl-th/0601099].


\end{thebibliography}
\end{document}